\documentclass[fleqn,usenatbib]{mnras}

\usepackage{newtxtext,newtxmath}
\usepackage{url}
\usepackage{float}

\usepackage[T1]{fontenc}

\DeclareRobustCommand{\VAN}[3]{#2}
\let\VANthebibliography\thebibliography
\def\thebibliography{\DeclareRobustCommand{\VAN}[3]{##3}\VANthebibliography}

\usepackage{graphicx}	
\usepackage{amsmath}	
\usepackage{booktabs}
\usepackage{adjustbox}
\usepackage{xcolor}

\title[Radio-selected FRB host candidates]{Low-frequency-selected Fast Radio Burst Host Galaxy Candidates}

\author[Y.-Z. Sun]{
Y.-Z. Sun,$^{1}$\thanks{E-mail: ys400@le.ac.uk}
R. L. C. Starling,$^{1}$
R. A. J. Eyles-Ferris,$^{1}$
A. Rowlinson,$^{2,3}$
R. A. M. J. Wijers,$^{2,3}$
N. R Tanvir$^{1}$
\\
$^{1}$School of Physics and Astronomy, University of Leicester, University Road, Leicester, LE1 7RH, UK\\
$^{2}$Anton Pannekoek Institute for Astronomy, University of Amsterdam, Science Park 904, 1098 XH, Amsterdam, The Netherlands\\
$^{3}$ASTRON, the Netherlands Institute for Radio Astronomy, Oude Hoogeveensedijk 4, 7991 PD, Dwingeloo, The Netherlands\\
}

\date{Accepted XXX. Received YYY; in original form ZZZ}

\pubyear{\the\year{}}

\begin{document}
\label{firstpage}
\pagerange{\pageref{firstpage}--\pageref{lastpage}}
\maketitle

\begin{abstract}
We present a pilot study on the host galaxy environments of CHIME/FRBs by cross-matching baseband-localised events with the LOFAR Two-metre Sky Survey Data Release 2 (LoTSS DR2) at 144 MHz. Unlike traditional methods reliant on optical imaging, our radio-based selection allows for the identification of dust-obscured or optically faint star-forming galaxies. Of the 140 CHIME FRBs considered, 33 lie within the LoTSS DR2 footprint, and 16 show potential radio counterparts. Through multi-wavelength analysis, spectral energy distribution (SED) fitting, and redshift constraints from the Macquart relation, we identify two secure and one tentative host candidates, all consistent with active star formation. However, their H$\alpha$-derived star formation rates appear underestimated, likely due to significant dust attenuation, as suggested by infrared colours and compact optical morphologies. Our results highlight the value of low-frequency radio data in complementing optical host searches and demonstrate the feasibility of host identification even in the absence of optical confirmation. With forthcoming data from LoTSS DR3 and the full CHIME/FRB baseband release, this method offers a promising path toward statistically robust studies of FRB host galaxies and their environments.
\end{abstract}




\section{Introduction}

Fast radio bursts (FRBs) are a class of millisecond-duration radio transients of extragalactic origin, as inferred from their large dispersion measures (DMs) along the line of sight. These events are extremely powerful, with typical bursts releasing $\sim10^{38}$–$10^{40}$ erg in just a few milliseconds, making FRBs among the most luminous radio phenomena in the Universe. The observed frequency range of FRBs spans widely, from about 110 MHz up to beyond 5.2 GHz \citep{2021ApJ...911L...3Pleunis, 2018ApJ...863....2Gajjar}. Over the past decade, large-scale surveys such as the Canadian Hydrogen Intensity Mapping Experiment (CHIME; \citealt{2018ApJ...863...48CHIME}), the Commensal Radio Astronomy FAST Survey (CRAFTS; \citealt{2021ApJ...909L...8Niu}), and the Commensal Real-time ASKAP Fast Transients Survey (CRAFT; \citealt{2019PASA...36....9James}) have detected more than thousand FRBs. 

The first FRB to be discovered, FRB 20010724A, was detected by \citet{2007Sci...318..777Lorimer} as a single bright pulse. The extreme luminosities and short durations of FRBs point to compact object progenitors, most notably magnetically powered neutron stars (magnetars). This connection is strongly supported by the detection of FRB 20200428A, an FRB-like burst from the Galactic magnetar SGR 1935+2154 \citep{2020Natur.587...59Bochenek}, supporting the idea that at least a fraction of FRBs are powered by magnetars. However, the localisation of FRB 20200120E to a globular cluster in M81 \citep{2022Natur.602..585Kirsten} challenges this paradigm, as such environments lack recent star formation and massive stars. Instead, FRB~20200120E is thought to originate from a newly formed neutron star within an old stellar population, produced through alternative channels such as the accretion-induced collapse of a white dwarf or the merger of compact objects. These contrasting cases suggest that multiple formation pathways, linked to both young and old stellar populations, may contribute to the observed FRB population. Roughly 2.6 per cent of FRBs have been confirmed to repeat, although statistical models suggest that the true repeating fraction may be higher, due to limitations in observational cadence and sensitivity \citep{2016Natur.531..202Spitler,2023ApJ...947...83Chime,2024NatAs...8..337Kirsten,2024MNRAS.52711158Yamasaki}.

Beyond their origin, FRBs also serve as powerful astrophysical probes. Their large DMs trace the integrated column of free electrons, offering unique insights into the ionised medium along the line of sight -- including the Milky Way, its halo, and the intergalactic medium (IGM). Notably, \citet{2020Natur.581..391Macquart} demonstrated that by using localised FRBs to estimate the average density of diffuse intergalactic gas, the long-standing problem of the Universe’s `missing baryons' can be resolved, underscoring the potential of FRBs as valuable cosmological probes. 

Despite their exceptional brightness and short durations in the radio band, confirmed counterparts of extragalactic FRBs at other wavelengths are extremely rare.
A number of multi-wavelength follow-up campaigns have targets FRBs, searching for associated emission in the optical, X-ray, and gamma-ray bands, either contemporaneously with or shortly after radio bursts, but have resulted in non-detections. 
For example, \citet{2017ApJ...846...80Scholz} conducted simultaneous radio and X-ray observations of the repeating FRB 20121102A, detecting multiple radio bursts without any associated X-ray signal. Similarly, \citet{2021A&A...653A.119Nunez} used the Las Cumbres Observatory global telescope network to search for optical transients associated with eight well-localised FRBs, but found no counterparts. 
Only two tentative associations between FRBs and other high-energy transients have been reported. One links FRB 20190425A to the gravitational-wave event GW190425, a binary neutron star merger detected by LIGO/Virgo, with a chance alignment probability of only 0.52 percent \citep{2023NatAs...7..579Moroianu}. However, this association has since been rebutted and is no longer considered viable \citep{2024ApJ...971L...5Maga}. Another connects FRB 20171209A to the host galaxy of GRB 110715A, with a matching redshift \citep{2020ApJ...894L..22Wang}. While intriguing, both cases lack firm confirmation and highlight the need for larger samples.

The physical origins of FRBs remain one of the most fundamental open questions in modern astrophysics. Due to the extremely short durations of the bursts and the lack of persistent multi-wavelength counterparts, directly identifying progenitor systems remains observationally unfeasible. As a result, studying the host galaxies of FRBs has become a crucial approach to constraining their nature. Host properties, such as star-formation rate (SFR), stellar mass, morphology, and environment, can provide indirect but valuable clues about the types of systems capable of producing FRBs.

One statistical approach involves investigating whether FRBs preferentially occur in galaxies with high SFRs or high stellar masses \citep{2017Natur.541...58Chatterjee,2021ApJ...917...75Mannings,2022AJ....163...69Bhandari,2023ApJ...954...80Gordon,2024Natur.635...61Sharma,2025arXiv250215566Loudas}. While not a definitive test of progenitor type, such correlations can offer insight into the typical ages and evolutionary histories of FRB-producing populations. A trend with SFR would favour progenitors linked to young, massive stars and their remnants (e.g. magnetars formed in core-collapse supernovae), whereas a stronger dependence on stellar mass would suggest origins tied to older stellar populations or dynamical channels, such as neutron star mergers or accretion-induced collapse. For a comprehensive review of potential progenitor models, see \citet{2023RvMP...95c5005Zhang}.

The diversity of FRB host galaxies has become increasingly apparent through case studies of individual localised sources. For example, the first repeating FRB 20121102A, is hosted by a low-metallicity, star-forming dwarf galaxy \citep{2017ApJ...834L...7Tendulkar}, while FRB 20180924B and FRB 20190523A reside in massive, quiescent galaxies with little or no current star formation \citep{2019Sci...365..565Bannister,2019Natur.572..352Ravi}. By contrast, the repeating FRB~20190711A is associated with a moderately star-forming disk galaxy \citep{2020ApJ...903..152Heintz}. These examples indicate that FRBs can occur in a wide range of galactic environments.

This diversity has been confirmed by recent statistical studies. \citet{2023ApJ...954...80Gordon} analysed 23 FRB host galaxies and found a diversity of properties, with most residing in star-forming environments but some hosted by more quiescent galaxies. Similarly, \citet{2025arXiv250408038Horowicz} argued that the FRB rate scales with both stellar mass and star formation, resembling trends seen in Type Ia supernovae. This view is supported by \citet{2025arXiv250215566Loudas}, who found that while many FRBs occur in star-forming galaxies, a non-negligible fraction are located in galaxies with relatively low star formation activity. Collectively, these findings suggest that FRBs may originate from multiple progenitor channels, and host galaxy studies will remain essential to disentangling their origins.

To date, over 100 FRB host galaxies have been identified \citep{2025arXiv250215566Loudas}, the vast majority of which were recognised via optical observations. Interestingly, in four cases FRBs have been found to coincide with a persistent radio source (PRS) -- a compact, long-lived radio counterpart, much smaller than the host galaxy and thus indicative of the FRB’s local environment -- that remains visible long after the burst, providing an additional clue to their origins \citep{2017ApJ...834L...8Marcote,2017Natur.541...58Chatterjee,2022Natur.606..873Niu,2023ApJ...958L..19Bhandari,2024arXiv241201478Bruni,2024Natur.632.1014Bruni,2024ApJ...961...44Dong}. While PRSs such as that at the location of FRB 20121102 have provided valuable insight into the local environments of some repeaters, most FRBs lack any detectable persistent emission. VLBI observations confirmed that the bursts and the PRS of FRB 20121102A are co-located within tens of parsecs, and follow-up studies disfavour an AGN origin, instead supporting interpretations such as a plerionic nebula \citep{2017ApJ...834L...8Marcote,2023ApJ...958..185Chen,2025arXiv250623861Bhardwaj}. Long-term monitoring further shows that the PRS is compact, stable, and not directly correlated with burst activity. 
Another example is FRB 20190520B, the source is localized to a star-forming dwarf galaxy at $z=0.241$ \citep{2022Natur.606..873Niu,2025ApJ...982..203Chen} and hosts a compact PRS constrained to a size of $<9$~pc, co-located with the FRB to within $\sim80$~pc \citep{2023ApJ...958L..19Bhandari}, and exhibits a flat GHz spectrum with significant long-term variability \citep{2023ApJ...959...89Zhang}. Recent low-frequency observations reveal a spectral break attributed to absorption, pointing to a dense and magnetized local environment. These properties distinguish the PRS from star-formation-powered radio emission and make FRB~20190520B a benchmark system for compact FRB-associated PRSs \citep{2025ApJ...995...51Balasubramanian}.
Furthermore, searches for radio emission from their host galaxies in existing continuum surveys have also frequently yielded non-detections, not necessarily because the galaxies lack radio emission, but likely due to insufficient survey depth or resolution. This motivates the use of deeper or lower-frequency radio data to explore whether more FRB hosts can be identified through their continuum radio emission.

In the absence of reliable radio detections, optical observations have become the primary method for host identification. However, this approach introduces its own selection biases. Faint or obscured host galaxies are less likely to be detected or confidently associated with FRBs, particularly when positional uncertainties are large. As a result, current host samples are biased toward more massive and actively star-forming galaxies. Additionally, \citet{2024Natur.634.1065Bhardwaj} demonstrated that FRBs are more often detected in face-on galaxies than in edge-on systems, likely due to lower line-of-sight scattering, further contributing to host population bias. These limitations may cause an underestimation of the true diversity of FRB host galaxies.

The bias against detecting faint hosts skews the inferred properties of FRB host galaxies, leading to an over-representation of massive, star-forming systems while underestimating the fraction of low-mass or quiescent galaxies. Radio-based host identification provides a complementary approach by being less affected by dust obscuration, allowing for more reliable estimates of star-formation activity. However, radio selection is itself not free from bias: it remains more sensitive to galaxies with significant synchrotron emission, and may still miss radio-faint, quiescent hosts. Nonetheless, combining optical and radio-selected samples offers a more complete view of the FRB host population and helps to better characterise their underlying diversity \citep{2021arXiv211207639Seebeck}.

In this study, we use the CHIME/FRB baseband localisation catalogue to identify candidate host galaxies through positional matches with LoTSS sources and assess their chance-coincidence probabilities (Section~\ref{sec:2}). We then estimate redshifts and derive physical properties using multi-wavelength data (Section~\ref{sec:3}), present the resulting host candidates, compare their properties with known samples (Section~\ref{sec:results}), discuss the limitations of our approach (Section~\ref{sec:discussion}), and conclude with a summary of our main findings (Section~\ref{sec:conclusion}). Throughout this paper we assume a flat $\Lambda$CDM cosmology with $H_{0}=70~\mathrm{km\,s^{-1}\,Mpc^{-1}}$, $\Omega_{m}=0.3$, and $\Omega_{\Lambda}=0.7$.

\section{Crossmatching and Candidate Identification}
\label{sec:2}
\subsection{Sample}
\label{sec:sample} 

\begin{figure*}
    \centering
    \includegraphics[width=0.9\textwidth]{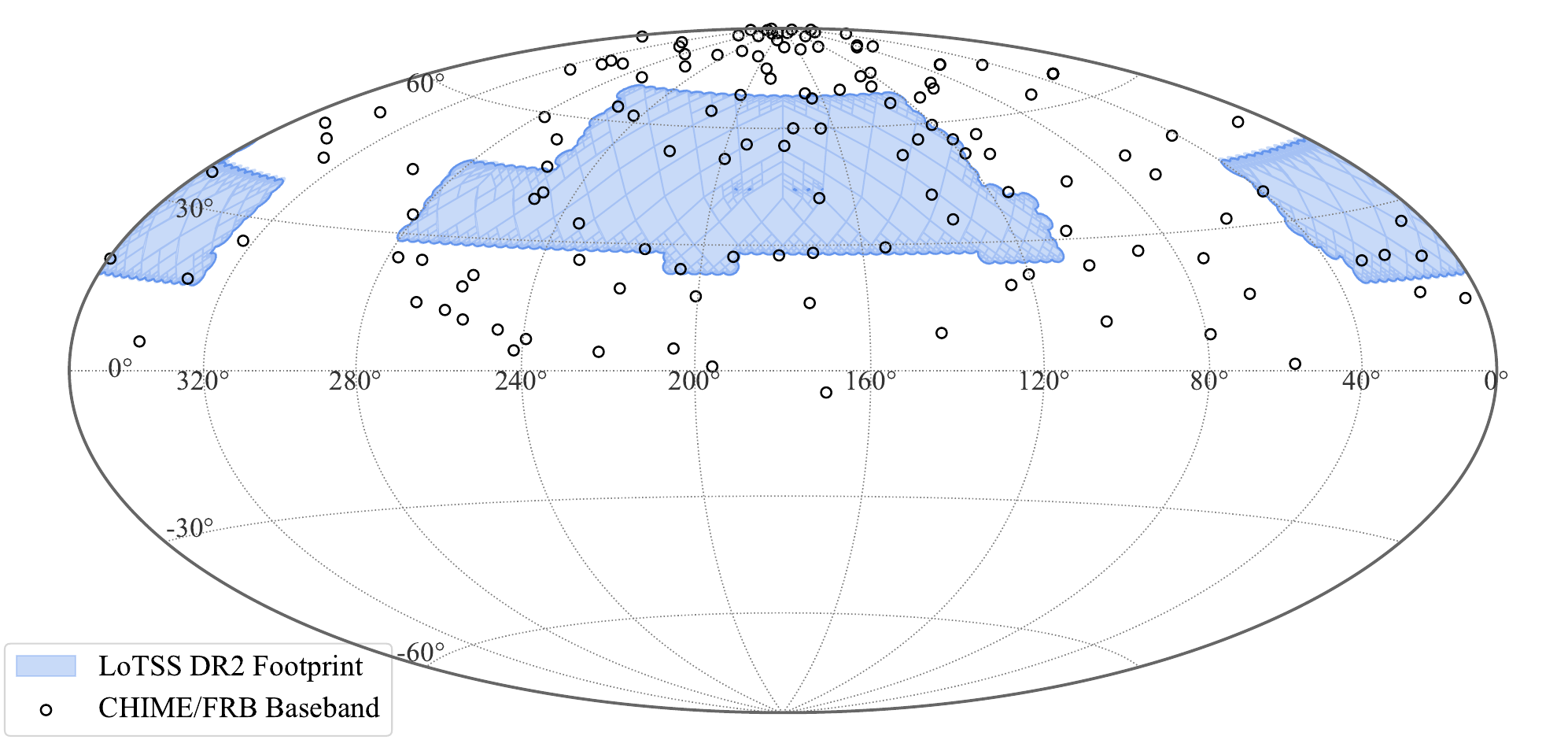}
    \caption{Sky distribution of 140 CHIME/FRB baseband-detected events (black circles) overlaid on the LoTSS DR2 survey footprint (shaded region). This figure illustrates which FRBs lie within the LoTSS coverage area based on the multi-order coverage (MOC) map. A total of 33 FRBs are found to fall within the footprint and are eligible for further cross-matching analysis.}
    \label{fig:chime-lotss}
\end{figure*}

Our sample of FRBs was taken from the CHIME/FRBs baseband data \citep[][]{2024ApJ...969..145Chime}. The Canadian Hydrogen Intensity Mapping Experiment Fast Radio Burst Project (CHIME/FRB) is a large-scale, real-time survey dedicated to the discovery and monitoring of fast radio bursts. Operating at frequencies between 400-800 MHz, CHIME/FRB utilises a stationary, wide-field interferometric radio telescope located in Penticton, British Columbia, Canada. The telescope's large field of view ($\sim$ 200 square degrees) and continuous monitoring capability make it particularly well-suited for detecting transient events like FRBs.

Since starting full operations in 2018, CHIME/FRB has rapidly become the leading instrument in terms of the number of FRB discoveries. The CHIME/FRB Collaboration has published catalogues containing hundreds to thousands of bursts, including both one-off and repeating sources \citep[][]{2021ApJS..257...59Chime,2023ApJ...947...83Chime}. In its most recent data release, the Baseband FRB catalogue \citep{2024ApJ...969..145Chime}, the collaboration reported 140 FRBs, 12 of which originate from seven repeaters. The measurements of burst properties including position, dispersion measure (DM), flux, fluence, and exposure, have been significantly improved compared to CHIME/FRB's first catalogue. 

Compared to Catalogue 1, which relied on real-time total intensity detection pipelines and typically achieved localisation uncertainties on the order of several arcminutes, the Baseband Catalogue utilises raw voltage (baseband) to coherently reconstruct the FRB wavefront. This allows for interferometric localisation with ten-arcsecond-level precision, representing an order-of-magnitude improvement in angular resolution. Such high spatial accuracy is crucial for unambiguous host galaxy associations, especially in crowded optical fields, and has enabled direct cross-identification with multi-wavelength surveys. 

In addition to enhanced localisation, the baseband data offer microsecond time resolution and finer frequency channels, which together allow for more accuratemeasurements of scattering timescales, burst sub-stucture, and polarisation. Overall, the Baseband Catalogue marks a significant advance in CHIME/FRB's ability to characterise both the intrinsic and propagation properties of FRBs.

\subsection{Crossmatching}
\label{sec:crossmatching} 

To identify potential host galaxies of FRBs detected by CHIME/FRB, we cross-match their positions with deep low-frequency radio surveys. In particular, the LOFAR Two-metre Sky Survey (LoTSS) provides a powerful complementary dataset. LoTSS is a deep radio continuum survey conducted at 120-168 MHz using the LOFAR High Band Antennas (HBA), with the aim of imaging the entire northern sky with unprecedented sensitivity and angular resolution at low frequencies \citep[][]{2017A&A...598A.104Shimwell,2019A&A...622A...1Shimwell}.

The second data release (LoTSS-DR2) covers over 5,700 square degrees and reaches a median sensitivity of $\sim 70\,\mu \rm{Jy/beam}$ with a resolution of 6 arcseconds, making it well-suited for detecting star-forming galaxies and AGN out to moderate redshifts. This resolution is comparable to, or better than the typical localisation accuracy of CHIME/FRB baseband-detected sources, thus enabling reliable cross-identification. By matching CHIME/FRB baseband positions with LoTSS radio sources, we can associate FRBs with persistent radio counterparts, which are crucial for confirming host galaxies, estimating star formation activity, or identifying AGN. 

We cross-matched the centroid positions of 140 CHIME/FRB baseband-detected events with the LoTSS DR2 source catalogue \footnote{\url{https://lofar-surveys.org/dr2_release.html}} to identify potential radio counterparts. Using the multi-order coverage (MOC) map provided by the LoTSS DR2 data release, we first determined which FRBs lie within the survey footprint. A total of 33 FRBs were found to be spatially covered by LoTSS (see Fig. \ref{fig:chime-lotss}).

For each FRB within the footprint, we queried the \texttt{lotss\_dr2.main\_sources} table using the \texttt{pyVO} package \citep{2014ascl.soft02004Grahma} to retrieve nearby LoTSS sources. The search region was defined as an ellipse centred on the FRB position, with semi-major and semi-minor axes equal to twice the reported $1\sigma$ uncertainties in right ascension and declination from the CHIME/FRB baseband catalogue, thereby corresponding to a $2\sigma$ localisation region. Any LoTSS source whose centroid lies within this elliptical $2\sigma$ region was recorded as a candidate counterpart.

A total of 27 LoTSS sources were identified within these search regions, associated with 16 unique FRBs. To exclude foreground stars, we cross-matched the radio source positions with the NASA/IPAC Extragalactic Database (NED) and SIMBAD using \texttt{astroquery} \citep{2019AJ....157...98Ginsburg}, adopting a $10''$ search radius centred on each LoTSS source. Any source classified as stellar (e.g. with type `*' or `Star') was removed from the sample. The final list comprises 24 non-stellar radio sources linked to 16 CHIME FRBs, of which 15 are non-repeating and one is a confirmed repeater. In cases where multiple LoTSS sources were found within the elliptical region of a single FRB, we retained all candidate matches for subsequent analysis without applying additional selection criteria at this stage.

\subsection{Chance Alignment Test}
\label{sec:chance}

To evaluate the likelihood of physical association between the CHIME/FRB events and their candidate LoTSS counterparts, we estimated the probability of chance alignment using a Monte Carlo approach. We generated a reference sample of 15,000 random sky positions and selected only those with at least one LoTSS DR2 source located within a radius of 5 arcminutes, thereby ensuring they lie within the survey footprint.

For each valid random postion, we measured the angular seperations to all nearby LoTSS DR2 sources within 5 arcminutes. To characterise the nature of these sources, we cross-matched them with the NASA/IPAC Extragalactic Database (NED), assuming the LoTSS source corresponds to the nearest object within a 10-arcsecond matching radius. Sources classified as stellar were excluded, and only extragalactic objects were retained for further analysis.

To assess the statistical significance of the observed cross-matching results, we employed a two-dimensional kernel density estimation (KDE) method in the separation-flux space. The KDE models were constructed for both the random sample and the FRB-matched sample using the same kernel bandwidth. For the choice of the KDE kernel bandwidth, we adopt Scott’s rule as the baseline and apply a scaling factor of 0.1 to reduce oversmoothing and preserve the tail structure of the random-match distribution, which is critical for identifying significant outliers. The null hypothesis of this analysis is that the FRB and radio source are not physically associated and that any apparent association arises purely from random alignment with the background LoTSS source population. By comparing the resulting density distributions, we quantified deviations from the null hypothesis of random alignment. 

\begin{figure}
    \centering
    \includegraphics[width=0.45\textwidth]{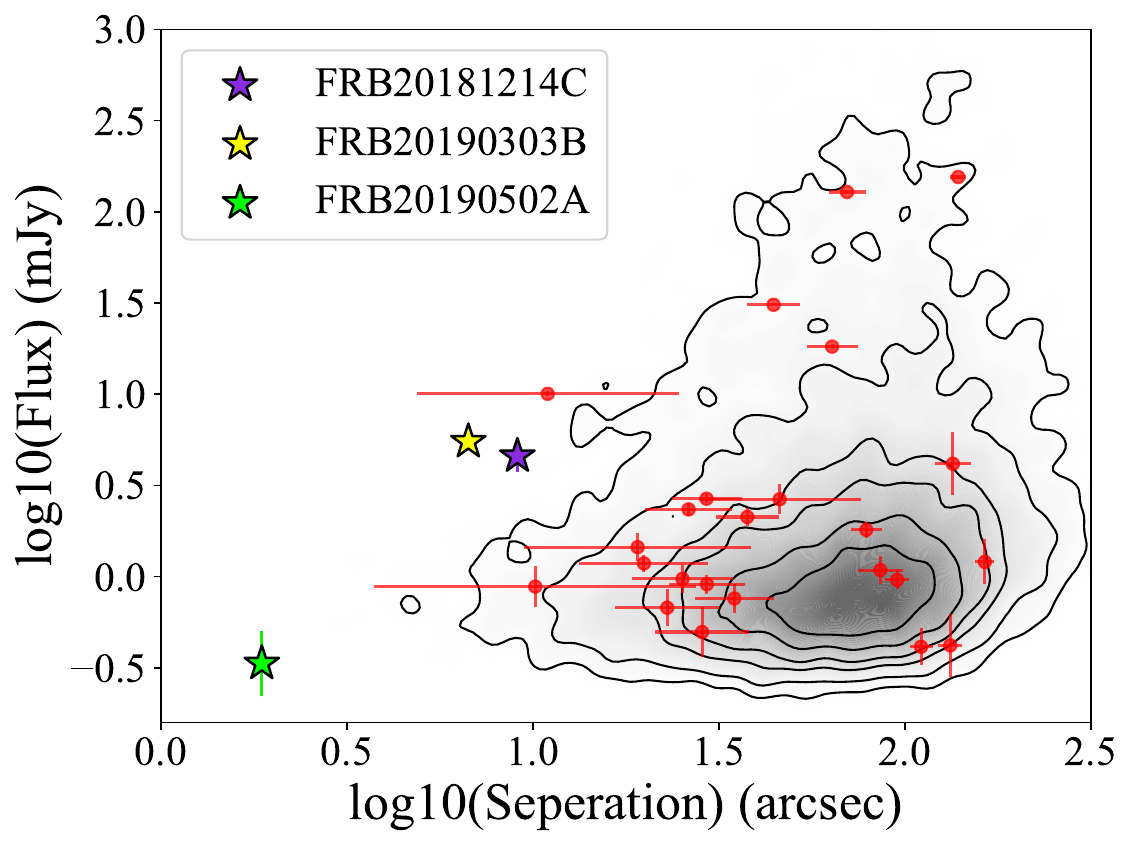}
    \caption{The correlation between FRB–radio-source separation and flux density for non-star objects in the LoTSS DR2 release. Coloured stars mark significant outliers from the random distribution, identified based on the KDE chance-alignment analysis and suggesting a higher likelihood of being the true host. Grey shade indicates density of matches to 15000 random points, and black contours, from the outermost to the innermost, represent the cumulative distributions of $\rm 99.9\%,\ 99\%,\ 95\%,\ 90\%,\ 75\%,\ and\ 50\%$ of the random points, respectively. Red points represent the matched FRBs, with error bars indicating the spatial extent of each radio galaxy.}
    \label{fig:chance_align}
\end{figure}

Fig. \ref{fig:chance_align} shows the KDE distributions for both datasets, with contours marking regions of increasing density. The outermost contour indicates the boundary enclosing 99.9 percent of random data. For each FRB-LoTSS pair, we evaluated its KDE-derived density under the random distribution model and calculated the probability of obtaining an equal or higher density by chance. This probability is a quantitative measure of statistical significance of each FRB-radio source association, with lower probabilities indicating a higher probability of physical association.

Based on the KDE chance-alignment analysis, we identified four candidate host galaxies with a chance probability below 0.1 percent.
One of these corresponds to the known repeater FRB~20191106C, which has already been studied in detail in the literature and shown to be associated with an optical host galaxy \citep{2024ApJ...961...99Ibik,2024ApJ...976..199Ibik}. We therefore do not include it in our analysis here. The remaining three sources were retained as plausible hosts. We further classified them into two secure identifications and one tentative association, which are highlighted in Fig. \ref{fig:chance_align} and will be discussed in Section \ref{sec:results}. Table \ref{tab:frb_lotss_matches} shows the basic properties of the three candidates for the FRB host. 

We also estimate the expected number of chance associations for our three candidates by randomly placing FRB error regions of the same size within the LoTSS DR2 footprint and counting how many LoTSS sources would fall within. This can be approximated using $N=\sum_{i=1}^{3}\pi\theta_i^2\,N_R/\Omega$, where $N_R$ is the total number of LoTSS DR2 sources, $\Omega$ is the sky area covered by LoTSS DR2, and $\sum_{i=1}^{3}\pi\theta_i^2$ represents the total area of the FRB error regions. All three candidates lie within a 1$\sigma$ region of the FRB centre positions. Assuming a 1$\sigma$ error region, the expected number of random crossmatches is $\sim 0.86$. This implies that, statistically, less than one of the three candidates is likely to be a false association.

\section{Host Analysis Methods}
\label{sec:3}
\subsection{Redshift Estimation}
\label{sec:refshift}

The dispersion measure (DM) quantifies the integrated column density of free electrons along the line of sight between the FRB source and the observer. It is defined as
$\mathrm{DM}=\int^D_0n_e(l)\mathrm{d}l $, where $n_e(l)$ is the free electron density (in $\mathrm{cm^{-3}}$) at a distance $l$ (in parsecs) along the line of sight. The resulting DM has units of $\mathrm{pc\ cm^{-3}}$, and higher DM values generally indicate longer path lengths through ionised media or regions with a higher plasma density.

For extragalactic FRBs, the total observed dispersion measure $\mathrm{DM}_{\rm FRB}$ is typically separated into three main contributions:

\begin{equation}
    \mathrm{DM}_{\rm FRB} = \mathrm{DM}_{\rm MW} + \mathrm{DM}_{\rm cosmic} + \frac{\mathrm{DM}_{\rm host}}{(1+z_{\rm FRB})},
    \label{eq:dm}
\end{equation}

where $\mathrm{DM}_{\rm MW}$ includes the Milky Way disk and halo contributions, $\mathrm{DM}_{\rm host}$ represents the host galaxy and local environment, and $\mathrm{DM}_{\rm cosmic}$ accounts for the contribution from the intergalactic medium (IGM). The $\mathrm{DM}_{\rm MW}$ component was estimated using the YMW16 Galactic electron density model \citep{2017ApJ...835...29YMW16} for the disk, and a fixed halo contribution of $30\ \mathrm{pc\ cm^{-3}}$ \citep{2015MNRAS.451.4277Dolag}, as adopted by the CHIME/FRB Collaboration \citep[][]{2024ApJ...969..145Chime}.

To estimate redshifts, we adopted the central DM–redshift relation between $\mathrm{DM}_{\rm cosmic}$ and redshift proposed by \citet{2020Natur.581..391Macquart} (see their Fig. 2), assuming a host contribution of $50(1+z)^{-1}\,\mathrm{pc\ cm^{-3}}$ \citep{2021MNRAS.501.5319Arcus, 2025A&A...696A..81Bernales}, with redshift uncertainties estimated from the intrinsic scatter (central 90 percent range) of the relation. In addition, we considered the probabilistic modelling of the DM–$z$ relation from \citet{2023ChPhC..47h5105Tang}, which accounts for IGM fluctuations and host uncertainties via a Bayesian inference framework calibrated with a sample of FRBs with known redshifts.

\subsection{Host Galaxy Characterisation}
\label{sec:model}
\subsubsection{Spectral Energy Distribution Fitting}

To investigate the physical nature of the candidate host galaxies, we performed spectral energy distribution (SED) fitting using publicly available optical and infrared photometry. The multi-wavelength data were retrieved via the VizieR catalogue service and compiled from several archival surveys, including Galaxy Evolution Explorer (GALEX) in the ultraviolet \citep{2011Ap&SS.335..161B_galex}, Sloan Digital Sky Survey (SDSS) DR16 in the optical \citep{2020ApJS..249....3A_sdss16}, Two Micron All-Sky Survey (2MASS) in the near-infrared \citep{2006AJ....131.1163S_2mass}, Wide-field Infrared Survey Explorer (WISE) in the mid-infrared \citep{2013wise.rept....1Cutri_wise}, and Herschel in the far-infrared \citep{2024yCat.8112....0Herschel} when accessible. A detailed description of the photometric measurements adopted from each survey is provided in Appendix~\ref{sec:appendixa}. These data were used to derive key galaxy properties such as stellar mass, star formation rate (SFR), dust attenuation, and stellar age. Not all sources are covered by all surveys; however, for each galaxy we include all available photometric measurements in the SED fitting. Prior to SED fitting, observed photometric fluxes were corrected for foreground Galactic extinction. The colour excess $E(B{-}V)$ at each source position was obtained from the Schlegel–Finkbeiner–Davis (SFD) dust map \citep{1998ApJ...500..525Schlegel}, and the extinction correction was applied using the extinction law of \citet{1989ApJ...345..245CCM} with $R_V=3.1$. 

The SED fitting was carried out using the \texttt{CIGALE} code \citep{2019A&A...622A.103Boquien}, adopting the \texttt{sfh2exp} star formation history module. This model consists of an old stellar population and a secondary burst, both characterised by exponentially declining star formation rates with independent ages and e-folding times. The mass fraction of the burst component was allowed to vary to account for recent star formation activity.

Stellar emission was modelled using the templates from \citet{2003MNRAS.344.1000Bruzual} with a \citet{1955ApJ...121..161Salpeter} initial mass function. 
The stellar metallicity was allowed to vary over two representative values (sub-solar and solar) to account for plausible variations among the host galaxies. Nebular emission, including both line and continuum components, was included with a fixed ionisation parameter and gas-phase metallicity. These nebular parameters were fixed to reduce degeneracies with other SED parameters, given the limited number of emission-line and ultraviolet constraints available for our sources, and because they have a negligible impact on the broadband-derived stellar masses and star formation rates. Dust attenuation was implemented via a modified power-law attenuation law, allowing for differential extinction between young and old stellar populations.
Infrared dust re-emission was modelled following \citet{2007ApJ...657..810Draine}, with free parameters for the polycyclic aromatic hydrocarbons (PAH) fraction and the interstellar radiation field intensity.

A Bayesian analysis was used to explore the parameter space and to constrain key physical properties, including stellar mass, SFR, dust attenuation, and mass-weighted age. The resulting posterior distributions were used for further interpretation in Section~\ref{sec:results}.

\subsubsection{Radio-based Star Formation Rate Estimation}

Star-forming galaxies emit both thermal (free-free) and non-thermal (synchrotron) radiation in the radio regime. Their total radio spectra are typically modelled as the superposition of two power-law components:

\begin{equation}
    S_{\rm tot}(\nu) = S_{\rm th}(\nu_0) \left( \frac{\nu}{\nu_0} \right)^{-0.1} + S_{\rm nth}(\nu_0) \left( \frac{\nu}{\nu_0} \right)^{-\alpha_{\rm nth}},
    \label{eq:powerlaw}
\end{equation}

where $S_{\rm th}$ and $S_{\rm nth}$ are the thermal and non-thermal components at a reference frequency $\nu_0$, and $\alpha_{\rm nth}$ is the non-thermal synchrotron spectral index. For frequencies below a few GHz, the synchrotron component dominates, especially at $\sim$150 MHz.

Assuming that the observed radio emission arises primarily from star formation processes, we estimated the SFRs using the empirical relation $L_{150} = L_1\,\psi^{\beta}$ derived by \citet{2018MNRAS.475.3010Gurkan}, where $L_{150}$ is the rest-frame 150 MHz monochromatic luminosity in $\mathrm{W\,Hz^{-1}}$, $\psi$ is the SFR in $M_\odot\,\mathrm{yr^{-1}}$. We adopted their best-fit parameters from \cite{2021A&A...648A...6Smith}: $\log_{10}L_{150\rm{MHz}}=(22.221\pm0.008)+(1.058\pm0.007)\log_{10}(\psi/M_\odot \rm{yr^{-1}})$.

The estimation of the star-formation rate from radio emission can also be affected by other factors, such as the stellar mass of the host galaxy \citep{2018MNRAS.475.3010Gurkan}. To improve the reliability of our estimates, we adopt a correlation that includes an explicit dependence on stellar mass, as presented by \citet{2021A&A...648A...6Smith}: $\log_{10}(L_{150\rm{MHz}}/\rm{W\,Hz^{-1}})=(0.90\pm0.01)\log_{10}(\psi/M_\odot yr^{-1})+(0.33\pm 0.04)\log_{10}(M/10^{10}M_\odot)+22.22\pm 0.02$. In our case, no archival stellar mass measurements are available for any of the three candidates. We therefore used the stellar masses estimated from our SED fitting in this relation and compared the resulting SFRs with those derived from the pure radio–flux relation. As the SED-based stellar mass estimates partly rely on optical data, they may also introduce some bias. For comparison, the resulting SFRs are listed in Table~\ref{tab:sed}, where the SFRs derived from the radio-flux-only relation are denoted as SFR${\rm_{radio}}$, and those obtained using the mass-dependent relation are denoted as SFR${\rm_{ radio\,w\,m}}$.

Monochromatic luminosities at rest-frame 150 MHz were derived by first extrapolating the observed LoTSS 144 MHz flux densities to 150 MHz using a standard synchrotron spectral slope ($\alpha_{\rm nth} = -0.7$), following $S_{150} = S_{144} \left(150/144 \right)^{\alpha_{\rm nth}}$ applying a full k-correction. These extrapolated fluxes were combined with redshift estimates, using spectroscopic values where available and best-fit photometric (SED) redshifts otherwise, to compute $L_{150}$.

\begin{figure*}
    \centering
    \includegraphics[width=0.8\textwidth]{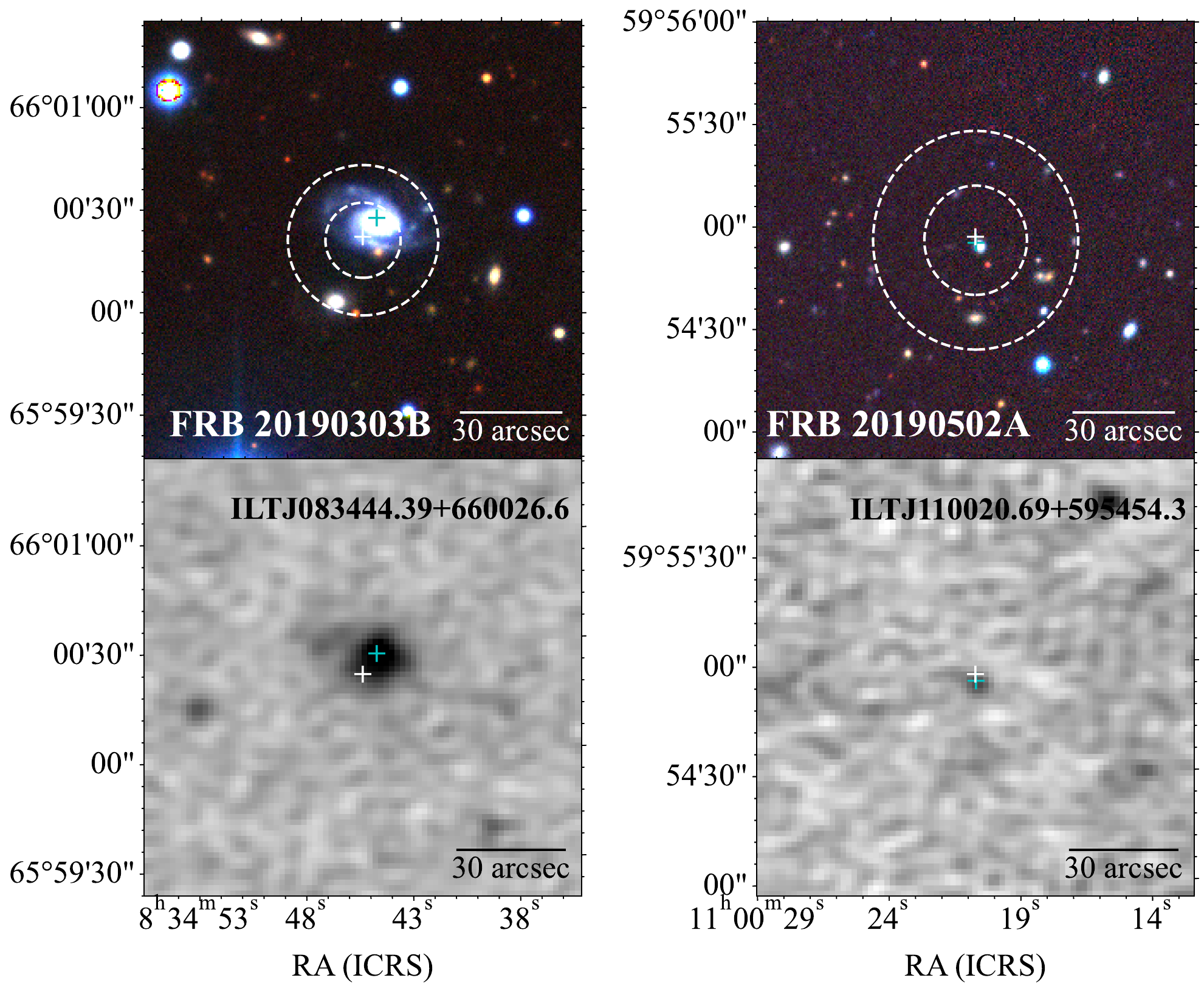}
    \caption{Optical and radio images of two secure candidates. \textit{Top panels}: False-colour RGB images constructed from the $g$, $r$, and $z$ bands of the Dark Energy Spectroscopic Instrument (DESI) Legacy Imaging Surveys for two secure candidates. The white ellipses represent the $1\sigma$ and $2\sigma$ localisation uncertainty regions of the FRB, with the white cross marking the FRB centroid and the cyan cross indicating the position of the matched LoTSS radio source. \textit{Bottom panels}: LoTSS DR2 radio continuum images of the same regions.}
    \label{fig:secure_candidates}
\end{figure*}

\subsubsection{H$\alpha$-derived Star Formation Rate}
\label{sec:spec_method}

For sources with optical spectra showing H$\alpha$ emission, SFRs were estimated based on the extinction-corrected H$\alpha$ luminosity, following the calibration of \citet{1998ApJ...498..541Kennicutt}:

\begin{equation}
    \mathrm{SFR}\,[M_\odot\,\mathrm{yr}^{-1}] = 7.9 \times 10^{-42}\, L(\mathrm{H}\alpha)\,[\mathrm{erg\,s}^{-1}].
    \label{eq:sfr_ha}
\end{equation}

In cases where the H$\alpha$ emission line is blended with [NII], we assumed a fixed [NII] contamination fraction of 29 percent based on \citet{2018ApJ...855..132Faisst}, and subtracted this contribution from the total blended flux.

Galactic extinction corrections were applied using the extinction curve of \citet{1989ApJ...345..245CCM} with $R_V = 3.1$, and colour excess $E(B{-}V)$ values were obtained from the SFD dust map \citep{1998ApJ...500..525Schlegel}. The attenuation at the wavelength of H$\alpha$ was computed as $A(\mathrm{H}\alpha) = 2.6\,E(B{-}V) $ \citep{2008ApJS..178..247Kennicutt}.

For sources without detected H$\beta$ emission, internal attenuation was estimated using the best-fit $A_V$ for young stellar populations obtained from \texttt{CIGALE} SED fitting. Following \citet{2000ApJ...533..682Calzetti}, gas attenuation was calculated as $A_V^{\mathrm{gas}} = 1.7\, A_V^{\mathrm{stars}}$, and converted to H$\alpha$ attenuation via $A(\mathrm{H}\alpha) = 0.81\, A_V^{\mathrm{gas}}$.

In cases where both H$\alpha$ and H$\beta$ are detected, the internal extinction was derived from the Balmer decrement. Assuming Case B recombination with an intrinsic flux ratio of $H\alpha/H\beta = 2.86$ \citep{1938ApJ....88...52Baker}, the colour excess was calculated as:

\begin{equation}
    E(B{-}V) = 1.97 \log_{10} \left( \frac{(H\alpha/H\beta)_{\mathrm{obs}}}{2.86} \right).
    \label{eq:attenuation_2}
\end{equation}

The corrected H$\alpha$ luminosities were then used to derive SFRs using Equation~\ref{eq:sfr_ha}.

\section{Results}
\label{sec:results}

\begin{table*}
\centering
\caption{Basic properties of the CHIME/FRB events and their matched LoTSS DR2 radio counterparts.
Dispersion measures (DMs) and the corresponding Galactic contributions at the FRB positions are taken from the CHIME/FRB baseband catalogue \citep[][]{2024ApJ...969..145Chime}. The extragalactic DM values, corrected for the Milky Way contribution using the NE2001 model \citep{2002astro.ph..7156Cordes_NE2001}, are adopted from \citet{2021ApJS..257...59Chime}.
The table reports four redshift estimates: $z_{\rm FRB}$, inferred from the central DM–redshift relation of \citet{2020Natur.581..391Macquart} (see their Fig.~2) following the method described in Section~\ref{sec:refshift}, with uncertainties derived from the intrinsic scatter (central 90 percent range) of the relation; $z_{\rm spec},$ the spectroscopic redshift from the DESI Extragalactic Science Project and from this work, respectively; $z_{\rm phot}$, the photometric redshift of the associated host galaxy from \citet{2021MNRAS.501.3309Zhou}; and $z_{\rm DM\,\rm Tang23}$, the DM–redshift estimate reported by \citet{2023ChPhC..47h5105Tang}.
The positions and flux densities of the LoTSS counterparts are taken from \citet{2022A&A...659A...1Shimwell}.}
\label{tab:frb_lotss_matches}
\begin{adjustbox}{width=\textwidth}
\begin{tabular}{lllllll}
\toprule
FRB Name  & FRB 20190303B& FRB 20190502A & FRB 20181214C\\
\midrule
LoTSS ID  & ILTJ083444.39+660026.6 & ILTJ110020.69+595454.3 & ILTJ114334.93+600331.4\\
Optical ID & SDSS J083444.42+660026.5&SDSS J110020.55+595454.4& SDSS J114335.11+600334.9\\
& (MCG+11-11-014)&  & \\
\midrule
$\rm RA_{FRB}$ (deg)  & 128.6877 & 165.0863& 175.8997\\
$\rm Dec_{FRB}$ (deg)   & 66.0059 & 59.9156& 60.0573\\
DM ($\rm pc\,cm^{-3}$)  & 193.4292& 625.738& 632.8323\\
$\rm DM_{MW}$ ($\rm pc\,cm^{-3}$) &41&44&41 \\
$\rm DM_{exc,MW}$ ($\rm pc\,cm^{-3}$)  & 146.4 & 590.8&599.5 \\
$\rm z_{FRB}$ &$0.1030^{+0.0364}_{-0.0405}$& $0.4933^{+0.1741}_{-0.1941}$& $0.5416^{+0.1912}_{-0.1583}$\\
$\rm z_{spec}$  &0.0656$\pm$0.0026 & 0.2803$\pm$0.0100& --\\
$z_{\rm phot}$&$\rm 0.063\pm0.011$ &$0.274\pm0.034$ & $0.447\pm0.261$\\
$z_{\rm DM\,Tang23}$&-- &$0.606^{+0.150}_{-0.321}$ & $0.616^{+0.152}_{-0.332}$\\
$\rm RA_{LoTSS}$ (deg)  & 128.6850 & 165.0862& 175.8955\\
$\rm Dec_{LoTSS}$ (deg)  & 66.0074 & 59.9151& 60.0587\\
Flux (mJy) &5.49$\pm$0.41&0.33$\pm$0.14&1.51$\pm$0.37$^{\dagger}$\\
Separation$_{\rm FRB\text{-}LoTSS}$ (arcsec) &6.7 & 1.9 & 9.1 \\
\midrule
Classification&\textbf{Secure}& \textbf{Secure}&\textit{Tentative}\\
\bottomrule
\end{tabular}
\end{adjustbox}
\vspace{2pt}
\raggedright
\footnotesize{
$^{\dagger}$ Flux derived in this work owing to the extended nature of the source and the low signal-to-noise ratio of the peak; the corresponding LoTSS DR2 catalogue value is 4.58$\pm$0.91 mJy.}
\end{table*}

\subsection{Candidate Host Galaxies}
\label{sec:host_candidates}

\subsubsection{Host galaxy of FRB 20190303B}

\begin{figure}
    \centering
    \includegraphics[width=0.4\textwidth]{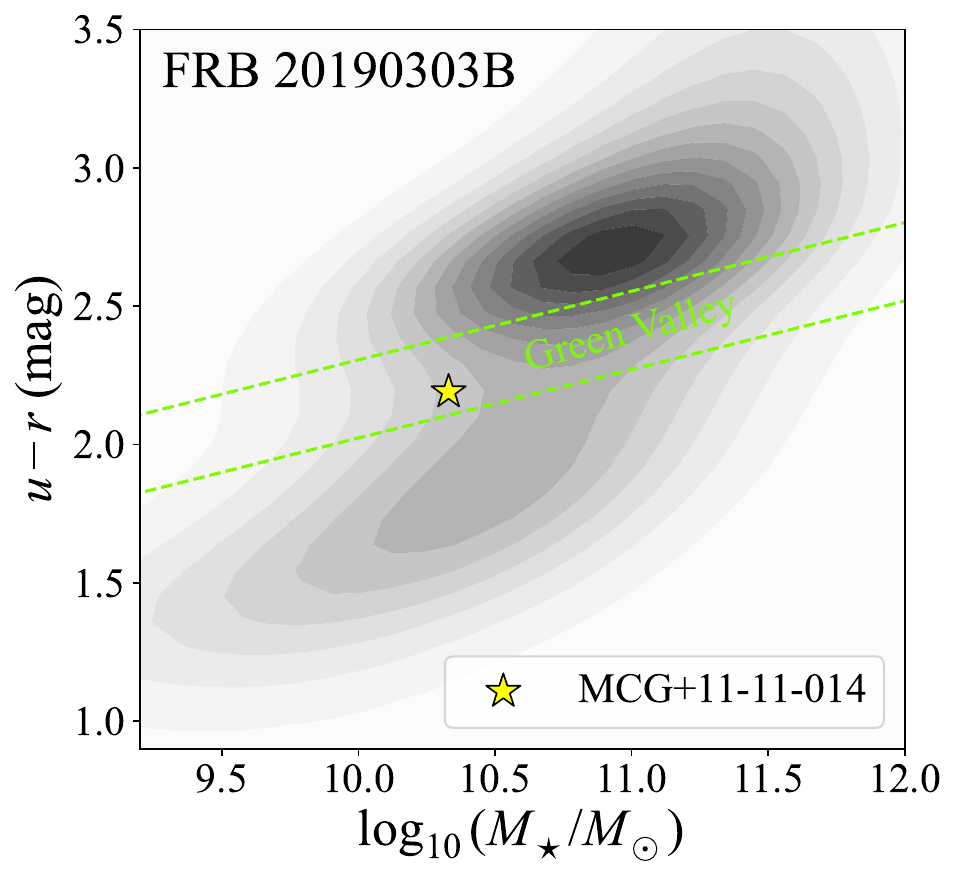}
    \caption{Colour–mass diagram, where green-valley galaxies lie between the two green lines. The yellow star marks the FRB 20190303B host-galaxy candidate MCG+11-11-014. Contours indicate the SDSS spectroscopic sample \citep{2015ApJS..219....8Chang}. Green lines show the green valley defined by \citet{2014MNRAS.440..889Schawinski}.}
    \label{fig:cmd}
\end{figure}

FRB~20190303B has a CHIME position centred at R.A., Dec.~=~128.6877, +66.0059, with a $1\sigma$ localisation uncertainty of 11 arcseconds. A spatially consistent radio source, ILTJ083444.39+660026.6, is detected in the LoTSS catalogue at R.A., Dec.~=~128.6850, +66.0074, with a separation of 6.7 arcseconds and an integrated flux density of 5.493~mJy at 144~MHz. The radio emission appears to be extended, with significant flux detected both in the compact central region and along one of the spiral arms of the galaxy, coincident with the FRB localisation. As shown in Fig. \ref{fig:secure_candidates}, this spatial alignment suggests ongoing star formation activity near the FRB site.

The candidate host galaxy, MCG+11-11-014, is clearly resolved in the Dark Energy Camera Plane Survey (DECaPS; \citealt{2018ApJS..234...39Schlafly_DECaPS}) optical images as a moderately inclined spiral galaxy, exhibiting a bright, reddened central core and loosely wound outer spiral arms. Its global $g-r$ colour is 0.66 mag. To better quantify the physical properties implied by this colour, we further used the integrated optical colours and photometric measurements from SDSS, which provides the u-band that is unavailable in DECaPS. Its rest-frame $u-r = 2.08$ mag (from SDSS) places it firmly within the ‘green-valley’ regime of the optical colour–mass diagram shown in Fig.~\ref{fig:cmd}, according to the definition of \citet{2014MNRAS.440..889Schawinski}. Such green-valley systems are commonly associated with galaxies transitioning from active star formation to quiescence.

\begin{figure*}
    \centering
    \includegraphics[width=\textwidth]{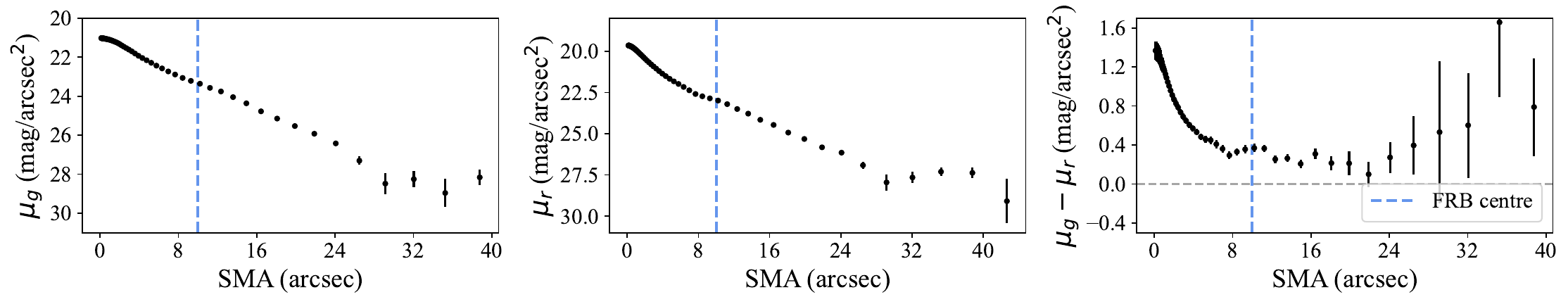}
    \caption{Radial surface brightness profiles of the candidate FRB~20190303B host galaxy MCG+11-11-014  in the $g$-band (left), $r$-band (middle), and the corresponding colour profile ($\mu_g - \mu_r$, right). Error bars represent photometric uncertainties propagated from the intensity measurements. The vertical dashed lines indicate the semi-major axis (SMA) radius corresponding to the FRB centre position. The colour profile shows a mild radial gradient, potentially reflecting variations in stellar population or internal dust attenuation.
    }
    \label{fig:profile}
\end{figure*}

To assess the galaxy morphology and colour gradient, we performed non-parametric elliptical isophotal fitting using the \texttt{Photutils} package \citep{larry_bradley_2023_7946442}. The fitting was carried out iteratively to determine stable geometric parameters of the galaxy. The ellipticity and inclination were first estimated from multiple fits and then fixed at their best-fitting values ($\epsilon = 0.34 \pm 0.05$, $i = 49 \pm 4^\circ$) to minimise degeneracies and ensure consistent radial sampling. Using these fixed geometric parameters, identical sets of elliptical isophotes were applied to both the $g$- and $r$-band images, allowing surface brightness profiles and colours to be measured at matched isophotal radii. The resulting $g$- and $r$-band surface brightness profiles, along with the radial $g{-}r$ colour profile, are shown in Fig.~\ref{fig:profile}. The colour profile reveals a reddened bulge, with $\mu_g - \mu_r$ reaching $\sim$1.5~mag in the inner region and declining to below $\sim$0.5~mag in the outer disk. The inferred inclination is consistent with the typical morphologies of FRB host galaxies reported by \citet{2024Natur.634.1065Bhardwaj}.

\begin{figure}
    \centering
	\includegraphics[width=0.5\textwidth]{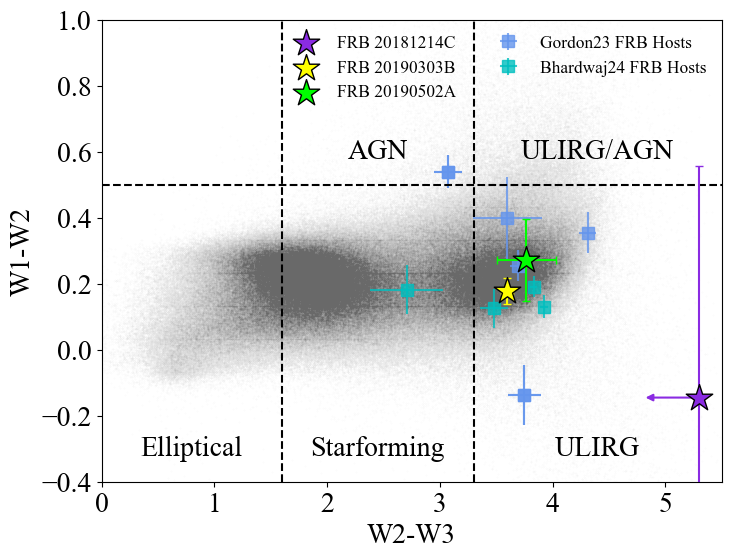}
    \caption{The W2–W3 versus W1–W2 diagram, where W1, W2, and W3 denote the WISE mid-infrared photometric bands, used to separate objects into different classes. The purple, yellow, and green stars indicate the host galaxies selected in this work and listed in Table~\ref{tab:frb_lotss_matches}. The arrow marks an upper limit for FRB 20181214C, as no W3 measurement is available. The background black dots are SDSS spectroscopic galaxy sample \citep{2015ApJS..219....8Chang}. The blue square indicates the FRB hosts in \citet{2023ApJ...954...80Gordon}, and the cyan square indicates the FRB hosts in \citet{2024ApJ...971L..51Bhardwaj}}
    \label{fig:wise_color}
\end{figure}

We additionally performed SED fitting using multi-band photometry from GALEX (FUV, NUV), SDSS ($u$, $g$, $r$, $i$, $z$)\citep{2020ApJS..249....3A_sdss16}, 2MASS ($J$, $H$, $K_s$), WISE (W1–W4)\citep{2013wise.rept....1Cutri_wise}, and Herschel/SPIRE (PMW, PSW). The best-fit stellar population model yields a stellar mass of $\rm \log_{10}M_{\star,SED}[M_\odot] =10.33\pm0.06 $ and a star formation rate of $\rm \log_{10}SFR_{SED}[M_\odot\,yr^{-1}]~=~0.52\pm0.09$. The host’s WISE mid-infrared colours also place it in the star-forming galaxy region, as defined by the WISE colour criteria described in Section~\ref{sec:comparison}(Fig.~\ref{fig:wise_color}), supporting its classification as a dust-rich, star-forming spiral.

We obtained an optical spectrum of the galaxy’s central region using the SPectrograph for the Rapid Acquisition of Transients (SPRAT) instrument on the Liverpool Telescope, as part of project~ID:PL24B06 (PI: R.A.J. Eyles-Ferris), with a total integration time of 2400 seconds. The wavelength coverage was 3711.36–8202.12~\AA, and the spectral resolution was 18~\AA. From this spectrum, we derive a spectroscopic redshift of $z = 0.0656 \pm 0.0026$. The fitting result is shown in Fig.~\ref{fig:ppxf}.

The H$\alpha$ and [N\,\textsc{ii}] lines in the optical spectrum of MCG+11-11-014 are not well resolved owing to the limited spectral resolution. We therefore estimated the H$\alpha$ flux following the method described in Section~\ref{sec:spec_method}. From the stellar attenuation derived via SED fitting, we obtained an internal dust attenuation of $A_V = 0.6$~mag, corresponding to a gas-phase attenuation of $A(\mathrm{H}\alpha) = 0.83$~mag.

After applying these corrections, the extinction-corrected H$\alpha$ luminosity yields a star formation rate of $\rm \log_{10}SFR_{H\alpha}[M_\odot\,yr^{-1}]=-0.84\pm0.07$. This value is significantly lower than the SED-derived SFR (see Table~\ref{tab:sed}), likely due to uncertainties in spectral flux calibration and line deblending, or to the slit capturing only the galaxy’s central region along with a small portion of its outer disk. Given these limitations, the SED-derived SFR is likely to provide a more representative estimate of the galaxy's global star-forming activity. The discrepancy between the SFRs inferred from H$\alpha$ emission and SED fitting will be further discussed in Section \ref{sec:comparison}.

To assess redshift consistency, we compared this spectroscopic measurement with redshifts estimated from FRB dispersion measures. Using the Macquart relation \citep{2020Natur.581..391Macquart}, and following the method in Section~\ref{sec:refshift}, the inferred redshift for FRB~20190303B is $z = 0.1030^{+0.0364}_{-0.0405}$, based on a total DM of 193.4292 $\rm pc\,cm^{-3}$ and Galactic contribution $\mathrm{DM}_{\mathrm{MW}} = 41\,\mathrm{pc\,cm^{-3}}$, as listed in Table~\ref{tab:frb_lotss_matches}. The spectroscopic redshift is somewhat lower than the DM-based estimate, but within 1-2$\sigma$ of the confidence interval. For completeness, we also note that the host galaxy has a photometric redshift of $z_{\rm phot} = 0.063 \pm 0.011$ from DESI DR8 \citep{2022MNRAS.512.3662Duncan}, which is broadly consistent with both the spectroscopic and DM-inferred redshifts within the uncertainties. Additionally, we searched for radio detections at other frequencies and found no counterparts in the NRAO VLA Sky Survey (NVSS; \citealt{1998AJ....115.1693Condon_NVSS};  1.4~GHz) or the Faint Images of the Radio Sky at Twenty-one centimeters (FIRST; \citealt{1995ApJ...450..559Becker_FIRST}; 1.5~GHz). 

Taken together, the small FRB-radio separation, alignment with the spiral arm, redshift consistency across DM, photometric, and spectroscopic methods, and multi-wavelength indications of active star formation, all support a robust association. We therefore propose FRB~20190303B and MCG+11-11-014 as a likely FRB–host association.

\subsubsection{Host galaxy of FRB 20190502A}
\begin{figure*}
    \centering
    \includegraphics[width=0.8\textwidth]{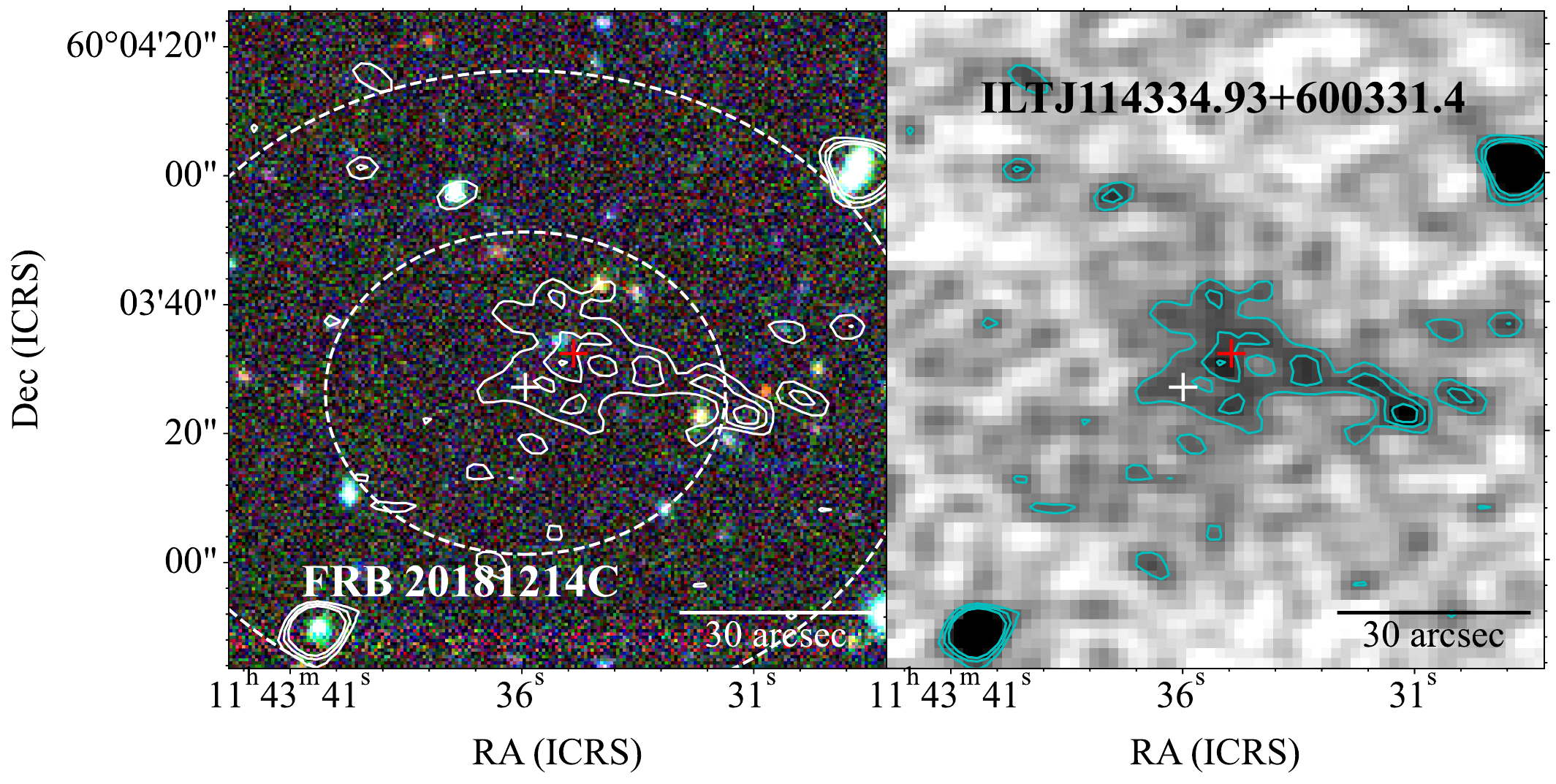}
    \caption{ Optical and radio images of the tentative candidiate. \textit{Left panel}: False-colour RGB images constructed from the $g$, $r$, and $z$ bands of the DESI Legacy Imaging Surveys for the tentative candidate. The white ellipses represent the $1\sigma$ and $2\sigma$ localisation uncertainty regions of the FRB, with the white cross marking the FRB centroid and the red cross indicating the position of the matched LoTSS radio source. \textit{Right panel}: LoTSS DR2 radio continuum images of the same regions. Cyan contours represent emission levels at 2, 3, and 4 times the local RMS noise.}
    \label{fig:tentative_candidates}
\end{figure*}
FRB~20190502A was detected by CHIME at R.A., Dec.~$=165.0818,+59.9146$ with a localisation uncertainty of $15^{\prime\prime}\times16^{\prime\prime}$ (1$\sigma$). The nearest radio source in the LoTSS catalogue is ILTJ110019.68+595455.5 at R.A., Dec.$=165.0862,+59.9151$, located just $1.87^{\prime\prime}$ away, with an integrated flux density of 0.3337mJy at 144~MHz, making it a strong positional match. 
The putative host galaxy, WISEA~J110020.52+595454.2, is detected in the DESI Extragalactic Science Project \footnote{\url{https://data.desi.lbl.gov/doc/releases/dr1/}} and has an available optical spectrum (target ID: 39633386943283235). No radio counterparts at higher frequencies are detected.

From the DESI spectrum we identify several prominent emission lines, yielding a best-fit spectroscopic redshift of $z = 0.2803 \pm 0.0100$. Although this is formally lower than the central Macquart relation estimate of $0.4933^{+0.1741}_{-0.1941}$, it remains statistically consistent once the large intrinsic scatter of the relation is considered. To further account for uncertainties in the host and IGM contributions, we adopt the Bayesian DM–$z$ estimate reported by \citet{2023ChPhC..47h5105Tang}, which assumes log-normal priors on $\mathrm{DM}_{\mathrm{host}}$. Their quoted value of $z = 0.606^{+0.150}_{-0.321}$ is marginally consistent with both the Macquart relation–based and spectroscopic estimates, lending support to a physical association between the FRB and this galaxy. Among the overlapping FRB sample, approximately 80 percent of Macquart-derived redshifts lie below $z \sim 0.8$, and two of our candidates are consistent with this dominant redshift range.

The host candidate exhibits moderately red optical colours, with $g - r \approx 0.92$, which might suggest a passive or quiescent galaxy. However, its infrared properties indicate otherwise. The WISE mid-infrared colour $W2 - W3 \approx 3.77$ points to a substantial infrared excess, consistent with active star formation. This discrepancy suggests the galaxy is a dusty star-forming system, where the red optical colours are likely due to internal dust attenuation rather than a lack of recent star formation.

The optical spectrum of WISEA~J110020.52+595454.2 from the DESI Extragalactic Sources Project includes several well-detected emission lines. Following the procedure described in Section~\ref{sec:spec_method}, we corrected for Galactic extinction and derived the internal dust attenuation using the Balmer decrement, based on the observed H$\alpha$/H$\beta$ flux ratio. Assuming Case~B recombination, we calculated a colour excess of $E(B{-}V) = \text{0.091 mag}$. The extinction-corrected H$\alpha$ luminosity yields a star formation rate of $\rm \log_{10}SFR_{H\alpha}[M_\odot yr^{-1}]=-0.06\pm0.04$, using the calibration from \citet{1998ApJ...498..541Kennicutt}. This value is substantially below the SFR inferred from the CIGALE modelling. A detailed discussion of the discrepancy between the SFRs inferred from H$\alpha$ emission and SED fitting is presented in Section \ref{sec:comparison}.

\begin{table*}
\centering
\caption{Stellar mass and star formation rates derived from different methods}
\label{tab:sed}
\begin{adjustbox}{width=0.65\textwidth}
\begin{tabular}{lcccccc}
\toprule
FRB Name  &FRB 20190303B& FRB 20190502A & FRB 20181214C\\
\midrule
$\log_{10} \mathrm{SFR_{SED}}\; [\mathrm{M}_\odot\,\mathrm{yr}^{-1}]
$  & $0.52\pm0.09$ & $0.61\pm0.12$& $0.31\pm0.36$\\

$\log_{10} \mathrm{SFR_{radio}}\; [\mathrm{M}_\odot\,\mathrm{yr}^{-1}]
$ & $0.51^{+0.05}_{-0.05}$ & $0.64^{+0.15}_{-0.23}$& $1.67^{+0.46}_{-0.80}$\\

$\log_{10} \mathrm{SFR_{radio\,w\,m}}\; [\mathrm{M}_\odot\,\mathrm{yr}^{-1}]$ & $0.48^{+0.06}_{-0.07}$ &  $0.47^{+0.18}_{-0.27}$
 & $2.16^{+0.54}_{-0.94}$ \\
$\log_{10} \mathrm{SFR_{H\alpha}}\; [\mathrm{M}_\odot\,\mathrm{yr}^{-1}]
$  & $-0.84\pm0.07$ & $-0.06\pm 0.04$& -\\
$\log_{10} \mathrm{M_{\star,SED}}\; [\mathrm{M}_\odot]
$  & $10.33\pm0.06$ & $10.77\pm0.07$& $9.45\pm0.23$\\
\bottomrule
\end{tabular}
\end{adjustbox}
\end{table*}

\subsubsection{Tentative host galaxy of FRB 20181214C}

FRB~20181214C is centred at  R.A., Dec.~=~175.8997, +60.0573, with a $1\sigma$ localisation uncertainty of $31''$ in right ascension and $25''$ in declination, according to the CHIME/FRB baseband catalogue. A corresponding radio source is detected in the LoTSS DR2 catalogue is ILTJ114334.93+600331.4 at  R.A., Dec.~=~175.8955 +60.0587 with a total flux of 4.575 mJy and a peak signal-to-noise ratio of 2.73. The source appears extended, with a deconvolved major axis of $33''$ and minor axis of $28''$.

We identified an optical counterpart to the LoTSS source in the DESI imaging survey. The counterpart is extremely blue, with magnitudes of $g = 22.59$, $r = 22.04$, and $z = 21.96$, and has no spectroscopic measurement available. \citet{2021MNRAS.501.3309Zhou} report a photometric redshift of $z = 0.447 \pm 0.261$, broadly consistent with other estimates in the literature, including $0.223 \pm 0.121$ \citep{2020ApJS..249....3A_sdss16} and $0.341 \pm 0.157$ \citep{2022MNRAS.512.3662Duncan}. The redshift inferred from the Macquart relation for FRB~20181214C is $z = 0.5416^{+0.1912}_{-0.1583}$, while \citet{2023ChPhC..47h5105Tang} reported $z = 0.616^{+0.152}_{-0.332}$. All of these estimates are consistent within uncertainties, supporting the plausibility of an association between the FRB and this galaxy.

Assuming a redshift of $\sim0.447$, the corresponding projected physical size of the radio emission is approximately $190 \times 161$~kpc, suggesting an extended source, possibly a star-forming galaxy with large-scale radio emission. Due to the very faint and extended nature of the radio source, as seen in the right panel of Fig.~\ref{fig:tentative_candidates}, we remeasured its integrated flux density using a custom aperture. Specifically, we defined an elliptical aperture centred at R.A. = 11$^{\mathrm{h}}$43$^{\mathrm{m}}$35.05$^{\mathrm{s}}$, Dec. = +60$^\circ$03$'$31.14$''$, with semi-major and semi-minor axes of 15$''$ and 10$''$, respectively. This aperture was chosen to enclose the emission within the 2$\sigma$ contour. The resulting integrated flux density is measured to be 1.51~mJy.

To estimate the uncertainty on the integrated flux density, we consider two primary sources of error: the background noise and the absolute flux calibration uncertainty. The total uncertainty is given by $\sigma_{\mathrm{total}} = \sqrt{ \left( \sigma_{\mathrm{rms}} \cdot \sqrt{N_{\mathrm{beam}}} \right)^2 + \left( f_{\mathrm{cal}} \cdot S_{\mathrm{int}} \right)^2 }$, where \( \sigma_{\mathrm{rms}} \) is the local background noise measured from the image in units of mJy/beam, \( S_{\mathrm{int}} \) is the measured integrated flux density, and \( f_{\mathrm{cal}} \) is the assumed fractional calibration uncertainty. The number of synthesized beams within the measurement aperture is computed as $N_{\mathrm{beam}} =A_{\mathrm{aperture}}/A_{\mathrm{beam}}$, with the beam area given by 
$A_{\mathrm{beam}} = 1.1331 \times \theta_{\mathrm{maj}} \times \theta_{\mathrm{min}}$, where \( \theta_{\mathrm{maj}} \) and \( \theta_{\mathrm{min}} \) are the full width at half maximum (FWHM) of the synthesized beam in arcseconds. In our case, the aperture corresponds to an ellipse with a major axis of $15''$ and a minor axis of $10''$, and the beam size is $6''\times6''$. A conservative 20 percent calibration uncertainty (\( f_{\mathrm{cal}} = 0.20 \)) is adopted following the LoTSS DR2 recommendations \citep{2022A&A...659A...1Shimwell}. The total uncertainty on the measured flux density is estimated to be 0.372 mJy. Considering the updated flux estimate, the source now falls within the 99.9 percent contour in Fig.~\ref{fig:chance_align}, indicating a lower likelihood of being intrinsically associated with the FRB. 

The infrared counterpart is detected in WISE, while no X-ray emission is detected at the corresponding position in existing X-ray catalogues (e.g. 4XMM–DR13; \citealt{2020A&A...641A.136Webb}, and the 2SXPS Swift catalogue; \citealt{2020ApJS..247...54Evans}).
In addition, the WISE colour diagram also argues against the presence of a bright active galactic nucleus (AGN). 
No radio counterparts at higher frequencies are detected, combined with the LoTSS detection, implies a steep radio spectral index, consistent with star formation as the dominant emission mechanism. Given the radio morphology, blue optical colours, the lack of AGN signatures and the photometric redshift consistent with the FRB DM-based distance estimate, we include this source as a tentative candidate host for FRB~20181214C. However, we cannot rule out the possibility that part of the emission arises from jet-related activity associated with a nearby AGN. In this scenario, any star formation inferred from the radio emission may be affected by jet–ISM interactions, and we therefore treat this source with caution.

We additionally performed SED fitting using multi-band photometry from SDSS (u, g, r, i, z) and WISE (W1–W2). The best-fit stellar population model yields a stellar mass of $\rm \log_{10}M_{\star,SED}[M_\odot]=9.45\pm0.23$ and a star formation rate of $\rm \log_{10}SFR_{SED}[M_\odot yr^{-1}]=0.31\pm0.36$. The redshift corresponding to the best-fitting model is 0.27.

\subsection{Comparison with Previously Identified FRB Host Galaxies}
\label{sec:comparison}

We analysed the infrared (IR) colour properties of our candidate host galaxies using data from the Wide-field Infrared Survey Explorer (WISE). Each candidate was cross-matched to the nearest counterpart in the AllWISE catalogue\footnote{\url{https://wise2.ipac.caltech.edu/docs/release/allwise/}}, and we manually inspected the corresponding optical images (from DESI and SDSS) to confirm that the infrared emission originated from the correct host candidate, avoiding contamination from nearby sources. 

To classify the galaxies, we adopted colour-based criteria similar to those proposed by \citet{2016MNRAS.462.2631Mingo,2019MNRAS.488.2701Mingo}. These diagnostics exploit the distinct infrared behaviours of different galaxy populations: AGN-dominated systems, luminous infrared galaxies (LIRGs), ultraluminous infrared galaxies (ULIRGs), and normal star-forming or quiescent galaxies occupy different regions in the WISE W1–W2 versus W2–W3 colour space, reflecting variations in dust heating, star-formation activity, and nuclear emission. The measured WISE magnitudes (W1, W2, W3) of the candidate hosts are listed in Table~\ref{tab:seddata1}, \ref{tab:seddata2} and \ref{tab:seddata3}, whereas the colour–colour diagram is shown in Fig.~\ref{fig:wise_color}.

\begin{figure*} 
    \centering
    \includegraphics[width=0.8\textwidth]{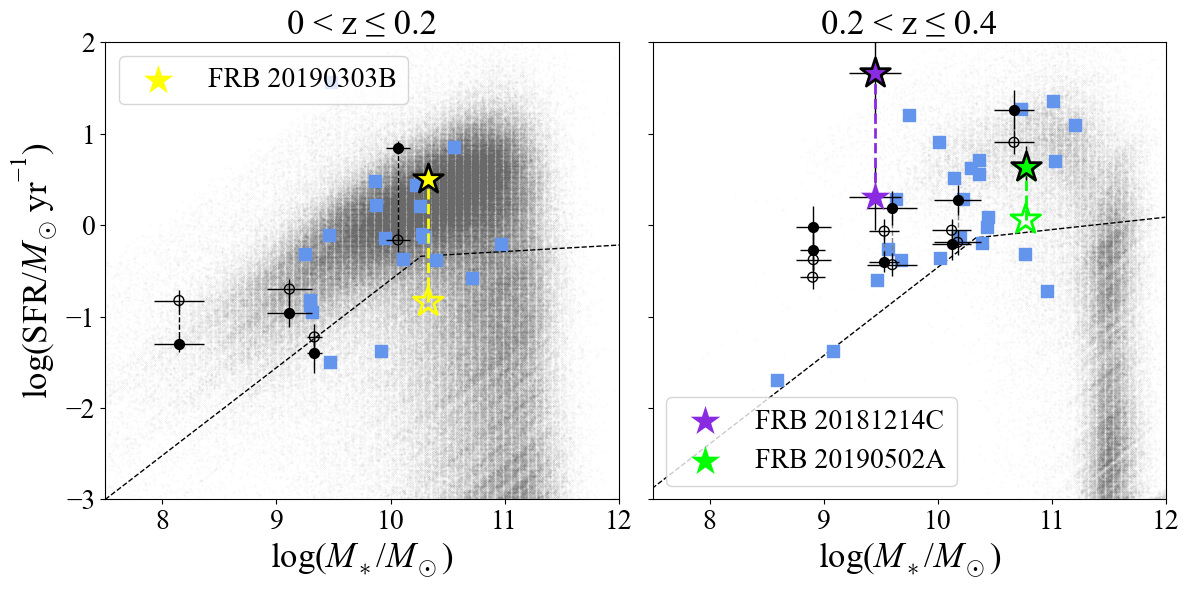}
    \caption{Stellar mass versus star-formation rate for the host galaxy candidates in two redshift bins. Filled stars indicate SFRs derived from SED fitting, while hollow stars represent those derived from H$\alpha$ fluxes. Purple, yellow, and green stars correspond to FRB~20190303B, FRB~20190502A, and FRB~20181214C, respectively. Stars with black outlines indicate the SFRs inferred from low-frequency radio emission. For both FRB~20190303B and FRB~20190502A, the radio-inferred SFRs overlap on the SED-derived values, indicating consistency between these two tracers. The background consists of the SDSS spectroscopic galaxy sample \citep{2015ApJS..219....8Chang}, while blue squares denote the sample from \citet{2025arXiv250215566Loudas}. Solid black circles represent the SED-derived SFRs from \citet{2023ApJ...954...80Gordon}, and hollow circles represent the H$\alpha$-derived SFRs from \citet{2020ApJ...903..152Heintz}. The dashed line indicates the division between star-forming and quiescent galaxies.}
    \label{fig:m_sfr}
\end{figure*}

We also included in Fig.~\ref{fig:wise_color} the WISE colours of FRB host galaxies previously analysed by \citet{2023ApJ...954...80Gordon} and \citet{2024ApJ...971L..51Bhardwaj} for comparison. According to our adopted classification scheme, two of our candidate hosts with reliable WISE detections fall in the region of the diagram typically occupied by dust-rich, infrared-luminous galaxies. These candidates also exhibit WISE colours consistent with those of previously confirmed FRB host galaxies, suggesting similar dusty star-forming environments. The WISE colour analysis provides complementary evidence for the active, possibly dust-obscured nature of the host galaxies, particularly where optical indicators (e.g. SFR from H$\alpha$) may be affected by internal extinction.

We further compared the star formation rates and stellar masses of our candidate host galaxies with the FRB host samples presented by \citet{2020ApJ...903..152Heintz,2023ApJ...954...80Gordon,2025arXiv250215566Loudas}. The star formation rates derived through multiple methods from this work, including spectral energy distribution (SED) fitting, H$\alpha$ emission, and radio luminosity -- do not show significant deviations from the distributions of previously identified FRB host galaxies. For clarity, the \citet{2023ApJ...954...80Gordon} sample is shown in both Fig.~\ref{fig:wise_color} and Fig.~\ref{fig:m_sfr} for comparison. Some hosts lack WISE photometry and are therefore excluded from Fig.~\ref{fig:wise_color}, while an additional redshift cut is applied in Fig.~\ref{fig:m_sfr}, leading to different sample sizes.

According to the SED fitting results, all three candidates lie on the star-forming main sequence, suggesting they are actively forming stars. However, as shown in Fig.~\ref{fig:m_sfr}, we find that the SFRs derived from H$\alpha$ fluxes are significantly lower than those inferred from SED modelling, particularly for FRB~20190303B and FRB~20190502A.
This discrepancy can be understood in terms of the different physical timescales and sensitivities probed by the two indicators: H$\alpha$ emission traces very recent star formation on timescales of $\lesssim$10~Myr, whereas SED-based SFRs reflect an average over longer timescales of $\sim$100~Myr and are therefore less sensitive to short-term variability.
In addition, both of these galaxies exhibit signs of internal dust attenuation: they have compact optical morphologies and reddened central regions. Specifically, the inner core of the spiral galaxy MCG+11-11-014 (associated with FRB~20190303B) displays substantially redder colours, while the host of FRB~20190502A also shows a compact, dusty profile. These features suggest that dust obscuration may play an important role, potentially leading to an underestimation of the H$\alpha$-derived SFRs.

Several statistical studies in the nearby universe have directly compared star formation rates from $\rm H\alpha$ emission to those from SED fitting. \cite{2007ApJS..173..267Salim} report that the UV/SED-based SFRs `compare remarkably well' with SFRs from SDSS $\rm H\alpha$ measurements, with any deviations mainly due to differing dust-extinction corrections. \cite{2015A&A...584A..87Catal}(Calar Alto Legacy Integral Field Area, CALIFA survey) further showed that with integral field spectroscopy one can robustly correct $\rm H\alpha$ for dust and achieve good agreement with hybrid tracers ($\rm H\alpha+IR$ or $\rm UV+IR$) across a wide SFR range (0.03-20 $\rm M_\odot\,yr^{-1}$). Local studies demonstrate a good calibration between $\rm H\alpha$ and SED-derived SFRs, and highlight the importance of applying consistent dust attenuation corrections in both methods to avoid systematic offsets.

H$\alpha$ traces star formation over timescales of approximately 0 to 3–10 Myr, but its reliability is affected by two major systematic uncertainties: dust attenuation and sensitivity to the upper end of the initial mass function (IMF), especially in regions with intrinsically low star formation rates \citep{2012ARA&A..50..531Kennicutt}. In regions with moderate extinction, Balmer decrements (the flux ratio of recombination lines such as $\rm H\alpha$ and $\rm H\beta$) can be used to estimate and correct for attenuation \citep{2002AJ....124.3135Kewley,2004MNRAS.351.1151Brinchmann,2006ApJ...642..775Moustakas}. However, this method break down in circumnuclear starbursts or other dusty galaxies (e.g.,\citealt{2006ApJ...642..775Moustakas}).

To investigate potential differences, we compared the SED-derived and H$\alpha$-derived SFRs reported in \citet{2023ApJ...954...80Gordon} and \citet{2020ApJ...903..152Heintz}, respectively. The comparison is shown in Fig.~\ref{fig:m_sfr}, where solid black circles indicate the SED-based SFRs, and hollow circles represent the H$\alpha$-based values. In some cases, the SED-derived SFRs are significantly higher than those from H$\alpha$, although the majority of sources show good agreement between the two methods. However, due to the limited sample size, no statistically significant conclusion can be drawn for the host sample.

\begin{figure}
    \centering
	\includegraphics[width=0.45\textwidth]{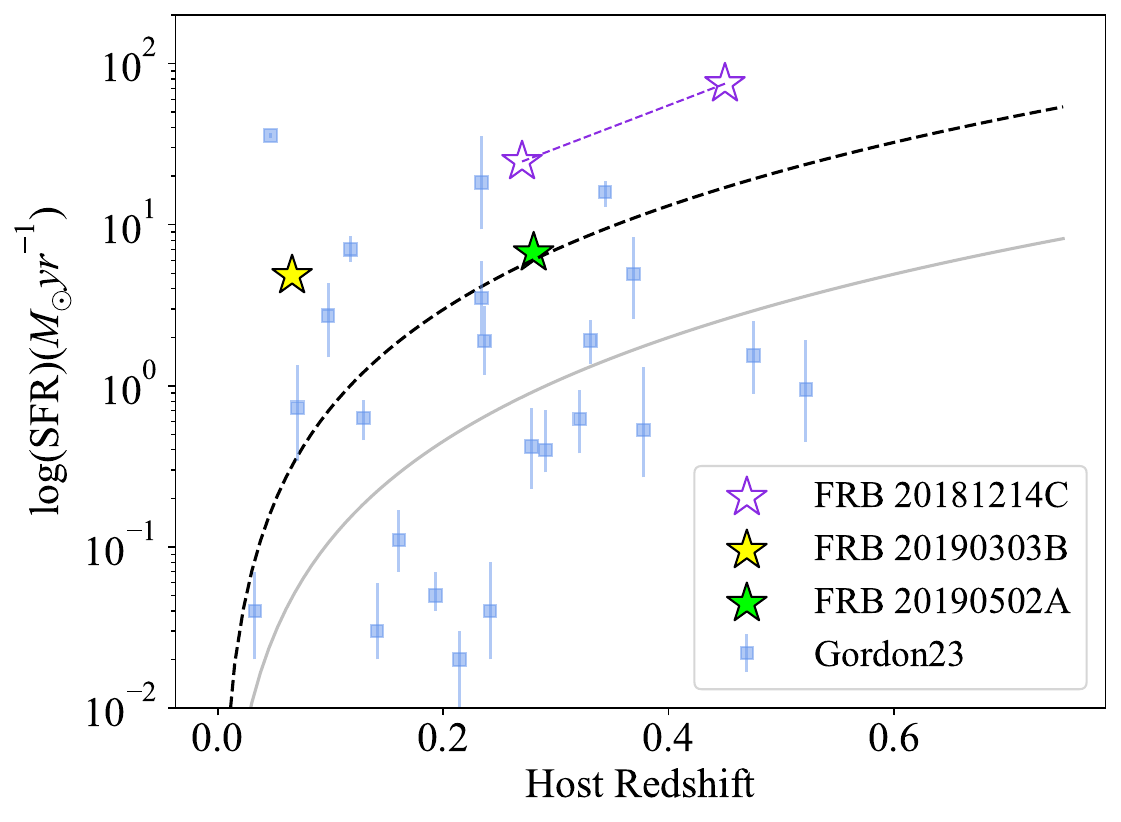}
    \caption{The required SFR to reach 0.30 mJy at 150 MHz for a given redshift (black dashed-line). The SFRs for other FRB hosts are taken from \citet{2023ApJ...954...80Gordon}. Filled stars indicate FRBs with spectroscopic redshift measurements of their candidate hosts; hollow stars represent candidates with only photometric redshift estimates; in cases of large uncertainty (i.e. FRB 20181214C), two points are shown to indicate the lower and upper limits for the same host galaxy. The grey solid line indicates the sensitivity limit achieved in the LOFAR Deep Fields.}
    \label{fig:sfr_z}
\end{figure}

This discrepancy reinforces the importance of multi-wavelength SFR indicators when assessing the nature of FRB host galaxies, particularly in dusty or edge-on systems where optical tracers alone may be insufficient. As shown in Fig.~\ref{fig:sfr_z}, a portion of the known FRB host galaxies could potentially be detected by the LOFAR survey, provided that the relevant regions of the sky are covered. Currently, none of the hosts reported in \citet{2023ApJ...954...80Gordon} fall within the LOFAR footprint. However, with the ongoing progress of the LoTSS survey, we anticipate identifying more FRB host candidates at low radio frequencies.

\section{Discussion}
\label{sec:discussion}
Our pilot study demonstrates the feasibility of identifying FRB host galaxy candidates using low-frequency radio continuum emission alone, offering a complementary approach to traditional optical association techniques. This method holds particular promise for uncovering dusty or optically faint host galaxies that may otherwise be missed. However, several limitations must be considered when interpreting these results.

First, the sample remains limited, with only two secure and one tentative host associations. This is due to the modest overlap between the CHIME/FRB baseband-localised catalogue and the current LoTSS DR2 footprint, as well as the still-small size of the CHIME/FRB baseband dataset and the depth limitations of the LOFAR survey. As such, our findings should be regarded as exploratory. Nevertheless, the successful identification of host candidates via radio emission highlights the potential of future large-scale, radio-selected FRB host samples. The forthcoming LOFAR 2.0 upgrade is expected to substantially enhance this potential. By enabling full-array operations, increased bandwidth, and stable, automated calibration pipelines, LOFAR 2.0 will facilitate commensal observations and broader sky coverage. In addition, the increased use of international stations will deliver sub-arcsecond astrometric precision, further reducing the probability of false positive associations in our cross-matching procedure. Although no specific RMS sensitivity targets have been quoted, the system is expected to approach the thermal-noise limit under a wider range of observing conditions \citep{LOFAR2.0WhitePaper2023}. Deep fields in LoTSS DR1 have already achieved $3\sigma$ detection limits of $\rm 40$-$\rm 60\, \mu Jy\,beam^{-1}$ in regions such as ELAIS‑N1 and the NEP \citep{2023MNRAS.523.1729Best}. With comparable or improved sensitivity and extended overlap with future CHIME/FRB baseband releases, the number of detectable host candidates could increase significantly, potentially doubling based on the projected detection limits (see Fig.~\ref{fig:sfr_z}). 

Looking further ahead, the Canadian Hydrogen Observatory and Radio-transient Detector (CHORD; \citealt{2019clrp.2020...28Vanderlinde}) will build on CHIME to provide a powerful next-generation facility for FRB studies. Operating over an ultra-wide 300–1500 MHz band, CHORD will be particularly sensitive to high-redshift FRBs, for which low-frequency observations maximise detectability and dispersion leverage. The inclusion of long-baseline outrigger stations will deliver arcsecond- to milliarcsecond-level localisations for large FRB samples, substantially reducing cross-matching uncertainties and enabling robust population-level studies of FRB host galaxies across cosmic time. In addition to CHORD, more immediate opportunities for improved FRB localisation and host identification are expected from the CHIME/FRB Outrigger system \citep{2025ApJ...993...55CHIME_outrigger}, which will provide arcsecond-level localisations for a growing sample of FRBs and enable the construction of new catalogues in the near future.
In concert with the depth and improved astrometry from LOFAR 2.0, this will make radio-selected FRB host samples both much larger and far more secure.

Second, although radio selection avoids the optical biases that disfavour dusty or high-redshift galaxies, it introduces its own selection effects. LoTSS is primarily sensitive to synchrotron-emitting galaxies and is therefore biased toward actively star-forming systems with sufficient radio brightness. Quiescent or low-SFR hosts may fall below the detection threshold, reducing the completeness of the radio-selected sample. This limitation is particularly relevant given that several well-localised FRBs have been associated with low-SFR or even early-type galaxies (e.g. FRB 20180924B; \citealt{2019Sci...365..565Bannister}, FRB 20190523; \citealt{2019Natur.572..352Ravi}, FRB 20240209A; \citealt{2025ApJ...979L..22Eftekhari,2025ApJ...979L..21Shah}).  Moreover, not all low-frequency radio emission necessarily traces star formation. Possible mechanisms for providing excess radio emission at low SFRs include pulsars, type Ia supernovae, residual contamination by active galactic nuclei (AGN), and varying magnetic field properties \citep{2021A&A...648A...6Smith, 2021PhRvD.103h3017Sudoh}. 
\cite{2025ApJS..280....6Chime_KKO} identified 21 new FRB host galaxies using the CHIME/FRB Outrigger KKO. Two of these candidates have radio counterparts in the LoTSS survey. FRB 20230926A is coincident with the extended 144 MHz source ILTJ75629.78+414836.4; however, the high probability of chance alignment disfavours a genuine association from a statistical perspective. In contrast, the repeating FRB 20231128A is associated with a luminous PRS detected in the FIRST survey, and a bright LoTSS counterpart (ILTJ131819.22+425958.9) is also found at the position of the FIRST source. The observed radio emission may arise either from a background AGN or from star-formation activity within the host galaxy. In our work, we mitigate these effects by combining radio data with independent multi-wavelength diagnostics. In particular, we use mid-infrared colour–colour information and available X-ray constraints to assess the presence of AGN activity independently of the radio emission. For nearby and spatially resolved galaxies, radio morphology provides an additional discriminator, allowing us to distinguish emission associated with the host galaxy from potential background AGN. While these approaches enable robust classification in most cases, we acknowledge that ambiguities remain for a small number of sources, which are therefore treated on a case-by-case basis in our analysis.

Within these limitations, the observed properties of our candidate hosts are consistent with the star-forming main sequence, in line with the hypothesis that at least some FRBs originate from young, short-lived progenitors such as magnetars formed via core-collapse supernovae (e.g. \citealt{2020Natur.587...59Bochenek,2023RvMP...95c5005Zhang}). However, this alignment may in part reflect the selection biases of LoTSS. As such, the current sample does not preclude contributions from alternative progenitor channels associated with older stellar populations or dynamical formation mechanisms. 
Disentangling these possibilities will require larger and more diverse samples, systematic comparisons across different FRB populations (including both repeaters and non-repeaters), as well as high-resolution imaging and multi-wavelength diagnostics to establish the physical origin of the detected radio emission.

To our knowledge, this is the first host-galaxy search conducted at low radio frequencies, whereas most FRB host galaxies to date have been identified using optical galaxy surveys. Having identified our host candidates through radio-selection, we now consider the association likelihood that would have been assigned using an optical method. We employed the commonly used Probabilistic Association of Transients to their Hosts
(\texttt{PATH})\footnote{\url{https://github.com/FRBs/astropath}} package \citep{2021ApJ...911...95A_path}, a Bayesian framework used for FRB host identification in optical. We constructed optical candidate lists using SDSS DR9 $g$-band magnitudes and Petrosian radii \citep{2012ApJS..203...21A_sdss9}. Each candidate was assigned a posterior probability based on its angular separation from the FRB, relative brightness, and assumed offset distribution. Using the default prior settings, our results show that in the optical $g$ band, MCG+11-11-014 has a $\sim 65$ percent probability of being the true host of FRB~20190303B, while SDSS~J110020.55+595454.4 has a $\sim 45$ percent probability of hosting FRB~20190502A. For FRB~20181214C, no optical counterpart attains a host probability above 30 percent. According to previous optical host-identification studies \citep{2021ApJ...919L..23Fong, 2023ApJ...954...80Gordon}, a secure host is typically required to have a chance-alignment probability of less than 10 percent, corresponding to a $>90$ percent probability of being the true host. The host galaxies selected in this work are more difficult to identify using optical-based methods, highlighting the need for larger samples of radio-selected host candidates to enable a robust comparison between radio and optical selection approaches.

Among our candidate hosts, we observe apparent discrepancies between different SFR indicators. In all three cases, the H$\alpha$-derived SFRs are systematically lower than those inferred from SED fitting or radio luminosity. 
While the small sample size precludes any general conclusion, these differences are plausibly attributable to a combination of internal dust attenuation and observational limitations, such as aperture effects or limited spectral resolution in the optical data. Similar discrepancies have been reported in other studies of dusty or compact galaxies (e.g. \citealt{2007ApJS..173..267Salim,2012ARA&A..50..531Kennicutt}). 
FRB~20171020A also has a radio counterpart, detected between 2.1 and 16.7~GHz. Using the measured spectral index and the 1.4~GHz luminosity–SFR relation, the inferred star formation rate is $0.18^{+2.54}_{-0.17}\,\mathrm{M\odot\,yr^{-1}}$. In comparison, the SFR derived from SED fitting is $0.09 \pm 0.01\,\mathrm{M_\odot\,yr^{-1}}$ \citep{2023PASA...40...29LeeWaddell}. The slight discrepancy between the two estimates can be attributed to the large intrinsic scatter in the radio–SFR relation, which provides only an order-of-magnitude approximation.  \citet{2020ApJ...901L..20Bhandari} reported radio detections of FRB~20191001A at 5.5 and 7.5~GHz, and estimated a star formation rate of $\sim 11.2\,\mathrm{M_\odot\,yr^{-1}}$, which is comparable to the SFR of $8.06 \pm 2.42\,\mathrm{M_\odot\,yr^{-1}}$ reported from the dust-corrected H$\alpha$ line fluxes 
by \citet{2020ApJ...903..152Heintz}.
In recent work on the host galaxy of FRB 20201124A \citep{2024ApJ...961...44Dong}, VLA observations at 1.5–6 GHz revealed that the host has a radio-inferred SFR approximately 2.5 times higher than that inferred from H$\alpha$ luminosity. As shown in their fig.~6, which compares optical and radio SFRs, most FRB hosts have only upper limits on their radio SFRs owing to sensitivity limitations. Although the frequency range of their observations differs from that used in this study, one possible explanation for our result is that we may be preferentially detecting outliers that are actively star-forming and heavily dust-obscured. Future work with larger samples and consistent multi-band observations will be essential to assess the prevalence and implications of such offsets among FRB host galaxies.

Overall, our results highlight both the promise and the current limitations of using low-frequency radio surveys to identify FRB host galaxies. As survey sensitivity, resolution, and sky coverage continue to improve, the methodologies developed here will serve as a foundation for future, larger-scale analyses of the FRB host population.

\section{Conclusions}
\label{sec:conclusion}

In this study, we investigated the host galaxy environments of CHIME/FRBs by cross-matching baseband-localised events with the LoTSS DR2 low-frequency radio source catalogue. Our selection strategy, which does not rely on optical detection, enables the identification of obscured or optically faint star-forming galaxies that may be overlooked by traditional methods.

Among 140 FRBs, 33 were found to lie within the LoTSS footprint, and 16 yielded potential extragalactic radio counterparts. Through multi-wavelength analysis, including photometric and spectroscopic data, spectral energy distribution (SED) fitting, and redshift constraints from the Macquart relation, we identified two secure and one tentative host candidates. All three exhibit signatures of star formation, though H$\alpha$-based SFR estimates fall below expectations, likely due to significant dust attenuation, as supported by infrared colours and compact morphologies.

Our results reinforce the utility of low-frequency radio data in complementing optical searches for FRB host galaxies. Despite the limited sample size and positional uncertainties, our approach demonstrates the potential of radio-based host identification and motivates future follow-up efforts combining deep imaging and spectroscopy to refine host associations and explore the diversity of FRB environments.

Looking ahead, LoTSS DR3 and the full CHIME/FRB baseband catalogue will together provide far greater opportunities to associate FRBs with their hosts. With LOFAR 2.0 offering deeper sensitivity, wider coverage, and sub-arcsecond astrometry, and with next-generation facilities such as CHORD delivering milli-arcsecond localisations, the prospects for robust, large-scale samples of FRB host galaxies are strong. Our pilot study demonstrates that radio continuum imaging can serve as an effective tool for identifying potential host galaxies, even in the absence of immediate optical confirmation. This approach offers a promising framework for future large-scale, statistically robust investigations of FRB environments and progenitors using upcoming radio survey data.

\section*{Acknowledgements}
We sincerely appreciate the anonymous referee’s constructive report on our work. We thank Dr. J.R. Callingham (ASTRON) and Dr. Inés Pastor-Marazuela (API/ASTRON) for their valuable suggestions on this work.

This research has made use of the NASA/IPAC Extragalactic Database, which is funded by the National Aeronautics and Space Administration and operated by the California Institute of Technology. This research has made use of the VizieR catalogue access tool, CDS,
 Strasbourg, France (DOI : 10.26093/cds/vizier). The original description of the VizieR service was published in 2000, A\&AS 143, 23. LOFAR data products were provided by the LOFAR Surveys Key Science project (LSKSP; https://lofar-surveys.org/) and were derived from observations with the International LOFAR Telescope (ILT). LOFAR \citep{2013A&A...556A...2VanHaaelem_LOFAR} is the Low Frequency Array designed and constructed by ASTRON. It has observing, data processing, and data storage facilities in several countries, which are owned by various parties (each with their own funding sources), and which are collectively operated by the ILT foundation under a joint scientific policy. The efforts of the LSKSP have benefited from funding from the European Research Council, NOVA, NWO, CNRS-INSU, the SURF Co-operative, the UK Science and Technology Funding Council and the Jülich Supercomputing Centre. The Liverpool Telescope is operated on the island of La Palma by Liverpool John Moores University in the Spanish Observatorio del Roque de los Muchachos of the Instituto de Astrofisica de Canarias with financial support from the UK Science and Technology Facilities Council. 

YZS is supported by a Chinese Scholarship Council award at the University of Leicester.
RLCS acknowledges support from the Leverhulme Trust grant RPG-2023-240.
AR acknowledges funding from the NWO Aspasia grant (number: 015.016.033 and funding from a Consolidator Grant from the Europe research and innovation programme (QuickBlitz, Project number 101170284). Views and opinions expressed are however those of the author(s) only and do not necessarily reflect those of the European Union or the European Research Council Executive Agency. Neither the European Union nor the granting authority can be held responsible for them. The authors acknowledge early testing of these ideas by Lee Tzabach through an undergraduate project at the University of Leicester. 

\section*{Data Availability}

The CHIME/FRB baseband localisation catalogue used in this study is publicly available at \url{https://www.chime-frb.ca/catalog}. The LoTSS DR2 radio source catalogue is available at \url{https://lofar-surveys.org}. Additional multi-wavelength data were obtained from public archives, including the Pan-STARRS and WISE surveys. 


\bibliographystyle{mnras}
\bibliography{frb}

@ARTICLE{2020ApJS..249....3A_sdss16,
       author = {{Ahumada}, Romina and {Allende Prieto}, Carlos and {Almeida}, Andr{\'e}s and {Anders}, Friedrich and {Anderson}, Scott F. and {Andrews}, Brett H. and {Anguiano}, Borja and {Arcodia}, Riccardo and {Armengaud}, Eric and {Aubert}, Marie and {Avila}, Santiago and {Avila-Reese}, Vladimir and {Badenes}, Carles and {Balland}, Christophe and {Barger}, Kat and {Barrera-Ballesteros}, Jorge K. and {Basu}, Sarbani and {Bautista}, Julian and {Beaton}, Rachael L. and {Beers}, Timothy C. and {Benavides}, B. Izamar T. and {Bender}, Chad F. and {Bernardi}, Mariangela and {Bershady}, Matthew and {Beutler}, Florian and {Bidin}, Christian Moni and {Bird}, Jonathan and {Bizyaev}, Dmitry and {Blanc}, Guillermo A. and {Blanton}, Michael R. and {Boquien}, M{\'e}d{\'e}ric and {Borissova}, Jura and {Bovy}, Jo and {Brandt}, W.~N. and {Brinkmann}, Jonathan and {Brownstein}, Joel R. and {Bundy}, Kevin and {Bureau}, Martin and {Burgasser}, Adam and {Burtin}, Etienne and {Cano-D{\'\i}az}, Mariana and {Capasso}, Raffaella and {Cappellari}, Michele and {Carrera}, Ricardo and {Chabanier}, Sol{\`e}ne and {Chaplin}, William and {Chapman}, Michael and {Cherinka}, Brian and {Chiappini}, Cristina and {Doohyun Choi}, Peter and {Chojnowski}, S. Drew and {Chung}, Haeun and {Clerc}, Nicolas and {Coffey}, Damien and {Comerford}, Julia M. and {Comparat}, Johan and {da Costa}, Luiz and {Cousinou}, Marie-Claude and {Covey}, Kevin and {Crane}, Jeffrey D. and {Cunha}, Katia and {Ilha}, Gabriele da Silva and {Dai}, Yu Sophia and {Damsted}, Sanna B. and {Darling}, Jeremy and {Davidson}, Jr., James W. and {Davies}, Roger and {Dawson}, Kyle and {De}, Nikhil and {de la Macorra}, Axel and {De Lee}, Nathan and {Queiroz}, Anna B{\'a}rbara de Andrade and {Deconto Machado}, Alice and {de la Torre}, Sylvain and {Dell'Agli}, Flavia and {du Mas des Bourboux}, H{\'e}lion and {Diamond-Stanic}, Aleksandar M. and {Dillon}, Sean and {Donor}, John and {Drory}, Niv and {Duckworth}, Chris and {Dwelly}, Tom and {Ebelke}, Garrett and {Eftekharzadeh}, Sarah and {Davis Eigenbrot}, Arthur and {Elsworth}, Yvonne P. and {Eracleous}, Mike and {Erfanianfar}, Ghazaleh and {Escoffier}, Stephanie and {Fan}, Xiaohui and {Farr}, Emily and {Fern{\'a}ndez-Trincado}, Jos{\'e} G. and {Feuillet}, Diane and {Finoguenov}, Alexis and {Fofie}, Patricia and {Fraser-McKelvie}, Amelia and {Frinchaboy}, Peter M. and {Fromenteau}, Sebastien and {Fu}, Hai and {Galbany}, Llu{\'\i}s and {Garcia}, Rafael A. and {Garc{\'\i}a-Hern{\'a}ndez}, D.~A. and {Garma Oehmichen}, Luis Alberto and {Ge}, Junqiang and {Geimba Maia}, Marcio Antonio and {Geisler}, Doug and {Gelfand}, Joseph and {Goddy}, Julian and {Gonzalez-Perez}, Violeta and {Grabowski}, Kathleen and {Green}, Paul and {Grier}, Catherine J. and {Guo}, Hong and {Guy}, Julien and {Harding}, Paul and {Hasselquist}, Sten and {Hawken}, Adam James and {Hayes}, Christian R. and {Hearty}, Fred and {Hekker}, S. and {Hogg}, David W. and {Holtzman}, Jon A. and {Horta}, Danny and {Hou}, Jiamin and {Hsieh}, Bau-Ching and {Huber}, Daniel and {Hunt}, Jason A.~S. and {Ider Chitham}, J. and {Imig}, Julie and {Jaber}, Mariana and {Jimenez Angel}, Camilo Eduardo and {Johnson}, Jennifer A. and {Jones}, Amy M. and {J{\"o}nsson}, Henrik and {Jullo}, Eric and {Kim}, Yerim and {Kinemuchi}, Karen and {Kirkpatrick}, IV, Charles C. and {Kite}, George W. and {Klaene}, Mark and {Kneib}, Jean-Paul and {Kollmeier}, Juna A. and {Kong}, Hui and {Kounkel}, Marina and {Krishnarao}, Dhanesh and {Lacerna}, Ivan and {Lan}, Ting-Wen and {Lane}, Richard R. and {Law}, David R. and {Le Goff}, Jean-Marc and {Leung}, Henry W. and {Lewis}, Hannah and {Li}, Cheng and {Lian}, Jianhui and {Lin}, Lihwai and {Long}, Dan and {Longa-Pe{\~n}a}, Pen{\'e}lope and {Lundgren}, Britt and {Lyke}, Brad W. and {Mackereth}, J. Ted and {MacLeod}, Chelsea L. and {Majewski}, Steven R. and {Manchado}, Arturo and {Maraston}, Claudia and {Martini}, Paul and {Masseron}, Thomas and {Masters}, Karen L. and {Mathur}, Savita and {McDermid}, Richard M. and {Merloni}, Andrea and {Merrifield}, Michael and {M{\'e}sz{\'a}ros}, Szabolcs and {Miglio}, Andrea and {Minniti}, Dante and {Minsley}, Rebecca and {Miyaji}, Takamitsu and {Mohammad}, Faizan Gohar and {Mosser}, Benoit and {Mueller}, Eva-Maria and {Muna}, Demitri and {Mu{\~n}oz-Guti{\'e}rrez}, Andrea and {Myers}, Adam D. and {Nadathur}, Seshadri and {Nair}, Preethi and {Nandra}, Kirpal and {Correa do Nascimento}, Janaina and {Nevin}, Rebecca Jean and {Newman}, Jeffrey A. and {Nidever}, David L. and {Nitschelm}, Christian and {Noterdaeme}, Pasquier and {O'Connell}, Julia E. and {Olmstead}, Matthew D. and {Oravetz}, Daniel and {Oravetz}, Audrey and {Osorio}, Yeisson and {Pace}, Zachary J. and {Padilla}, Nelson and {Palanque-Delabrouille}, Nathalie and {Palicio}, Pedro A.},
        title = "{The 16th Data Release of the Sloan Digital Sky Surveys: First Release from the APOGEE-2 Southern Survey and Full Release of eBOSS Spectra}",
      journal = {\apjs},
         year = 2020,
        month = jul,
       volume = {249},
       number = {1},
          eid = {3},
        pages = {3},
          doi = {10.3847/1538-4365/ab929e},
archivePrefix = {arXiv},
       eprint = {1912.02905},
 primaryClass = {astro-ph.GA},
       adsurl = {https://ui.adsabs.harvard.edu/abs/2020ApJS..249....3A}
}

@ARTICLE{2021MNRAS.501.5319Arcus,
       author = {{Arcus}, W.~R. and {Macquart}, J.-P. and {Sammons}, M.~W. and {James}, C.~W. and {Ekers}, R.~D.},
        title = "{The fast radio burst dispersion measure distribution}",
      journal = {\mnras},
         year = 2021,
        month = mar,
       volume = {501},
       number = {4},
        pages = {5319-5329},
          doi = {10.1093/mnras/staa3948},
archivePrefix = {arXiv},
       eprint = {2012.15051},
 primaryClass = {astro-ph.CO},
       adsurl = {https://ui.adsabs.harvard.edu/abs/2021MNRAS.501.5319A}
}

@ARTICLE{2012ApJS..203...21A_sdss9,
       author = {{Ahn}, Christopher P. and {Alexandroff}, Rachael and {Allende Prieto}, Carlos and {Anderson}, Scott F. and {Anderton}, Timothy and {Andrews}, Brett H. and {Aubourg}, {\'E}ric and {Bailey}, Stephen and {Balbinot}, Eduardo and {Barnes}, Rory and {Bautista}, Julian and {Beers}, Timothy C. and {Beifiori}, Alessandra and {Berlind}, Andreas A. and {Bhardwaj}, Vaishali and {Bizyaev}, Dmitry and {Blake}, Cullen H. and {Blanton}, Michael R. and {Blomqvist}, Michael and {Bochanski}, John J. and {Bolton}, Adam S. and {Borde}, Arnaud and {Bovy}, Jo and {Brandt}, W.~N. and {Brinkmann}, J. and {Brown}, Peter J. and {Brownstein}, Joel R. and {Bundy}, Kevin and {Busca}, N.~G. and {Carithers}, William and {Carnero}, Aurelio R. and {Carr}, Michael A. and {Casetti-Dinescu}, Dana I. and {Chen}, Yanmei and {Chiappini}, Cristina and {Comparat}, Johan and {Connolly}, Natalia and {Crepp}, Justin R. and {Cristiani}, Stefano and {Croft}, Rupert A.~C. and {Cuesta}, Antonio J. and {da Costa}, Luiz N. and {Davenport}, James R.~A. and {Dawson}, Kyle S. and {de Putter}, Roland and {De Lee}, Nathan and {Delubac}, Timoth{\'e}e and {Dhital}, Saurav and {Ealet}, Anne and {Ebelke}, Garrett L. and {Edmondson}, Edward M. and {Eisenstein}, Daniel J. and {Escoffier}, S. and {Esposito}, Massimiliano and {Evans}, Michael L. and {Fan}, Xiaohui and {Femen{\'\i}a Castell{\'a}}, Bruno and {Fern{\'a}ndez Alvar}, Emma and {Ferreira}, Leticia D. and {Filiz Ak}, N. and {Finley}, Hayley and {Fleming}, Scott W. and {Font-Ribera}, Andreu and {Frinchaboy}, Peter M. and {Garc{\'\i}a-Hern{\'a}ndez}, D.~A. and {Garc{\'\i}a P{\'e}rez}, A.~E. and {Ge}, Jian and {G{\'e}nova-Santos}, R. and {Gillespie}, Bruce A. and {Girardi}, L{\'e}o and {Gonz{\'a}lez Hern{\'a}ndez}, Jonay I. and {Grebel}, Eva K. and {Gunn}, James E. and {Guo}, Hong and {Haggard}, Daryl and {Hamilton}, Jean-Christophe and {Harris}, David W. and {Hawley}, Suzanne L. and {Hearty}, Frederick R. and {Ho}, Shirley and {Hogg}, David W. and {Holtzman}, Jon A. and {Honscheid}, Klaus and {Huehnerhoff}, J. and {Ivans}, Inese I. and {Ivezi{\'c}}, {\v{Z}}eljko and {Jacobson}, Heather R. and {Jiang}, Linhua and {Johansson}, Jonas and {Johnson}, Jennifer A. and {Kauffmann}, Guinevere and {Kirkby}, David and {Kirkpatrick}, Jessica A. and {Klaene}, Mark A. and {Knapp}, Gillian R. and {Kneib}, Jean-Paul and {Le Goff}, Jean-Marc and {Leauthaud}, Alexie and {Lee}, Khee-Gan and {Lee}, Young Sun and {Long}, Daniel C. and {Loomis}, Craig P. and {Lucatello}, Sara and {Lundgren}, Britt and {Lupton}, Robert H. and {Ma}, Bo and {Ma}, Zhibo and {MacDonald}, Nicholas and {Mack}, Claude E. and {Mahadevan}, Suvrath and {Maia}, Marcio A.~G. and {Majewski}, Steven R. and {Makler}, Martin and {Malanushenko}, Elena and {Malanushenko}, Viktor and {Manchado}, A. and {Mandelbaum}, Rachel and {Manera}, Marc and {Maraston}, Claudia and {Margala}, Daniel and {Martell}, Sarah L. and {McBride}, Cameron K. and {McGreer}, Ian D. and {McMahon}, Richard G. and {M{\'e}nard}, Brice and {Meszaros}, Sz. and {Miralda-Escud{\'e}}, Jordi and {Montero-Dorta}, Antonio D. and {Montesano}, Francesco and {Morrison}, Heather L. and {Muna}, Demitri and {Munn}, Jeffrey A. and {Murayama}, Hitoshi and {Myers}, Adam D. and {Neto}, A.~F. and {Nguyen}, Duy Cuong and {Nichol}, Robert C. and {Nidever}, David L. and {Noterdaeme}, Pasquier and {Nuza}, Sebasti{\'a}n E. and {Ogando}, Ricardo L.~C. and {Olmstead}, Matthew D. and {Oravetz}, Daniel J. and {Owen}, Russell and {Padmanabhan}, Nikhil and {Palanque-Delabrouille}, Nathalie and {Pan}, Kaike and {Parejko}, John K. and {Parihar}, Prachi and {P{\^a}ris}, Isabelle and {Pattarakijwanich}, Petchara and {Pepper}, Joshua and {Percival}, Will J. and {P{\'e}rez-Fournon}, Ismael and {P{\'e}rez-R{\`a}fols}, Ignasi and {Petitjean}, Patrick and {Pforr}, Janine and {Pieri}, Matthew M. and {Pinsonneault}, Marc H. and {Porto de Mello}, G.~F. and {Prada}, Francisco and {Price-Whelan}, Adrian M. and {Raddick}, M. Jordan and {Rebolo}, Rafael and {Rich}, James and {Richards}, Gordon T. and {Robin}, Annie C. and {Rocha-Pinto}, Helio J. and {Rockosi}, Constance M. and {Roe}, Natalie A. and {Ross}, Ashley J. and {Ross}, Nicholas P. and {Rossi}, Graziano and {Rubi{\~n}o-Martin}, J.~A. and {Samushia}, Lado and {Sanchez Almeida}, J. and {S{\'a}nchez}, Ariel G. and {Santiago}, Bas{\'\i}lio and {Sayres}, Conor and {Schlegel}, David J. and {Schlesinger}, Katharine J. and {Schmidt}, Sarah J. and {Schneider}, Donald P. and {Schultheis}, Mathias and {Schwope}, Axel D. and {Sc{\'o}ccola}, C.~G. and {Seljak}, Uros and {Sheldon}, Erin and {Shen}, Yue and {Shu}, Yiping and {Simmerer}, Jennifer and {Simmons}, Audrey E. and {Skibba}, Ramin A. and {Skrutskie}, M.~F. and {Slosar}, A. and {Sobreira}, Flavia and {Sobeck}, Jennifer S. and {Stassun}, Keivan G. and {Steele}, Oliver and {Steinmetz}, Matthias},
        title = "{The Ninth Data Release of the Sloan Digital Sky Survey: First Spectroscopic Data from the SDSS-III Baryon Oscillation Spectroscopic Survey}",
      journal = {\apjs},
         year = 2012,
        month = dec,
       volume = {203},
       number = {2},
          eid = {21},
        pages = {21},
          doi = {10.1088/0067-0049/203/2/21},
archivePrefix = {arXiv},
       eprint = {1207.7137},
 primaryClass = {astro-ph.IM},
       adsurl = {https://ui.adsabs.harvard.edu/abs/2012ApJS..203...21A}
}

@ARTICLE{2021ApJ...911...95A_path,
       author = {{Aggarwal}, Kshitij and {Budav{\'a}ri}, Tam{\'a}s and {Deller}, Adam T. and {Eftekhari}, Tarraneh and {James}, Clancy W. and {Prochaska}, J. Xavier and {Tendulkar}, Shriharsh P.},
        title = "{Probabilistic Association of Transients to their Hosts (PATH)}",
      journal = {\apj},
         year = 2021,
        month = apr,
       volume = {911},
       number = {2},
          eid = {95},
        pages = {95},
          doi = {10.3847/1538-4357/abe8d2},
archivePrefix = {arXiv},
       eprint = {2102.10627},
 primaryClass = {astro-ph.HE},
       adsurl = {https://ui.adsabs.harvard.edu/abs/2021ApJ...911...95A}
}

@ARTICLE{2004MNRAS.351.1151Brinchmann,
       author = {{Brinchmann}, J. and {Charlot}, S. and {White}, S.~D.~M. and {Tremonti}, C. and {Kauffmann}, G. and {Heckman}, T. and {Brinkmann}, J.},
        title = "{The physical properties of star-forming galaxies in the low-redshift Universe}",
      journal = {\mnras},
         year = 2004,
        month = jul,
       volume = {351},
       number = {4},
        pages = {1151-1179},
          doi = {10.1111/j.1365-2966.2004.07881.x},
archivePrefix = {arXiv},
       eprint = {astro-ph/0311060},
 primaryClass = {astro-ph},
       adsurl = {https://ui.adsabs.harvard.edu/abs/2004MNRAS.351.1151B}
}

@ARTICLE{2003MNRAS.344.1000Bruzual,
       author = {{Bruzual}, G. and {Charlot}, S.},
        title = "{Stellar population synthesis at the resolution of 2003}",
      journal = {\mnras},
         year = 2003,
        month = oct,
       volume = {344},
       number = {4},
        pages = {1000-1028},
          doi = {10.1046/j.1365-8711.2003.06897.x},
archivePrefix = {arXiv},
       eprint = {astro-ph/0309134},
 primaryClass = {astro-ph},
       adsurl = {https://ui.adsabs.harvard.edu/abs/2003MNRAS.344.1000B}
}

@ARTICLE{2011Ap&SS.335..161B_galex,
       author = {{Bianchi}, L. and {Herald}, J. and {Efremova}, B. and {Girardi}, L. and {Zabot}, A. and {Marigo}, P. and {Conti}, A. and {Shiao}, B.},
        title = "{GALEX catalogs of UV sources: statistical properties and sample science applications: hot white dwarfs in the Milky Way}",
      journal = {\apss},
         year = 2011,
        month = sep,
       volume = {335},
       number = {1},
        pages = {161-169},
          doi = {10.1007/s10509-010-0581-x},
       adsurl = {https://ui.adsabs.harvard.edu/abs/2011Ap&SS.335..161B}
}

@ARTICLE{2025ApJ...995...51Balasubramanian,
       author = {{Balasubramanian}, Arvind and {Bhardwaj}, Mohit and {Tendulkar}, Shriharsh P.},
        title = "{Continued Radio Observations of the Persistent Radio Source Associated with FRB 20190520B Provide Insights into Its Origin}",
      journal = {\apj},
         year = 2025,
        month = dec,
       volume = {995},
       number = {1},
          eid = {51},
        pages = {51},
          doi = {10.3847/1538-4357/ae1cc4},
archivePrefix = {arXiv},
       eprint = {2507.03113},
 primaryClass = {astro-ph.HE},
       adsurl = {https://ui.adsabs.harvard.edu/abs/2025ApJ...995...51B}
}

@ARTICLE{2025A&A...696A..81Bernales,
       author = {{Bernales-Cortes}, Lucas and {Tejos}, Nicolas and {Prochaska}, J. Xavier and {Khrykin}, Ilya S. and {Marnoch}, Lachlan and {Ryder}, Stuart D. and {Shannon}, Ryan M.},
        title = "{Empirical estimation of host galaxy dispersion measure toward well-localized fast radio bursts}",
      journal = {\aap},
         year = 2025,
        month = apr,
       volume = {696},
          eid = {A81},
        pages = {A81},
          doi = {10.1051/0004-6361/202452026},
archivePrefix = {arXiv},
       eprint = {2501.14063},
 primaryClass = {astro-ph.GA},
       adsurl = {https://ui.adsabs.harvard.edu/abs/2025A&A...696A..81B}
}

@ARTICLE{2020ApJ...901L..20Bhandari,
       author = {{Bhandari}, Shivani and {Bannister}, Keith W. and {Lenc}, Emil and {Cho}, Hyerin and {Ekers}, Ron and {Day}, Cherie K. and {Deller}, Adam T. and {Flynn}, Chris and {James}, Clancy W. and {Macquart}, Jean-Pierre and {Mahony}, Elizabeth K. and {Marnoch}, Lachlan and {Moss}, Vanessa A. and {Phillips}, Chris and {Prochaska}, J. Xavier and {Qiu}, Hao and {Ryder}, Stuart D. and {Shannon}, Ryan M. and {Tejos}, Nicolas and {Wong}, O. Ivy},
        title = "{Limits on Precursor and Afterglow Radio Emission from a Fast Radio Burst in a Star-forming Galaxy}",
      journal = {\apjl},
         year = 2020,
        month = oct,
       volume = {901},
       number = {2},
          eid = {L20},
        pages = {L20},
          doi = {10.3847/2041-8213/abb462},
archivePrefix = {arXiv},
       eprint = {2008.12488},
 primaryClass = {astro-ph.HE},
       adsurl = {https://ui.adsabs.harvard.edu/abs/2020ApJ...901L..20B}
}

@ARTICLE{2022AJ....163...69Bhandari,
       author = {{Bhandari}, Shivani and {Heintz}, Kasper E. and {Aggarwal}, Kshitij and {Marnoch}, Lachlan and {Day}, Cherie K. and {Sydnor}, Jessica and {Burke-Spolaor}, Sarah and {Law}, Casey J. and {Xavier Prochaska}, J. and {Tejos}, Nicolas and {Bannister}, Keith W. and {Butler}, Bryan J. and {Deller}, Adam T. and {Ekers}, R.~D. and {Flynn}, Chris and {Fong}, Wen-fai and {James}, Clancy W. and {Lazio}, T. Joseph W. and {Luo}, Rui and {Mahony}, Elizabeth K. and {Ryder}, Stuart D. and {Sadler}, Elaine M. and {Shannon}, Ryan M. and {Han}, JinLin and {Lee}, Kejia and {Zhang}, Bing},
        title = "{Characterizing the Fast Radio Burst Host Galaxy Population and its Connection to Transients in the Local and Extragalactic Universe}",
      journal = {\aj},
         year = 2022,
        month = feb,
       volume = {163},
       number = {2},
          eid = {69},
        pages = {69},
          doi = {10.3847/1538-3881/ac3aec},
archivePrefix = {arXiv},
       eprint = {2108.01282},
 primaryClass = {astro-ph.HE},
       adsurl = {https://ui.adsabs.harvard.edu/abs/2022AJ....163...69B}
}

@ARTICLE{2024Natur.632.1014Bruni,
       author = {{Bruni}, Gabriele and {Piro}, Luigi and {Yang}, Yuan-Pei and {Quai}, Salvatore and {Zhang}, Bing and {Palazzi}, Eliana and {Nicastro}, Luciano and {Feruglio}, Chiara and {Tripodi}, Roberta and {O'Connor}, Brendan and {Gardini}, Angela and {Savaglio}, Sandra and {Rossi}, Andrea and {Nicuesa Guelbenzu}, Ana M. and {Paladino}, Rosita},
        title = "{A nebular origin for the persistent radio emission of fast radio bursts}",
      journal = {\nat},
         year = 2024,
        month = aug,
       volume = {632},
       number = {8027},
        pages = {1014-1016},
          doi = {10.1038/s41586-024-07782-6},
archivePrefix = {arXiv},
       eprint = {2312.15296},
 primaryClass = {astro-ph.HE},
       adsurl = {https://ui.adsabs.harvard.edu/abs/2024Natur.632.1014B}
}

@ARTICLE{2024Natur.634.1065Bhardwaj,
       author = {{Bhardwaj}, Mohit and {Lee}, Jimin and {Ji}, Kevin},
        title = "{Selection bias obfuscates the discovery of fast radio burst sources}",
      journal = {\nat},
         year = 2024,
        month = oct,
       volume = {634},
       number = {8036},
        pages = {1065-1069},
          doi = {10.1038/s41586-024-08065-w},
archivePrefix = {arXiv},
       eprint = {2408.01876},
 primaryClass = {astro-ph.HE},
       adsurl = {https://ui.adsabs.harvard.edu/abs/2024Natur.634.1065B}
}

@ARTICLE{1995ApJ...450..559Becker_FIRST,
       author = {{Becker}, Robert H. and {White}, Richard L. and {Helfand}, David J.},
        title = "{The FIRST Survey: Faint Images of the Radio Sky at Twenty Centimeters}",
      journal = {\apj},
         year = 1995,
        month = sep,
       volume = {450},
        pages = {559},
          doi = {10.1086/176166},
       adsurl = {https://ui.adsabs.harvard.edu/abs/1995ApJ...450..559B}
}

@ARTICLE{2024ApJ...971L..51Bhardwaj,
       author = {{Bhardwaj}, Mohit and {Michilli}, Daniele and {Kirichenko}, Aida Yu. and {Modilim}, Obinna and {Shin}, Kaitlyn and {Kaspi}, Victoria M. and {Andersen}, Bridget C. and {Cassanelli}, Tomas and {Brar}, Charanjot and {Chatterjee}, Shami and {Cook}, Amanda M. and {Dong}, Fengqiu Adam and {Fonseca}, Emmanuel and {Gaensler}, B.~M. and {Ibik}, Adaeze L. and {Kaczmarek}, J.~F. and {Lanman}, Adam E. and {Leung}, Calvin and {Masui}, K.~W. and {Pandhi}, Ayush and {Pearlman}, Aaron B. and {Petroff}, Emily and {Pleunis}, Ziggy and {Prochaska}, J. Xavier and {Rafiei-Ravandi}, Masoud and {Sand}, Ketan R. and {Scholz}, Paul and {Smith}, Kendrick M.},
        title = "{Host Galaxies for Four Nearby CHIME/FRB Sources and the Local Universe FRB Host Galaxy Population}",
      journal = {\apjl},
         year = 2024,
        month = aug,
       volume = {971},
       number = {2},
          eid = {L51},
        pages = {L51},
          doi = {10.3847/2041-8213/ad64d1},
archivePrefix = {arXiv},
       eprint = {2310.10018},
 primaryClass = {astro-ph.HE},
       adsurl = {https://ui.adsabs.harvard.edu/abs/2024ApJ...971L..51B}
}

@ARTICLE{2025arXiv250623861Bhardwaj,
       author = {{Bhardwaj}, Mohit and {Balasubramanian}, Arvind and {Kaushal}, Yasha and {Tendulkar}, Shriharsh P.},
        title = "{Constraining the Origin of FRB 20121102A's Persistent Radio Source with Long-Term Radio Observations}",
      journal = {arXiv e-prints},
         year = 2025,
        month = jun,
          eid = {arXiv:2506.23861},
        pages = {arXiv:2506.23861},
          doi = {10.48550/arXiv.2506.23861},
archivePrefix = {arXiv},
       eprint = {2506.23861},
 primaryClass = {astro-ph.HE},
       adsurl = {https://ui.adsabs.harvard.edu/abs/2025arXiv250623861B}
}

@ARTICLE{1938ApJ....88...52Baker,
       author = {{Baker}, James G. and {Menzel}, Donald H.},
        title = "{Physical Processes in Gaseous Nebulae. III. The Balmer Decrement.}",
      journal = {\apj},
         year = 1938,
        month = jul,
       volume = {88},
        pages = {52},
          doi = {10.1086/143959},
       adsurl = {https://ui.adsabs.harvard.edu/abs/1938ApJ....88...52B}
}

@ARTICLE{2019A&A...622A.103Boquien,
       author = {{Boquien}, M. and {Burgarella}, D. and {Roehlly}, Y. and {Buat}, V. and {Ciesla}, L. and {Corre}, D. and {Inoue}, A.~K. and {Salas}, H.},
        title = "{CIGALE: a python Code Investigating GALaxy Emission}",
      journal = {\aap},
         year = 2019,
        month = feb,
       volume = {622},
          eid = {A103},
        pages = {A103},
          doi = {10.1051/0004-6361/201834156},
archivePrefix = {arXiv},
       eprint = {1811.03094},
 primaryClass = {astro-ph.GA},
       adsurl = {https://ui.adsabs.harvard.edu/abs/2019A&A...622A.103B}
}

@ARTICLE{2020Natur.587...59Bochenek,
       author = {{Bochenek}, C.~D. and {Ravi}, V. and {Belov}, K.~V. and {Hallinan}, G. and {Kocz}, J. and {Kulkarni}, S.~R. and {McKenna}, D.~L.},
        title = "{A fast radio burst associated with a Galactic magnetar}",
      journal = {\nat},
         year = 2020,
        month = nov,
       volume = {587},
       number = {7832},
        pages = {59-62},
          doi = {10.1038/s41586-020-2872-x},
archivePrefix = {arXiv},
       eprint = {2005.10828},
 primaryClass = {astro-ph.HE},
       adsurl = {https://ui.adsabs.harvard.edu/abs/2020Natur.587...59B}
}

@ARTICLE{2024arXiv241201478Bruni,
       author = {{Bruni}, G. and {Piro}, L. and {Yang}, Y. -P. and {Palazzi}, E. and {Nicastro}, L. and {Rossi}, A. and {Savaglio}, S. and {Maiorano}, E. and {Zhang}, B.},
        title = "{Discovery of a PRS associated with FRB 20240114A}",
      journal = {arXiv e-prints},
         year = 2024,
        month = dec,
          eid = {arXiv:2412.01478},
        pages = {arXiv:2412.01478},
          doi = {10.48550/arXiv.2412.01478},
archivePrefix = {arXiv},
       eprint = {2412.01478},
 primaryClass = {astro-ph.HE},
       adsurl = {https://ui.adsabs.harvard.edu/abs/2024arXiv241201478B}
}

@ARTICLE{2023ApJ...958L..19Bhandari,
       author = {{Bhandari}, Shivani and {Marcote}, Benito and {Sridhar}, Navin and {Eftekhari}, Tarraneh and {Hessels}, Jason W.~T. and {Hewitt}, Dant{\'e} M. and {Kirsten}, Franz and {Ould-Boukattine}, Omar S. and {Paragi}, Zsolt and {Snelders}, Mark P.},
        title = "{Constraints on the Persistent Radio Source Associated with FRB 20190520B Using the European VLBI Network}",
      journal = {\apjl},
         year = 2023,
        month = dec,
       volume = {958},
       number = {2},
          eid = {L19},
        pages = {L19},
          doi = {10.3847/2041-8213/ad083f},
archivePrefix = {arXiv},
       eprint = {2308.12801},
 primaryClass = {astro-ph.HE},
       adsurl = {https://ui.adsabs.harvard.edu/abs/2023ApJ...958L..19B}
}

@ARTICLE{2019Sci...365..565Bannister,
       author = {{Bannister}, K.~W. and {Deller}, A.~T. and {Phillips}, C. and {Macquart}, J. -P. and {Prochaska}, J.~X. and {Tejos}, N. and {Ryder}, S.~D. and {Sadler}, E.~M. and {Shannon}, R.~M. and {Simha}, S. and {Day}, C.~K. and {McQuinn}, M. and {North-Hickey}, F.~O. and {Bhandari}, S. and {Arcus}, W.~R. and {Bennert}, V.~N. and {Burchett}, J. and {Bouwhuis}, M. and {Dodson}, R. and {Ekers}, R.~D. and {Farah}, W. and {Flynn}, C. and {James}, C.~W. and {Kerr}, M. and {Lenc}, E. and {Mahony}, E.~K. and {O'Meara}, J. and {Os{\l}owski}, S. and {Qiu}, H. and {Treu}, T. and {U}, V. and {Bateman}, T.~J. and {Bock}, D.~C. -J. and {Bolton}, R.~J. and {Brown}, A. and {Bunton}, J.~D. and {Chippendale}, A.~P. and {Cooray}, F.~R. and {Cornwell}, T. and {Gupta}, N. and {Hayman}, D.~B. and {Kesteven}, M. and {Koribalski}, B.~S. and {MacLeod}, A. and {McClure-Griffiths}, N.~M. and {Neuhold}, S. and {Norris}, R.~P. and {Pilawa}, M.~A. and {Qiao}, R. -Y. and {Reynolds}, J. and {Roxby}, D.~N. and {Shimwell}, T.~W. and {Voronkov}, M.~A. and {Wilson}, C.~D.},
        title = "{A single fast radio burst localized to a massive galaxy at cosmological distance}",
      journal = {Science},
         year = 2019,
        month = aug,
       volume = {365},
       number = {6453},
        pages = {565-570},
          doi = {10.1126/science.aaw5903},
archivePrefix = {arXiv},
       eprint = {1906.11476},
 primaryClass = {astro-ph.HE},
       adsurl = {https://ui.adsabs.harvard.edu/abs/2019Sci...365..565B}
}

@ARTICLE{2023MNRAS.523.1729Best,
       author = {{Best}, P.~N. and {Kondapally}, R. and {Williams}, W.~L. and {Cochrane}, R.~K. and {Duncan}, K.~J. and {Hale}, C.~L. and {Haskell}, P. and {Ma{\l}ek}, K. and {McCheyne}, I. and {Smith}, D.~J.~B. and {Wang}, L. and {Botteon}, A. and {Bonato}, M. and {Bondi}, M. and {Calistro Rivera}, G. and {Gao}, F. and {G{\"u}rkan}, G. and {Hardcastle}, M.~J. and {Jarvis}, M.~J. and {Mingo}, B. and {Miraghaei}, H. and {Morabito}, L.~K. and {Nisbet}, D. and {Prandoni}, I. and {R{\"o}ttgering}, H.~J.~A. and {Sabater}, J. and {Shimwell}, T. and {Tasse}, C. and {van Weeren}, R.},
        title = "{The LOFAR Two-metre Sky Survey: Deep Fields data release 1. V. Survey description, source classifications, and host galaxy properties}",
      journal = {\mnras},
         year = 2023,
        month = aug,
       volume = {523},
       number = {2},
        pages = {1729-1755},
          doi = {10.1093/mnras/stad1308},
archivePrefix = {arXiv},
       eprint = {2305.05782},
 primaryClass = {astro-ph.GA},
       adsurl = {https://ui.adsabs.harvard.edu/abs/2023MNRAS.523.1729B}
}

@MISC{2013wise.rept....1Cutri_wise,
       author = {{Cutri}, R.~M. and {Wright}, E.~L. and {Conrow}, T. and {Fowler}, J.~W. and {Eisenhardt}, P.~R.~M. and {Grillmair}, C. and {Kirkpatrick}, J.~D. and {Masci}, F. and {McCallon}, H.~L. and {Wheelock}, S.~L. and {Fajardo-Acosta}, S. and {Yan}, L. and {Benford}, D. and {Harbut}, M. and {Jarrett}, T. and {Lake}, S. and {Leisawitz}, D. and {Ressler}, M.~E. and {Stanford}, S.~A. and {Tsai}, C.~W. and {Liu}, F. and {Helou}, G. and {Mainzer}, A. and {Gettings}, D. and {Gonzalez}, A. and {Hoffman}, D. and {Marsh}, K.~A. and {Padgett}, D. and {Skrutskie}, M.~F. and {Beck}, R.~P. and {Papin}, M. and {Wittman}, M.},
        title = "{Explanatory Supplement to the AllWISE Data Release Products}",
 howpublished = {Explanatory Supplement to the AllWISE Data Release Products, by R. M. Cutri et al.},
         year = 2013,
        month = nov,
        pages = {1},
       adsurl = {https://ui.adsabs.harvard.edu/abs/2013wise.rept....1C}
}

@ARTICLE{1989ApJ...345..245CCM,
       author = {{Cardelli}, Jason A. and {Clayton}, Geoffrey C. and {Mathis}, John S.},
        title = "{The Relationship between Infrared, Optical, and Ultraviolet Extinction}",
      journal = {\apj},
         year = 1989,
        month = oct,
       volume = {345},
        pages = {245},
          doi = {10.1086/167900},
       adsurl = {https://ui.adsabs.harvard.edu/abs/1989ApJ...345..245C}
}

@ARTICLE{1998AJ....115.1693Condon_NVSS,
       author = {{Condon}, J.~J. and {Cotton}, W.~D. and {Greisen}, E.~W. and {Yin}, Q.~F. and {Perley}, R.~A. and {Taylor}, G.~B. and {Broderick}, J.~J.},
        title = "{The NRAO VLA Sky Survey}",
      journal = {\aj},
         year = 1998,
        month = may,
       volume = {115},
       number = {5},
        pages = {1693-1716},
          doi = {10.1086/300337},
       adsurl = {https://ui.adsabs.harvard.edu/abs/1998AJ....115.1693C}
}

@ARTICLE{2017Natur.541...58Chatterjee,
       author = {{Chatterjee}, S. and {Law}, C.~J. and {Wharton}, R.~S. and {Burke-Spolaor}, S. and {Hessels}, J.~W.~T. and {Bower}, G.~C. and {Cordes}, J.~M. and {Tendulkar}, S.~P. and {Bassa}, C.~G. and {Demorest}, P. and {Butler}, B.~J. and {Seymour}, A. and {Scholz}, P. and {Abruzzo}, M.~W. and {Bogdanov}, S. and {Kaspi}, V.~M. and {Keimpema}, A. and {Lazio}, T.~J.~W. and {Marcote}, B. and {McLaughlin}, M.~A. and {Paragi}, Z. and {Ransom}, S.~M. and {Rupen}, M. and {Spitler}, L.~G. and {van Langevelde}, H.~J.},
        title = "{A direct localization of a fast radio burst and its host}",
      journal = {\nat},
         year = 2017,
        month = jan,
       volume = {541},
       number = {7635},
        pages = {58-61},
          doi = {10.1038/nature20797},
archivePrefix = {arXiv},
       eprint = {1701.01098},
 primaryClass = {astro-ph.HE},
       adsurl = {https://ui.adsabs.harvard.edu/abs/2017Natur.541...58C}
}

@ARTICLE{2018ApJ...863...48CHIME,
       author = {{CHIME/FRB Collaboration} and {Amiri}, M. and {Bandura}, K. and {Berger}, P. and {Bhardwaj}, M. and {Boyce}, M.~M. and {Boyle}, P.~J. and {Brar}, C. and {Burhanpurkar}, M. and {Chawla}, P. and {Chowdhury}, J. and {Cliche}, J. -F. and {Cranmer}, M.~D. and {Cubranic}, D. and {Deng}, M. and {Denman}, N. and {Dobbs}, M. and {Fandino}, M. and {Fonseca}, E. and {Gaensler}, B.~M. and {Giri}, U. and {Gilbert}, A.~J. and {Good}, D.~C. and {Guliani}, S. and {Halpern}, M. and {Hinshaw}, G. and {H{\"o}fer}, C. and {Josephy}, A. and {Kaspi}, V.~M. and {Landecker}, T.~L. and {Lang}, D. and {Liao}, H. and {Masui}, K.~W. and {Mena-Parra}, J. and {Naidu}, A. and {Newburgh}, L.~B. and {Ng}, C. and {Patel}, C. and {Pen}, U. -L. and {Pinsonneault-Marotte}, T. and {Pleunis}, Z. and {Rafiei Ravandi}, M. and {Ransom}, S.~M. and {Renard}, A. and {Scholz}, P. and {Sigurdson}, K. and {Siegel}, S.~R. and {Smith}, K.~M. and {Stairs}, I.~H. and {Tendulkar}, S.~P. and {Vanderlinde}, K. and {Wiebe}, D.~V.},
        title = "{The CHIME Fast Radio Burst Project: System Overview}",
      journal = {\apj},
         year = 2018,
        month = aug,
       volume = {863},
       number = {1},
          eid = {48},
        pages = {48},
          doi = {10.3847/1538-4357/aad188},
archivePrefix = {arXiv},
       eprint = {1803.11235},
 primaryClass = {astro-ph.IM},
       adsurl = {https://ui.adsabs.harvard.edu/abs/2018ApJ...863...48C}
}

@ARTICLE{2025ApJ...982..203Chen,
       author = {{Chen}, Xiang-Lei and {Tsai}, Chao-Wei and {Stern}, Daniel and {Bochenek}, Christopher D. and {Chatterjee}, Shami and {Law}, Casey and {Li}, Di and {Niu}, Chen-hui and {Niino}, Yuu and {Feng}, Yi and {Wang}, Pei and {Assef}, Roberto J. and {Li}, Guo-dong and {Lake}, Sean E. and {Luo}, Gan and {Liao}, Mai},
        title = "{The Host Galaxy of FRB 20190520B and Its Unique Ionized Gas Distribution}",
      journal = {\apj},
         year = 2025,
        month = apr,
       volume = {982},
       number = {2},
          eid = {203},
        pages = {203},
          doi = {10.3847/1538-4357/adb84d},
archivePrefix = {arXiv},
       eprint = {2503.01740},
 primaryClass = {astro-ph.GA},
       adsurl = {https://ui.adsabs.harvard.edu/abs/2025ApJ...982..203C}
}

@ARTICLE{2002astro.ph..7156Cordes_NE2001,
       author = {{Cordes}, J.~M. and {Lazio}, T.~J.~W.},
        title = "{NE2001.I. A New Model for the Galactic Distribution of Free Electrons and its Fluctuations}",
      journal = {arXiv e-prints},
         year = 2002,
        month = jul,
          eid = {astro-ph/0207156},
        pages = {astro-ph/0207156},
          doi = {10.48550/arXiv.astro-ph/0207156},
archivePrefix = {arXiv},
       eprint = {astro-ph/0207156},
 primaryClass = {astro-ph},
       adsurl = {https://ui.adsabs.harvard.edu/abs/2002astro.ph..7156C}
}

@ARTICLE{2021ApJS..257...59Chime,
       author = {{CHIME/FRB Collaboration} and {Amiri}, Mandana and {Andersen}, Bridget C. and {Bandura}, Kevin and {Berger}, Sabrina and {Bhardwaj}, Mohit and {Boyce}, Michelle M. and {Boyle}, P.~J. and {Brar}, Charanjot and {Breitman}, Daniela and {Cassanelli}, Tomas and {Chawla}, Pragya and {Chen}, Tianyue and {Cliche}, J. -F. and {Cook}, Amanda and {Cubranic}, Davor and {Curtin}, Alice P. and {Deng}, Meiling and {Dobbs}, Matt and {Dong}, Fengqiu Adam and {Eadie}, Gwendolyn and {Fandino}, Mateus and {Fonseca}, Emmanuel and {Gaensler}, B.~M. and {Giri}, Utkarsh and {Good}, Deborah C. and {Halpern}, Mark and {Hill}, Alex S. and {Hinshaw}, Gary and {Josephy}, Alexander and {Kaczmarek}, Jane F. and {Kader}, Zarif and {Kania}, Joseph W. and {Kaspi}, Victoria M. and {Landecker}, T.~L. and {Lang}, Dustin and {Leung}, Calvin and {Li}, Dongzi and {Lin}, Hsiu-Hsien and {Masui}, Kiyoshi W. and {McKinven}, Ryan and {Mena-Parra}, Juan and {Merryfield}, Marcus and {Meyers}, Bradley W. and {Michilli}, Daniele and {Milutinovic}, Nikola and {Mirhosseini}, Arash and {M{\"u}nchmeyer}, Moritz and {Naidu}, Arun and {Newburgh}, Laura and {Ng}, Cherry and {Patel}, Chitrang and {Pen}, Ue-Li and {Petroff}, Emily and {Pinsonneault-Marotte}, Tristan and {Pleunis}, Ziggy and {Rafiei-Ravandi}, Masoud and {Rahman}, Mubdi and {Ransom}, Scott M. and {Renard}, Andre and {Sanghavi}, Pranav and {Scholz}, Paul and {Shaw}, J. Richard and {Shin}, Kaitlyn and {Siegel}, Seth R. and {Sikora}, Andrew E. and {Singh}, Saurabh and {Smith}, Kendrick M. and {Stairs}, Ingrid and {Tan}, Chia Min and {Tendulkar}, S.~P. and {Vanderlinde}, Keith and {Wang}, Haochen and {Wulf}, Dallas and {Zwaniga}, A.~V.},
        title = "{The First CHIME/FRB Fast Radio Burst Catalog}",
      journal = {\apjs},
         year = 2021,
        month = dec,
       volume = {257},
       number = {2},
          eid = {59},
        pages = {59},
          doi = {10.3847/1538-4365/ac33ab},
archivePrefix = {arXiv},
       eprint = {2106.04352},
 primaryClass = {astro-ph.HE},
       adsurl = {https://ui.adsabs.harvard.edu/abs/2021ApJS..257...59C}
}

@ARTICLE{2023ApJ...947...83Chime,
       author = {{Chime/Frb Collaboration} and {Andersen}, Bridget C. and {Bandura}, Kevin and {Bhardwaj}, Mohit and {Boyle}, P.~J. and {Brar}, Charanjot and {Cassanelli}, Tomas and {Chatterjee}, S. and {Chawla}, Pragya and {Cook}, Amanda M. and {Curtin}, Alice P. and {Dobbs}, Matt and {Dong}, Fengqiu Adam and {Faber}, Jakob T. and {Fandino}, Mateus and {Fonseca}, Emmanuel and {Gaensler}, B.~M. and {Giri}, Utkarsh and {Herrera-Martin}, Antonio and {Hill}, Alex S. and {Ibik}, Adaeze and {Josephy}, Alexander and {Kaczmarek}, Jane F. and {Kader}, Zarif and {Kaspi}, Victoria and {Landecker}, T.~L. and {Lanman}, Adam E. and {Lazda}, Mattias and {Leung}, Calvin and {Lin}, Hsiu-Hsien and {Masui}, Kiyoshi W. and {McKinven}, Ryan and {Mena-Parra}, Juan and {Meyers}, Bradley W. and {Michilli}, D. and {Ng}, Cherry and {Pandhi}, Ayush and {Pearlman}, Aaron B. and {Pen}, Ue-Li and {Petroff}, Emily and {Pleunis}, Ziggy and {Rafiei-Ravandi}, Masoud and {Rahman}, Mubdi and {Ransom}, Scott M. and {Renard}, Andre and {Sand}, Ketan R. and {Sanghavi}, Pranav and {Scholz}, Paul and {Shah}, Vishwangi and {Shin}, Kaitlyn and {Siegel}, Seth and {Smith}, Kendrick and {Stairs}, Ingrid and {Su}, Jianing and {Tendulkar}, Shriharsh P. and {Vanderlinde}, Keith and {Wang}, Haochen and {Wulf}, Dallas and {Zwaniga}, Andrew},
        title = "{CHIME/FRB Discovery of 25 Repeating Fast Radio Burst Sources}",
      journal = {\apj},
         year = 2023,
        month = apr,
       volume = {947},
       number = {2},
          eid = {83},
        pages = {83},
          doi = {10.3847/1538-4357/acc6c1},
archivePrefix = {arXiv},
       eprint = {2301.08762},
 primaryClass = {astro-ph.HE},
       adsurl = {https://ui.adsabs.harvard.edu/abs/2023ApJ...947...83C}
}

@ARTICLE{2025ApJS..280....6Chime_KKO,
       author = {{Chime/Frb Collaboration} and {Amiri}, Mandana and {Amouyal}, Daniel and {Andersen}, Bridget C. and {Andrew}, Shion and {Bandura}, Kevin and {Bhardwaj}, Mohit and {Boyle}, P.~J. and {Brar}, Charanjot and {Cassity}, Alyssa and {Chatterjee}, Shami and {Curtin}, Alice P. and {Dobbs}, Matt and {Dong}, Fengqiu Adam and {Dong}, Yuxin and {Eadie}, Gwendolyn M. and {Eftekhari}, Tarraneh and {Fong}, Wen-Fai and {Fonseca}, Emmanuel and {Gaensler}, B.~M. and {Halpern}, Mark and {Hessels}, Jason W.~T. and {Hopkins}, Hans and {Ibik}, Adaeze L. and {Joseph}, Ronniy C. and {Kaczmarek}, Jane and {Kahinga}, Lordrick and {Kaspi}, Victoria and {Khairy}, Kholoud and {Kilpatrick}, Charles D. and {Lanman}, Adam E. and {Lazda}, Mattias and {Leung}, Calvin and {Main}, Robert and {Mas-Ribas}, Lluis and {Masui}, Kiyoshi W. and {McKinven}, Ryan and {Mena-Parra}, Juan and {Meyers}, Bradley W. and {Michilli}, Daniele and {Milutinovic}, Nikola and {Nimmo}, Kenzie and {Noble}, Gavin and {Pandhi}, Ayush and {Patil}, Swarali Shivraj and {Pearlman}, Aaron B. and {Petroff}, Emily and {Pleunis}, Ziggy and {Prochaska}, J. Xavier and {Rafiei-Ravandi}, Masoud and {Rahman}, Mubdi and {Renard}, Andre and {Sammons}, Mawson W. and {Sand}, Ketan R. and {Scholz}, Paul and {Shah}, Vishwangi and {Shin}, Kaitlyn and {Siegel}, Seth R. and {Simha}, Sunil and {Smith}, Kendrick and {Stairs}, Ingrid and {Vanderlinde}, Keith and {Wang}, Haochen and {Wulf}, Dallas and {Zegmott}, Tarik J.},
        title = "{A Catalog of Local Universe Fast Radio Bursts from CHIME/FRB and the KKO}",
      journal = {\apjs},
         year = 2025,
        month = sep,
       volume = {280},
       number = {1},
          eid = {6},
        pages = {6},
          doi = {10.3847/1538-4365/addbda},
archivePrefix = {arXiv},
       eprint = {2502.11217},
 primaryClass = {astro-ph.HE},
       adsurl = {https://ui.adsabs.harvard.edu/abs/2025ApJS..280....6C}
}

@ARTICLE{2000ApJ...533..682Calzetti,
       author = {{Calzetti}, Daniela and {Armus}, Lee and {Bohlin}, Ralph C. and {Kinney}, Anne L. and {Koornneef}, Jan and {Storchi-Bergmann}, Thaisa},
        title = "{The Dust Content and Opacity of Actively Star-forming Galaxies}",
      journal = {\apj},
         year = 2000,
        month = apr,
       volume = {533},
       number = {2},
        pages = {682-695},
          doi = {10.1086/308692},
archivePrefix = {arXiv},
       eprint = {astro-ph/9911459},
 primaryClass = {astro-ph},
       adsurl = {https://ui.adsabs.harvard.edu/abs/2000ApJ...533..682C}
}

@ARTICLE{2015A&A...584A..87Catal,
       author = {{Catal{\'a}n-Torrecilla}, C. and {Gil de Paz}, A. and {Castillo-Morales}, A. and {Iglesias-P{\'a}ramo}, J. and {S{\'a}nchez}, S.~F. and {Kennicutt}, R.~C. and {P{\'e}rez-Gonz{\'a}lez}, P.~G. and {Marino}, R.~A. and {Walcher}, C.~J. and {Husemann}, B. and {Garc{\'\i}a-Benito}, R. and {Mast}, D. and {Gonz{\'a}lez Delgado}, R.~M. and {Mu{\~n}oz-Mateos}, J.~C. and {Bland-Hawthorn}, J. and {Bomans}, D.~J. and {Del Olmo}, A. and {Galbany}, L. and {Gomes}, J.~M. and {Kehrig}, C. and {L{\'o}pez-S{\'a}nchez}, {\'A}. R. and {Mendoza}, M.~A. and {Monreal-Ibero}, A. and {P{\'e}rez-Torres}, M. and {S{\'a}nchez-Bl{\'a}zquez}, P. and {Vilchez}, J.~M. and {CALIFA Collaboration}},
        title = "{Star formation in the local Universe from the CALIFA sample. I. Calibrating the SFR using integral field spectroscopy data}",
      journal = {\aap},
         year = 2015,
        month = dec,
       volume = {584},
          eid = {A87},
        pages = {A87},
          doi = {10.1051/0004-6361/201526023},
archivePrefix = {arXiv},
       eprint = {1507.03801},
 primaryClass = {astro-ph.GA},
       adsurl = {https://ui.adsabs.harvard.edu/abs/2015A&A...584A..87C}
}

@ARTICLE{2024ApJ...969..145Chime,
       author = {{CHIME/FRB Collaboration} and {Amiri}, Mandana and {Andersen}, Bridget C. and {Andrew}, Shion and {Bandura}, Kevin and {Bhardwaj}, Mohit and {Boyle}, P.~J. and {Brar}, Charanjot and {Breitman}, Daniela and {Cassanelli}, Tomas and {Chawla}, Pragya and {Cook}, Amanda M. and {Curtin}, Alice P. and {Dobbs}, Matt and {Dong}, Fengqiu Adam and {Eadie}, Gwendolyn and {Fonseca}, Emmanuel and {Gaensler}, B.~M. and {Giri}, Utkarsh and {Herrera-Martin}, Antonio and {Hopkins}, Hans and {Ibik}, Adaeze L. and {Joseph}, Ronniy C. and {Kaczmarek}, J.~F. and {Kader}, Zarif and {Kaspi}, Victoria M. and {Lanman}, Adam E. and {Lazda}, Mattias and {Leung}, Calvin and {Liu}, Siqi and {Masui}, Kiyoshi W. and {McKinven}, Ryan and {Mena-Parra}, Juan and {Merryfield}, Marcus and {Michilli}, Daniele and {Ng}, Cherry and {Nimmo}, Kenzie and {Noble}, Gavin and {Pandhi}, Ayush and {Patel}, Chitrang and {Pearlman}, Aaron B. and {Pen}, Ue-Li and {Petroff}, Emily and {Pleunis}, Ziggy and {Rafiei-Ravandi}, Masoud and {Rahman}, Mubdi and {Ransom}, Scott M. and {Sand}, Ketan R. and {Scholz}, Paul and {Shah}, Vishwangi and {Shin}, Kaitlyn and {Shpunarska}, Yuliya and {Siegel}, Seth R. and {Smith}, Kendrick and {Stairs}, Ingrid and {Stenning}, David C. and {Vanderlinde}, Keith and {Wang}, Haochen and {White}, Henry and {Wulf}, Dallas},
        title = "{Updating the First CHIME/FRB Catalog of Fast Radio Bursts with Baseband Data}",
      journal = {\apj},
         year = 2024,
        month = jul,
       volume = {969},
       number = {2},
          eid = {145},
        pages = {145},
          doi = {10.3847/1538-4357/ad464b},
archivePrefix = {arXiv},
       eprint = {2311.00111},
 primaryClass = {astro-ph.HE},
       adsurl = {https://ui.adsabs.harvard.edu/abs/2024ApJ...969..145C}
}

@ARTICLE{2015ApJS..219....8Chang,
       author = {{Chang}, Yu-Yen and {van der Wel}, Arjen and {da Cunha}, Elisabete and {Rix}, Hans-Walter},
        title = "{Stellar Masses and Star Formation Rates for 1M Galaxies from SDSS+WISE}",
      journal = {\apjs},
         year = 2015,
        month = jul,
       volume = {219},
       number = {1},
          eid = {8},
        pages = {8},
          doi = {10.1088/0067-0049/219/1/8},
archivePrefix = {arXiv},
       eprint = {1506.00648},
 primaryClass = {astro-ph.GA},
       adsurl = {https://ui.adsabs.harvard.edu/abs/2015ApJS..219....8C}
}

@ARTICLE{2025ApJ...993...55CHIME_outrigger,
       author = {{CHIME/FRB Collaboration} and {Amiri}, Mandana and {Andersen}, Bridget C. and {Andrew}, Shion and {Bandura}, Kevin and {Bhardwaj}, Mohit and {Bhopi}, Kalyani and {Bidula}, Vadym and {Boyle}, P.~J. and {Brar}, Charanjot and {Carlson}, Mark and {Cassanelli}, Tomas and {Cassity}, Alyssa and {Chatterjee}, Shami and {Cliche}, Jean-Fran{\c{c}}ois and {Curtin}, Alice P. and {Darlinger}, Rachel and {Deboer}, David R. and {Dobbs}, Matt and {Dong}, Fengqiu Adam and {Eadie}, Gwendolyn and {Fonseca}, Emmanuel and {Gaensler}, B.~M. and {Gusinskaia}, Nina and {Halpern}, Mark and {Hendricksen}, Ian and {Hessels}, Jason and {Joseph}, Ronniy C. and {Kaczmarek}, Jane and {Kaspi}, Victoria M. and {Khairy}, Kholoud and {Landecker}, T.~L. and {Lanman}, Adam E. and {Lau}, Albert Wai Kit and {Lazda}, Mattias and {Leung}, Calvin and {Main}, Robert A. and {Masui}, Kiyoshi W. and {McKinven}, Ryan and {Mena-Parra}, Juan and {Meyers}, Bradley W. and {Michilli}, Daniele and {Milutinovic}, Nikola and {Nimmo}, Kenzie and {Noble}, Gavin and {Pandhi}, Ayush and {Pearlman}, Aaron B. and {Peterson}, Jeffrey B. and {Petroff}, Emily and {Pleunis}, Ziggy and {Pollak}, Alexander W. and {Rafiei-Ravandi}, Masoud and {Renard}, Andre and {Sammons}, Mawson W. and {Sand}, Ketan R. and {Sanghavi}, Pranav and {Scholz}, Paul and {Shah}, Vishwangi and {Shin}, Kaitlyn and {Siegel}, Seth R. and {Siemion}, Andrew and {Sievers}, Jonathan L. and {Smith}, Kendrick and {Spear}, David and {Stairs}, Ingrid and {Vanderlinde}, Keith and {Wang}, Haochen and {Willis}, Jacob P. and {Zegmott}, Tarik J.},
        title = "{CHIME/FRB Outriggers: Design Overview}",
      journal = {\apj},
         year = 2025,
        month = nov,
       volume = {993},
       number = {1},
          eid = {55},
        pages = {55},
          doi = {10.3847/1538-4357/adfdcc},
archivePrefix = {arXiv},
       eprint = {2504.05192},
 primaryClass = {astro-ph.HE},
       adsurl = {https://ui.adsabs.harvard.edu/abs/2025ApJ...993...55C}
}

@ARTICLE{2023ApJ...958..185Chen,
       author = {{Chen}, Ge and {Ravi}, Vikram and {Hallinan}, Gregg W.},
        title = "{A Comprehensive Observational Study of the FRB 121102 Persistent Radio Source}",
      journal = {\apj},
         year = 2023,
        month = dec,
       volume = {958},
       number = {2},
          eid = {185},
        pages = {185},
          doi = {10.3847/1538-4357/ad02f3},
archivePrefix = {arXiv},
       eprint = {2201.00999},
 primaryClass = {astro-ph.HE},
       adsurl = {https://ui.adsabs.harvard.edu/abs/2023ApJ...958..185C}
}

@software{2012ascl.soft10002Cappellari,
       author = {{Cappellari}, Michele},
        title = "{pPXF: Penalized Pixel-Fitting stellar kinematics extraction}",
 howpublished = {Astrophysics Source Code Library, record ascl:1210.002},
         year = 2012,
        month = oct,
          eid = {ascl:1210.002},
       adsurl = {https://ui.adsabs.harvard.edu/abs/2012ascl.soft10002C}
}

@ARTICLE{2015MNRAS.451.4277Dolag,
       author = {{Dolag}, K. and {Gaensler}, B.~M. and {Beck}, A.~M. and {Beck}, M.~C.},
        title = "{Constraints on the distribution and energetics of fast radio bursts using cosmological hydrodynamic simulations}",
      journal = {\mnras},
         year = 2015,
        month = aug,
       volume = {451},
       number = {4},
        pages = {4277-4289},
          doi = {10.1093/mnras/stv1190},
archivePrefix = {arXiv},
       eprint = {1412.4829},
 primaryClass = {astro-ph.CO},
       adsurl = {https://ui.adsabs.harvard.edu/abs/2015MNRAS.451.4277D}
}

@ARTICLE{2007ApJ...657..810Draine,
       author = {{Draine}, B.~T. and {Li}, Aigen},
        title = "{Infrared Emission from Interstellar Dust. IV. The Silicate-Graphite-PAH Model in the Post-Spitzer Era}",
      journal = {\apj},
         year = 2007,
        month = mar,
       volume = {657},
       number = {2},
        pages = {810-837},
          doi = {10.1086/511055},
archivePrefix = {arXiv},
       eprint = {astro-ph/0608003},
 primaryClass = {astro-ph},
       adsurl = {https://ui.adsabs.harvard.edu/abs/2007ApJ...657..810D}
}

@ARTICLE{2024ApJ...961...44Dong,
       author = {{Dong}, Yuxin and {Eftekhari}, Tarraneh and {Fong}, Wen-fai and {Deller}, Adam T. and {Mannings}, Alexandra G. and {Simha}, Sunil and {Sridhar}, Navin and {Rafelski}, Marc and {Gordon}, Alexa C. and {Bhandari}, Shivani and {Day}, Cherie K. and {Heintz}, Kasper E. and {Hessels}, Jason W.~T. and {Leja}, Joel and {James}, Clancy W. and {Kilpatrick}, Charles D. and {Mahony}, Elizabeth K. and {Marcote}, Benito and {Margalit}, Ben and {Nimmo}, Kenzie and {Prochaska}, J. Xavier and {Escorial}, Alicia Rouco and {Ryder}, Stuart D. and {Schroeder}, Genevieve and {Shannon}, Ryan M. and {Tejos}, Nicolas},
        title = "{Mapping Obscured Star Formation in the Host Galaxy of FRB 20201124A}",
      journal = {\apj},
         year = 2024,
        month = jan,
       volume = {961},
       number = {1},
          eid = {44},
        pages = {44},
          doi = {10.3847/1538-4357/ad0cbd},
archivePrefix = {arXiv},
       eprint = {2307.06995},
 primaryClass = {astro-ph.HE},
       adsurl = {https://ui.adsabs.harvard.edu/abs/2024ApJ...961...44D}
}

@ARTICLE{2022MNRAS.512.3662Duncan,
       author = {{Duncan}, Kenneth J.},
        title = "{All-purpose, all-sky photometric redshifts for the Legacy Imaging Surveys Data Release 8}",
      journal = {\mnras},
         year = 2022,
        month = may,
       volume = {512},
       number = {3},
        pages = {3662-3683},
          doi = {10.1093/mnras/stac608},
archivePrefix = {arXiv},
       eprint = {2203.01949},
 primaryClass = {astro-ph.GA},
       adsurl = {https://ui.adsabs.harvard.edu/abs/2022MNRAS.512.3662D}
}

@ARTICLE{2020ApJS..247...54Evans,
       author = {{Evans}, P.~A. and {Page}, K.~L. and {Osborne}, J.~P. and {Beardmore}, A.~P. and {Willingale}, R. and {Burrows}, D.~N. and {Kennea}, J.~A. and {Perri}, M. and {Capalbi}, M. and {Tagliaferri}, G. and {Cenko}, S.~B.},
        title = "{2SXPS: An Improved and Expanded Swift X-Ray Telescope Point-source Catalog}",
      journal = {\apjs},
         year = 2020,
        month = apr,
       volume = {247},
       number = {2},
          eid = {54},
        pages = {54},
          doi = {10.3847/1538-4365/ab7db9},
archivePrefix = {arXiv},
       eprint = {1911.11710},
 primaryClass = {astro-ph.IM},
       adsurl = {https://ui.adsabs.harvard.edu/abs/2020ApJS..247...54E}
}

@ARTICLE{2025ApJ...979L..22Eftekhari,
       author = {{Eftekhari}, T. and {Dong}, Y. and {Fong}, W. and {Shah}, V. and {Simha}, S. and {Andersen}, B.~C. and {Andrew}, S. and {Bhardwaj}, M. and {Cassanelli}, T. and {Chatterjee}, S. and {Coulter}, D.~A. and {Fonseca}, E. and {Gaensler}, B.~M. and {Gordon}, A.~C. and {Hessels}, J.~W.~T. and {Ibik}, A.~L. and {Joseph}, R.~C. and {Kahinga}, L.~A. and {Kaspi}, V. and {Kharel}, B. and {Kilpatrick}, C.~D. and {Lanman}, A.~E. and {Lazda}, M. and {Leung}, C. and {Liu}, C. and {Mas-Ribas}, L. and {Masui}, K.~W. and {Mckinven}, R. and {Mena-Parra}, J. and {Miller}, A.~A. and {Nimmo}, K. and {Pandhi}, A. and {Patil}, S.~S. and {Pearlman}, A.~B. and {Pleunis}, Z. and {Prochaska}, J.~X. and {Rafiei-Ravandi}, M. and {Sammons}, M. and {Scholz}, P. and {Shin}, K. and {Smith}, K. and {Stairs}, I.},
        title = "{The Massive and Quiescent Elliptical Host Galaxy of the Repeating Fast Radio Burst FRB 20240209A}",
      journal = {\apjl},
         year = 2025,
        month = feb,
       volume = {979},
       number = {2},
          eid = {L22},
        pages = {L22},
          doi = {10.3847/2041-8213/ad9de2},
archivePrefix = {arXiv},
       eprint = {2410.23336},
 primaryClass = {astro-ph.HE},
       adsurl = {https://ui.adsabs.harvard.edu/abs/2025ApJ...979L..22E}
}

@ARTICLE{2018ApJ...855..132Faisst,
       author = {{Faisst}, Andreas L. and {Masters}, Daniel and {Wang}, Yun and {Merson}, Alexander and {Capak}, Peter and {Malhotra}, Sangeeta and {Rhoads}, James E.},
        title = "{Empirical Modeling of the Redshift Evolution of the [\{\textbackslashrm\{N\}\}\textbackslash,\{\textbackslashrm\{II\}\}]/H{\ensuremath{\alpha}} Ratio for Galaxy Redshift Surveys}",
      journal = {\apj},
         year = 2018,
        month = mar,
       volume = {855},
       number = {2},
          eid = {132},
        pages = {132},
          doi = {10.3847/1538-4357/aab1fc},
archivePrefix = {arXiv},
       eprint = {1710.00834},
 primaryClass = {astro-ph.GA},
       adsurl = {https://ui.adsabs.harvard.edu/abs/2018ApJ...855..132F}
}

@ARTICLE{2021ApJ...919L..23Fong,
       author = {{Fong}, Wen-fai and {Dong}, Yuxin and {Leja}, Joel and {Bhandari}, Shivani and {Day}, Cherie K. and {Deller}, Adam T. and {Kumar}, Pravir and {Prochaska}, J. Xavier and {Scott}, Danica R. and {Bannister}, Keith W. and {Eftekhari}, Tarraneh and {Gordon}, Alexa C. and {Heintz}, Kasper E. and {James}, Clancy W. and {Kilpatrick}, Charles D. and {Mahony}, Elizabeth K. and {Rouco Escorial}, Alicia and {Ryder}, Stuart D. and {Shannon}, Ryan M. and {Tejos}, Nicolas},
        title = "{Chronicling the Host Galaxy Properties of the Remarkable Repeating FRB 20201124A}",
      journal = {\apjl},
         year = 2021,
        month = oct,
       volume = {919},
       number = {2},
          eid = {L23},
        pages = {L23},
          doi = {10.3847/2041-8213/ac242b},
archivePrefix = {arXiv},
       eprint = {2106.11993},
 primaryClass = {astro-ph.GA},
       adsurl = {https://ui.adsabs.harvard.edu/abs/2021ApJ...919L..23F}
}

@ARTICLE{2023ApJ...954...80Gordon,
       author = {{Gordon}, Alexa C. and {Fong}, Wen-fai and {Kilpatrick}, Charles D. and {Eftekhari}, Tarraneh and {Leja}, Joel and {Prochaska}, J. Xavier and {Nugent}, Anya E. and {Bhandari}, Shivani and {Blanchard}, Peter K. and {Caleb}, Manisha and {Day}, Cherie K. and {Deller}, Adam T. and {Dong}, Yuxin and {Glowacki}, Marcin and {Gourdji}, Kelly and {Mannings}, Alexandra G. and {Mahoney}, Elizabeth K. and {Marnoch}, Lachlan and {Miller}, Adam A. and {Paterson}, Kerry and {Rastinejad}, Jillian C. and {Ryder}, Stuart D. and {Sadler}, Elaine M. and {Scott}, Danica R. and {Sears}, Huei and {Shannon}, Ryan M. and {Simha}, Sunil and {Stappers}, Benjamin W. and {Tejos}, Nicolas},
        title = "{The Demographics, Stellar Populations, and Star Formation Histories of Fast Radio Burst Host Galaxies: Implications for the Progenitors}",
      journal = {\apj},
         year = 2023,
        month = sep,
       volume = {954},
       number = {1},
          eid = {80},
        pages = {80},
          doi = {10.3847/1538-4357/ace5aa},
archivePrefix = {arXiv},
       eprint = {2302.05465},
 primaryClass = {astro-ph.GA},
       adsurl = {https://ui.adsabs.harvard.edu/abs/2023ApJ...954...80G}
}

@ARTICLE{2018MNRAS.475.3010Gurkan,
       author = {{G{\"u}rkan}, G. and {Hardcastle}, M.~J. and {Smith}, D.~J.~B. and {Best}, P.~N. and {Bourne}, N. and {Calistro-Rivera}, G. and {Heald}, G. and {Jarvis}, M.~J. and {Prandoni}, I. and {R{\"o}ttgering}, H.~J.~A. and {Sabater}, J. and {Shimwell}, T. and {Tasse}, C. and {Williams}, W.~L.},
        title = "{LOFAR/H-ATLAS: the low-frequency radio luminosity-star formation rate relation}",
      journal = {\mnras},
         year = 2018,
        month = apr,
       volume = {475},
       number = {3},
        pages = {3010-3028},
          doi = {10.1093/mnras/sty016},
archivePrefix = {arXiv},
       eprint = {1801.02629},
 primaryClass = {astro-ph.GA},
       adsurl = {https://ui.adsabs.harvard.edu/abs/2018MNRAS.475.3010G}
}

@ARTICLE{2018ApJ...863....2Gajjar,
       author = {{Gajjar}, V. and {Siemion}, A.~P.~V. and {Price}, D.~C. and {Law}, C.~J. and {Michilli}, D. and {Hessels}, J.~W.~T. and {Chatterjee}, S. and {Archibald}, A.~M. and {Bower}, G.~C. and {Brinkman}, C. and {Burke-Spolaor}, S. and {Cordes}, J.~M. and {Croft}, S. and {Enriquez}, J. Emilio and {Foster}, G. and {Gizani}, N. and {Hellbourg}, G. and {Isaacson}, H. and {Kaspi}, V.~M. and {Lazio}, T.~J.~W. and {Lebofsky}, M. and {Lynch}, R.~S. and {MacMahon}, D. and {McLaughlin}, M.~A. and {Ransom}, S.~M. and {Scholz}, P. and {Seymour}, A. and {Spitler}, L.~G. and {Tendulkar}, S.~P. and {Werthimer}, D. and {Zhang}, Y.~G.},
        title = "{Highest Frequency Detection of FRB 121102 at 4-8 GHz Using the Breakthrough Listen Digital Backend at the Green Bank Telescope}",
      journal = {\apj},
         year = 2018,
        month = aug,
       volume = {863},
       number = {1},
          eid = {2},
        pages = {2},
          doi = {10.3847/1538-4357/aad005},
archivePrefix = {arXiv},
       eprint = {1804.04101},
 primaryClass = {astro-ph.HE},
       adsurl = {https://ui.adsabs.harvard.edu/abs/2018ApJ...863....2G}
}

@software{2014ascl.soft02004Grahma,
       author = {{Graham}, Matthew and {Plante}, Ray and {Tody}, Doug and {Fitzpatrick}, Mike},
        title = "{PyVO: Python access to the Virtual Observatory}",
 howpublished = {Astrophysics Source Code Library, record ascl:1402.004},
         year = 2014,
        month = feb,
          eid = {ascl:1402.004},
       adsurl = {https://ui.adsabs.harvard.edu/abs/2014ascl.soft02004G}
}

@ARTICLE{2019AJ....157...98Ginsburg,
       author = {{Ginsburg}, Adam and {Sip{\H{o}}cz}, Brigitta M. and {Brasseur}, C.~E. and {Cowperthwaite}, Philip S. and {Craig}, Matthew W. and {Deil}, Christoph and {Guillochon}, James and {Guzman}, Giannina and {Liedtke}, Simon and {Lian Lim}, Pey and {Lockhart}, Kelly E. and {Mommert}, Michael and {Morris}, Brett M. and {Norman}, Henrik and {Parikh}, Madhura and {Persson}, Magnus V. and {Robitaille}, Thomas P. and {Segovia}, Juan-Carlos and {Singer}, Leo P. and {Tollerud}, Erik J. and {de Val-Borro}, Miguel and {Valtchanov}, Ivan and {Woillez}, Julien and {Astroquery Collaboration} and {a subset of astropy Collaboration}},
        title = "{astroquery: An Astronomical Web-querying Package in Python}",
      journal = {\aj},
         year = 2019,
        month = mar,
       volume = {157},
       number = {3},
          eid = {98},
        pages = {98},
          doi = {10.3847/1538-3881/aafc33},
archivePrefix = {arXiv},
       eprint = {1901.04520},
 primaryClass = {astro-ph.IM},
       adsurl = {https://ui.adsabs.harvard.edu/abs/2019AJ....157...98G}
}

@dataset{2024yCat.8112....0Herschel,
       author = {{Herschel Team} and {Schulz}, B. and {Marton}, G. and {Valtchanov}, I. and {Perez Garcia}, A.~M. and {Pinter}, S. and {Appleton}, P. and {Kiss}, C. and {Lim}, T. and {Lu}, N. and {Papageorgiou}, A. and {Pearson}, C. and {Rector}, J. and {Sanchez Portal}, M. and {Shupe}, D. and {Toth}, V.~L. and {van Dyk}, S. and {Varga-Verebelyi}, E. and {Xu}, K.},
        title = "{VizieR Online Data Catalog: Herschel/SPIRE point source catalog (HSPSC) (Herschel team+, 2017)}",
 howpublished = {VizieR On-line Data Catalog: VIII/112.  Originally published in: Herschel catalogs (2017)},
         year = 2024,
        month = jun,
          eid = {VIII/112},
       adsurl = {https://ui.adsabs.harvard.edu/abs/2024yCat.8112....0H}
}

@ARTICLE{2025arXiv250408038Horowicz,
       author = {{Horowicz}, Asaf and {Margalit}, Ben},
        title = "{The Host Galaxies of Fast Radio Bursts Track a Combination of Stellar Mass and Star Formation, Similar to Type Ia Supernovae}",
      journal = {arXiv e-prints},
         year = 2025,
        month = apr,
          eid = {arXiv:2504.08038},
        pages = {arXiv:2504.08038},
          doi = {10.48550/arXiv.2504.08038},
archivePrefix = {arXiv},
       eprint = {2504.08038},
 primaryClass = {astro-ph.HE},
       adsurl = {https://ui.adsabs.harvard.edu/abs/2025arXiv250408038H}
}

@ARTICLE{2020ApJ...903..152Heintz,
       author = {{Heintz}, Kasper E. and {Prochaska}, J. Xavier and {Simha}, Sunil and {Platts}, Emma and {Fong}, Wen-fai and {Tejos}, Nicolas and {Ryder}, Stuart D. and {Aggerwal}, Kshitij and {Bhandari}, Shivani and {Day}, Cherie K. and {Deller}, Adam T. and {Kilpatrick}, Charles D. and {Law}, Casey J. and {Macquart}, Jean-Pierre and {Mannings}, Alexandra and {Marnoch}, Lachlan J. and {Sadler}, Elaine M. and {Shannon}, Ryan M.},
        title = "{Host Galaxy Properties and Offset Distributions of Fast Radio Bursts: Implications for Their Progenitors}",
      journal = {\apj},
         year = 2020,
        month = nov,
       volume = {903},
       number = {2},
          eid = {152},
        pages = {152},
          doi = {10.3847/1538-4357/abb6fb},
archivePrefix = {arXiv},
       eprint = {2009.10747},
 primaryClass = {astro-ph.GA},
       adsurl = {https://ui.adsabs.harvard.edu/abs/2020ApJ...903..152H}
}

@ARTICLE{2024ApJ...961...99Ibik,
       author = {{Ibik}, Adaeze L. and {Drout}, Maria R. and {Gaensler}, B.~M. and {Scholz}, Paul and {Michilli}, Daniele and {Bhardwaj}, Mohit and {Kaspi}, Victoria M. and {Pleunis}, Ziggy and {Cassanelli}, Tomas and {Cook}, Amanda M. and {Dong}, Fengqiu A. and {Kaczmarek}, Jane F. and {Leung}, Calvin and {Lu}, Katherine J. and {Masui}, Kiyoshi W. and {Pearlman}, Aaron B. and {Rafiei-Ravandi}, Masoud and {Sand}, Ketan R. and {Shin}, Kaitlyn and {Smith}, Kendrick M. and {Stairs}, Ingrid H.},
        title = "{Proposed Host Galaxies of Repeating Fast Radio Burst Sources Detected by CHIME/FRB}",
      journal = {\apj},
         year = 2024,
        month = jan,
       volume = {961},
       number = {1},
          eid = {99},
        pages = {99},
          doi = {10.3847/1538-4357/ad0893},
archivePrefix = {arXiv},
       eprint = {2304.02638},
 primaryClass = {astro-ph.HE},
       adsurl = {https://ui.adsabs.harvard.edu/abs/2024ApJ...961...99I}
}

@ARTICLE{2024ApJ...976..199Ibik,
       author = {{Ibik}, Adaeze L. and {Drout}, Maria R. and {Gaensler}, B.~M. and {Scholz}, Paul and {Sridhar}, Navin and {Margalit}, Ben and {Clarke}, T.~E. and {Law}, Casey J. and {Tendulkar}, Shriharsh P. and {Michilli}, Daniele and {Eftekhari}, Tarraneh and {Bhardwaj}, Mohit and {Burke-Spolaor}, Sarah and {Chatterjee}, Shami and {Cook}, Amanda M. and {Hessels}, Jason W.~T. and {Kirsten}, Franz and {Joseph}, Ronniy C. and {Kaspi}, Victoria M. and {Lazda}, Mattias and {Masui}, Kiyoshi W. and {Nimmo}, Kenzie and {Pandhi}, Ayush and {Pearlman}, Aaron B. and {Pleunis}, Ziggy and {Rafiei-Ravandi}, Masoud and {Shin}, Kaitlyn and {Smith}, Kendrick M.},
        title = "{A Search for Persistent Radio Sources toward Repeating Fast Radio Bursts Discovered by CHIME/FRB}",
      journal = {\apj},
         year = 2024,
        month = dec,
       volume = {976},
       number = {2},
          eid = {199},
        pages = {199},
          doi = {10.3847/1538-4357/ad808e},
archivePrefix = {arXiv},
       eprint = {2409.11533},
 primaryClass = {astro-ph.HE},
       adsurl = {https://ui.adsabs.harvard.edu/abs/2024ApJ...976..199I}
}

@ARTICLE{2019PASA...36....9James,
       author = {{James}, C.~W. and {Bannister}, K.~W. and {Macquart}, J. -P. and {Ekers}, R.~D. and {Oslowski}, S. and {Shannon}, R.~M. and {Allison}, J.~R. and {Chippendale}, A.~P. and {Collier}, J.~D. and {Franzen}, T. and {Hotan}, A.~W. and {Leach}, M. and {McConnell}, D. and {Pilawa}, M.~A. and {Voronkov}, M.~A. and {Whiting}, M.~T.},
        title = "{The performance and calibration of the CRAFT fly's eye fast radio burst survey}",
      journal = {\pasa},
         year = 2019,
        month = feb,
       volume = {36},
          eid = {e009},
        pages = {e009},
          doi = {10.1017/pasa.2019.1},
archivePrefix = {arXiv},
       eprint = {1810.04356},
 primaryClass = {astro-ph.HE},
       adsurl = {https://ui.adsabs.harvard.edu/abs/2019PASA...36....9J}
}

@ARTICLE{1998ApJ...498..541Kennicutt,
       author = {{Kennicutt}, Jr., Robert C.},
        title = "{The Global Schmidt Law in Star-forming Galaxies}",
      journal = {\apj},
         year = 1998,
        month = may,
       volume = {498},
       number = {2},
        pages = {541-552},
          doi = {10.1086/305588},
archivePrefix = {arXiv},
       eprint = {astro-ph/9712213},
 primaryClass = {astro-ph},
       adsurl = {https://ui.adsabs.harvard.edu/abs/1998ApJ...498..541K}
}

@ARTICLE{2008ApJS..178..247Kennicutt,
       author = {{Kennicutt}, Jr., Robert C. and {Lee}, Janice C. and {Funes}, Jos{\'e} G. and {J.}, S. and {Sakai}, Shoko and {Akiyama}, Sanae},
        title = "{An H{\ensuremath{\alpha}} Imaging Survey of Galaxies in the Local 11 Mpc Volume}",
      journal = {\apjs},
         year = 2008,
        month = oct,
       volume = {178},
       number = {2},
        pages = {247-279},
          doi = {10.1086/590058},
archivePrefix = {arXiv},
       eprint = {0807.2035},
 primaryClass = {astro-ph},
       adsurl = {https://ui.adsabs.harvard.edu/abs/2008ApJS..178..247K}
}

@ARTICLE{2012ARA&A..50..531Kennicutt,
       author = {{Kennicutt}, Robert C. and {Evans}, Neal J.},
        title = "{Star Formation in the Milky Way and Nearby Galaxies}",
      journal = {\araa},
         year = 2012,
        month = sep,
       volume = {50},
        pages = {531-608},
          doi = {10.1146/annurev-astro-081811-125610},
archivePrefix = {arXiv},
       eprint = {1204.3552},
 primaryClass = {astro-ph.GA},
       adsurl = {https://ui.adsabs.harvard.edu/abs/2012ARA&A..50..531K}
}

@ARTICLE{2002AJ....124.3135Kewley,
       author = {{Kewley}, Lisa J. and {Geller}, Margaret J. and {Jansen}, Rolf A. and {Dopita}, Michael A.},
        title = "{The H{\ensuremath{\alpha}} and Infrared Star Formation Rates for the Nearby Field Galaxy Survey}",
      journal = {\aj},
         year = 2002,
        month = dec,
       volume = {124},
       number = {6},
        pages = {3135-3143},
          doi = {10.1086/344487},
archivePrefix = {arXiv},
       eprint = {astro-ph/0208508},
 primaryClass = {astro-ph},
       adsurl = {https://ui.adsabs.harvard.edu/abs/2002AJ....124.3135K}
}

@ARTICLE{2024NatAs...8..337Kirsten,
       author = {{Kirsten}, F. and {Ould-Boukattine}, O.~S. and {Herrmann}, W. and {Gawro{\'n}ski}, M.~P. and {Hessels}, J.~W.~T. and {Lu}, W. and {Snelders}, M.~P. and {Chawla}, P. and {Yang}, J. and {Blaauw}, R. and {Nimmo}, K. and {Puchalska}, W. and {Wolak}, P. and {van Ruiten}, R.},
        title = "{A link between repeating and non-repeating fast radio bursts through their energy distributions}",
      journal = {Nature Astronomy},
         year = 2024,
        month = mar,
       volume = {8},
        pages = {337-346},
          doi = {10.1038/s41550-023-02153-z},
archivePrefix = {arXiv},
       eprint = {2306.15505},
 primaryClass = {astro-ph.HE},
       adsurl = {https://ui.adsabs.harvard.edu/abs/2024NatAs...8..337K}
}

@ARTICLE{2022Natur.602..585Kirsten,
       author = {{Kirsten}, F. and {Marcote}, B. and {Nimmo}, K. and {Hessels}, J.~W.~T. and {Bhardwaj}, M. and {Tendulkar}, S.~P. and {Keimpema}, A. and {Yang}, J. and {Snelders}, M.~P. and {Scholz}, P. and {Pearlman}, A.~B. and {Law}, C.~J. and {Peters}, W.~M. and {Giroletti}, M. and {Paragi}, Z. and {Bassa}, C. and {Hewitt}, D.~M. and {Bach}, U. and {Bezrukovs}, V. and {Burgay}, M. and {Buttaccio}, S.~T. and {Conway}, J.~E. and {Corongiu}, A. and {Feiler}, R. and {Forss{\'e}n}, O. and {Gawro{\'n}ski}, M.~P. and {Karuppusamy}, R. and {Kharinov}, M.~A. and {Lindqvist}, M. and {Maccaferri}, G. and {Melnikov}, A. and {Ould-Boukattine}, O.~S. and {Possenti}, A. and {Surcis}, G. and {Wang}, N. and {Yuan}, J. and {Aggarwal}, K. and {Anna-Thomas}, R. and {Bower}, G.~C. and {Blaauw}, R. and {Burke-Spolaor}, S. and {Cassanelli}, T. and {Clarke}, T.~E. and {Fonseca}, E. and {Gaensler}, B.~M. and {Gopinath}, A. and {Kaspi}, V.~M. and {Kassim}, N. and {Lazio}, T.~J.~W. and {Leung}, C. and {Li}, D.~Z. and {Lin}, H.~H. and {Masui}, K.~W. and {Mckinven}, R. and {Michilli}, D. and {Mikhailov}, A.~G. and {Ng}, C. and {Orbidans}, A. and {Pen}, U.~L. and {Petroff}, E. and {Rahman}, M. and {Ransom}, S.~M. and {Shin}, K. and {Smith}, K.~M. and {Stairs}, I.~H. and {Vlemmings}, W.},
        title = "{A repeating fast radio burst source in a globular cluster}",
      journal = {\nat},
         year = 2022,
        month = feb,
       volume = {602},
       number = {7898},
        pages = {585-589},
          doi = {10.1038/s41586-021-04354-w},
archivePrefix = {arXiv},
       eprint = {2105.11445},
 primaryClass = {astro-ph.HE},
       adsurl = {https://ui.adsabs.harvard.edu/abs/2022Natur.602..585K}
}

@ARTICLE{2025arXiv250215566Loudas,
       author = {{Loudas}, Nick and {Li}, Dongzi and {Strauss}, Michael A. and {Leja}, Joel},
        title = "{Unveiling the origin of fast radio bursts by modeling the stellar mass and star formation distributions of their host galaxies}",
      journal = {arXiv e-prints},
         year = 2025,
        month = feb,
          eid = {arXiv:2502.15566},
        pages = {arXiv:2502.15566},
          doi = {10.48550/arXiv.2502.15566},
archivePrefix = {arXiv},
       eprint = {2502.15566},
 primaryClass = {astro-ph.HE},
       adsurl = {https://ui.adsabs.harvard.edu/abs/2025arXiv250215566L}
}

@ARTICLE{2007Sci...318..777Lorimer,
       author = {{Lorimer}, D.~R. and {Bailes}, M. and {McLaughlin}, M.~A. and {Narkevic}, D.~J. and {Crawford}, F.},
        title = "{A Bright Millisecond Radio Burst of Extragalactic Origin}",
      journal = {Science},
         year = 2007,
        month = nov,
       volume = {318},
       number = {5851},
        pages = {777},
          doi = {10.1126/science.1147532},
archivePrefix = {arXiv},
       eprint = {0709.4301},
 primaryClass = {astro-ph},
       adsurl = {https://ui.adsabs.harvard.edu/abs/2007Sci...318..777L}
}

@misc{LOFAR2.0WhitePaper2023,
  author       = {{LOFAR2.0 Science Working Group}},
  title        = {LOFAR2.0 White Paper (v2023.1)},
  howpublished = {\url{https://www.lofar.eu/wp-content/uploads/2023/04/LOFAR2_0_White_Paper_v2023.1.pdf}},
  year         = 2023,
  note         = {Accessed: 2025-08-04}
}

@ARTICLE{2023PASA...40...29LeeWaddell,
       author = {{Lee-Waddell}, Karen and {James}, Clancy W. and {Ryder}, Stuart D. and {Mahony}, Elizabeth K. and {Bahramian}, Arash and {Koribalski}, B{\"a}rbel S. and {Kumar}, Pravir and {Marnoch}, Lachlan and {North-Hickey}, Freya O. and {Sadler}, Elaine M. and {Shannon}, Ryan and {Tejos}, Nicolas and {Thorne}, Jessica E. and {Wang}, Jing and {Wayth}, Randall},
        title = "{The host galaxy of FRB 20171020A revisited}",
      journal = {\pasa},
         year = 2023,
        month = jul,
       volume = {40},
          eid = {e029},
        pages = {e029},
          doi = {10.1017/pasa.2023.27},
archivePrefix = {arXiv},
       eprint = {2305.17960},
 primaryClass = {astro-ph.GA},
       adsurl = {https://ui.adsabs.harvard.edu/abs/2023PASA...40...29L}
}

@ARTICLE{2024ApJ...971L...5Maga,
       author = {{Maga{\~n}a Hernandez}, Ignacio and {d'Emilio}, Virginia and {Morisaki}, Soichiro and {Bhardwaj}, Mohit and {Palmese}, Antonella},
        title = "{On the Association of GW190425 with Its Potential Electromagnetic Counterpart FRB 20190425A}",
      journal = {\apjl},
         year = 2024,
        month = aug,
       volume = {971},
       number = {1},
          eid = {L5},
        pages = {L5},
          doi = {10.3847/2041-8213/ad5b4c},
archivePrefix = {arXiv},
       eprint = {2404.02496},
 primaryClass = {astro-ph.HE},
       adsurl = {https://ui.adsabs.harvard.edu/abs/2024ApJ...971L...5M}
}

@ARTICLE{2017ApJ...834L...8Marcote,
       author = {{Marcote}, B. and {Paragi}, Z. and {Hessels}, J.~W.~T. and {Keimpema}, A. and {van Langevelde}, H.~J. and {Huang}, Y. and {Bassa}, C.~G. and {Bogdanov}, S. and {Bower}, G.~C. and {Burke-Spolaor}, S. and {Butler}, B.~J. and {Campbell}, R.~M. and {Chatterjee}, S. and {Cordes}, J.~M. and {Demorest}, P. and {Garrett}, M.~A. and {Ghosh}, T. and {Kaspi}, V.~M. and {Law}, C.~J. and {Lazio}, T.~J.~W. and {McLaughlin}, M.~A. and {Ransom}, S.~M. and {Salter}, C.~J. and {Scholz}, P. and {Seymour}, A. and {Siemion}, A. and {Spitler}, L.~G. and {Tendulkar}, S.~P. and {Wharton}, R.~S.},
        title = "{The Repeating Fast Radio Burst FRB 121102 as Seen on Milliarcsecond Angular Scales}",
      journal = {\apjl},
         year = 2017,
        month = jan,
       volume = {834},
       number = {2},
          eid = {L8},
        pages = {L8},
          doi = {10.3847/2041-8213/834/2/L8},
archivePrefix = {arXiv},
       eprint = {1701.01099},
 primaryClass = {astro-ph.HE},
       adsurl = {https://ui.adsabs.harvard.edu/abs/2017ApJ...834L...8M}
}

@ARTICLE{2020Natur.581..391Macquart,
       author = {{Macquart}, J. -P. and {Prochaska}, J.~X. and {McQuinn}, M. and {Bannister}, K.~W. and {Bhandari}, S. and {Day}, C.~K. and {Deller}, A.~T. and {Ekers}, R.~D. and {James}, C.~W. and {Marnoch}, L. and {Os{\l}owski}, S. and {Phillips}, C. and {Ryder}, S.~D. and {Scott}, D.~R. and {Shannon}, R.~M. and {Tejos}, N.},
        title = "{A census of baryons in the Universe from localized fast radio bursts}",
      journal = {\nat},
         year = 2020,
        month = may,
       volume = {581},
       number = {7809},
        pages = {391-395},
          doi = {10.1038/s41586-020-2300-2},
archivePrefix = {arXiv},
       eprint = {2005.13161},
 primaryClass = {astro-ph.CO},
       adsurl = {https://ui.adsabs.harvard.edu/abs/2020Natur.581..391M}
}

@ARTICLE{2021ApJ...917...75Mannings,
       author = {{Mannings}, Alexandra G. and {Fong}, Wen-fai and {Simha}, Sunil and {Prochaska}, J. Xavier and {Rafelski}, Marc and {Kilpatrick}, Charles D. and {Tejos}, Nicolas and {Heintz}, Kasper E. and {Bannister}, Keith W. and {Bhandari}, Shivani and {Day}, Cherie K. and {Deller}, Adam T. and {Ryder}, Stuart D. and {Shannon}, Ryan M. and {Tendulkar}, Shriharsh P.},
        title = "{A High-resolution View of Fast Radio Burst Host Environments}",
      journal = {\apj},
         year = 2021,
        month = aug,
       volume = {917},
       number = {2},
          eid = {75},
        pages = {75},
          doi = {10.3847/1538-4357/abff56},
archivePrefix = {arXiv},
       eprint = {2012.11617},
 primaryClass = {astro-ph.GA},
       adsurl = {https://ui.adsabs.harvard.edu/abs/2021ApJ...917...75M}
}

@ARTICLE{2016MNRAS.462.2631Mingo,
       author = {{Mingo}, B. and {Watson}, M.~G. and {Rosen}, S.~R. and {Hardcastle}, M.~J. and {Ruiz}, A. and {Blain}, A. and {Carrera}, F.~J. and {Mateos}, S. and {Pineau}, F. -X. and {Stewart}, G.~C.},
        title = "{The MIXR sample: AGN activity versus star formation across the cross-correlation of WISE, 3XMM, and FIRST/NVSS}",
      journal = {\mnras},
         year = 2016,
        month = nov,
       volume = {462},
       number = {3},
        pages = {2631-2667},
          doi = {10.1093/mnras/stw1826},
archivePrefix = {arXiv},
       eprint = {1607.06471},
 primaryClass = {astro-ph.GA},
       adsurl = {https://ui.adsabs.harvard.edu/abs/2016MNRAS.462.2631M}
}

@ARTICLE{2019MNRAS.488.2701Mingo,
       author = {{Mingo}, B. and {Croston}, J.~H. and {Hardcastle}, M.~J. and {Best}, P.~N. and {Duncan}, K.~J. and {Morganti}, R. and {Rottgering}, H.~J.~A. and {Sabater}, J. and {Shimwell}, T.~W. and {Williams}, W.~L. and {Brienza}, M. and {Gurkan}, G. and {Mahatma}, V.~H. and {Morabito}, L.~K. and {Prandoni}, I. and {Bondi}, M. and {Ineson}, J. and {Mooney}, S.},
        title = "{Revisiting the Fanaroff-Riley dichotomy and radio-galaxy morphology with the LOFAR Two-Metre Sky Survey (LoTSS)}",
      journal = {\mnras},
         year = 2019,
        month = sep,
       volume = {488},
       number = {2},
        pages = {2701-2721},
          doi = {10.1093/mnras/stz1901},
archivePrefix = {arXiv},
       eprint = {1907.03726},
 primaryClass = {astro-ph.GA},
       adsurl = {https://ui.adsabs.harvard.edu/abs/2019MNRAS.488.2701M}
}

@ARTICLE{2006ApJ...642..775Moustakas,
       author = {{Moustakas}, John and {Kennicutt}, Jr., Robert C. and {Tremonti}, Christy A.},
        title = "{Optical Star Formation Rate Indicators}",
      journal = {\apj},
         year = 2006,
        month = may,
       volume = {642},
       number = {2},
        pages = {775-796},
          doi = {10.1086/500964},
archivePrefix = {arXiv},
       eprint = {astro-ph/0511730},
 primaryClass = {astro-ph},
       adsurl = {https://ui.adsabs.harvard.edu/abs/2006ApJ...642..775M}
}

@ARTICLE{2023NatAs...7..579Moroianu,
       author = {{Moroianu}, Alexandra and {Wen}, Linqing and {James}, Clancy W. and {Ai}, Shunke and {Kovalam}, Manoj and {Panther}, Fiona H. and {Zhang}, Bing},
        title = "{An assessment of the association between a fast radio burst and binary neutron star merger}",
      journal = {Nature Astronomy},
         year = 2023,
        month = may,
       volume = {7},
        pages = {579-589},
          doi = {10.1038/s41550-023-01917-x},
archivePrefix = {arXiv},
       eprint = {2212.00201},
 primaryClass = {astro-ph.HE},
       adsurl = {https://ui.adsabs.harvard.edu/abs/2023NatAs...7..579M}
}

@ARTICLE{2022Natur.606..873Niu,
       author = {{Niu}, C.-H. and {Aggarwal}, K. and {Li}, D. and {Zhang}, X. and {Chatterjee}, S. and {Tsai}, C.-W. and {Yu}, W. and {Law}, C.~J. and {Burke-Spolaor}, S. and {Cordes}, J.~M. and {Zhang}, Y.-K. and {Ocker}, S.~K. and {Yao}, J.-M. and {Wang}, P. and {Feng}, Y. and {Niino}, Y. and {Bochenek}, C. and {Cruces}, M. and {Connor}, L. and {Jiang}, J.-A. and {Dai}, S. and {Luo}, R. and {Li}, G.-D. and {Miao}, C.-C. and {Niu}, J.-R. and {Anna-Thomas}, R. and {Sydnor}, J. and {Stern}, D. and {Wang}, W.-Y. and {Yuan}, M. and {Yue}, Y.-L. and {Zhou}, D.-J. and {Yan}, Z. and {Zhu}, W.-W. and {Zhang}, B.},
        title = "{A repeating fast radio burst associated with a persistent radio source}",
      journal = {\nat},
         year = 2022,
        month = jun,
       volume = {606},
       number = {7916},
        pages = {873-877},
          doi = {10.1038/s41586-022-04755-5},
archivePrefix = {arXiv},
       eprint = {2110.07418},
 primaryClass = {astro-ph.HE},
       adsurl = {https://ui.adsabs.harvard.edu/abs/2022Natur.606..873N}
}

@ARTICLE{2021A&A...653A.119Nunez,
       author = {{N{\'u}{\~n}ez}, C. and {Tejos}, N. and {Pignata}, G. and {Kilpatrick}, C.~D. and {Prochaska}, J.~X. and {Heintz}, K.~E. and {Bannister}, K.~W. and {Bhandari}, S. and {Day}, C.~K. and {Deller}, A.~T. and {Flynn}, C. and {Mahony}, E.~K. and {Majewski}, D. and {Marnoch}, L. and {Qiu}, H. and {Ryder}, S.~D. and {Shannon}, R.~M.},
        title = "{Constraining bright optical counterparts of fast radio bursts}",
      journal = {\aap},
         year = 2021,
        month = sep,
       volume = {653},
          eid = {A119},
        pages = {A119},
          doi = {10.1051/0004-6361/202141110},
archivePrefix = {arXiv},
       eprint = {2104.09727},
 primaryClass = {astro-ph.HE},
       adsurl = {https://ui.adsabs.harvard.edu/abs/2021A&A...653A.119N}
}

@ARTICLE{2021ApJ...909L...8Niu,
       author = {{Niu}, Chen-Hui and {Li}, Di and {Luo}, Rui and {Wang}, Wei-Yang and {Yao}, Jumei and {Zhang}, Bing and {Zhu}, Wei-Wei and {Wang}, Pei and {Ye}, Haoyang and {Zhang}, Yong-Kun and {Niu}, Jia-rui and {Tang}, Ning-yu and {Duan}, Ran and {Krco}, Marko and {Dai}, Shi and {Feng}, Yi and {Miao}, Chenchen and {Pan}, Zhichen and {Qian}, Lei and {Xue}, Mengyao and {Yuan}, Mao and {Yue}, Youling and {Zhang}, Lei and {Zhang}, Xinxin},
        title = "{CRAFTS for Fast Radio Bursts: Extending the Dispersion-Fluence Relation with New FRBs Detected by FAST}",
      journal = {\apjl},
         year = 2021,
        month = mar,
       volume = {909},
       number = {1},
          eid = {L8},
        pages = {L8},
          doi = {10.3847/2041-8213/abe7f0},
archivePrefix = {arXiv},
       eprint = {2102.10546},
 primaryClass = {astro-ph.HE},
       adsurl = {https://ui.adsabs.harvard.edu/abs/2021ApJ...909L...8N}
}

@ARTICLE{2021ApJ...911L...3Pleunis,
       author = {{Pleunis}, Z. and {Michilli}, D. and {Bassa}, C.~G. and {Hessels}, J.~W.~T. and {Naidu}, A. and {Andersen}, B.~C. and {Chawla}, P. and {Fonseca}, E. and {Gopinath}, A. and {Kaspi}, V.~M. and {Kondratiev}, V.~I. and {Li}, D.~Z. and {Bhardwaj}, M. and {Boyle}, P.~J. and {Brar}, C. and {Cassanelli}, T. and {Gupta}, Y. and {Josephy}, A. and {Karuppusamy}, R. and {Keimpema}, A. and {Kirsten}, F. and {Leung}, C. and {Marcote}, B. and {Masui}, K.~W. and {Mckinven}, R. and {Meyers}, B.~W. and {Ng}, C. and {Nimmo}, K. and {Paragi}, Z. and {Rahman}, M. and {Scholz}, P. and {Shin}, K. and {Smith}, K.~M. and {Stairs}, I.~H. and {Tendulkar}, S.~P.},
        title = "{LOFAR Detection of 110-188 MHz Emission and Frequency-dependent Activity from FRB 20180916B}",
      journal = {\apjl},
         year = 2021,
        month = apr,
       volume = {911},
       number = {1},
          eid = {L3},
        pages = {L3},
          doi = {10.3847/2041-8213/abec72},
archivePrefix = {arXiv},
       eprint = {2012.08372},
 primaryClass = {astro-ph.HE},
       adsurl = {https://ui.adsabs.harvard.edu/abs/2021ApJ...911L...3P}
}

@ARTICLE{2019Natur.572..352Ravi,
       author = {{Ravi}, V. and {Catha}, M. and {D'Addario}, L. and {Djorgovski}, S.~G. and {Hallinan}, G. and {Hobbs}, R. and {Kocz}, J. and {Kulkarni}, S.~R. and {Shi}, J. and {Vedantham}, H.~K. and {Weinreb}, S. and {Woody}, D.~P.},
        title = "{A fast radio burst localized to a massive galaxy}",
      journal = {\nat},
         year = 2019,
        month = aug,
       volume = {572},
       number = {7769},
        pages = {352-354},
          doi = {10.1038/s41586-019-1389-7},
archivePrefix = {arXiv},
       eprint = {1907.01542},
 primaryClass = {astro-ph.HE},
       adsurl = {https://ui.adsabs.harvard.edu/abs/2019Natur.572..352R}
}

@ARTICLE{2018ApJS..234...39Schlafly_DECaPS,
       author = {{Schlafly}, E.~F. and {Green}, G.~M. and {Lang}, D. and {Daylan}, T. and {Finkbeiner}, D.~P. and {Lee}, A. and {Meisner}, A.~M. and {Schlegel}, D. and {Valdes}, F.},
        title = "{The DECam Plane Survey: Optical Photometry of Two Billion Objects in the Southern Galactic Plane}",
      journal = {\apjs},
         year = 2018,
        month = feb,
       volume = {234},
       number = {2},
          eid = {39},
        pages = {39},
          doi = {10.3847/1538-4365/aaa3e2},
archivePrefix = {arXiv},
       eprint = {1710.01309},
 primaryClass = {astro-ph.GA},
       adsurl = {https://ui.adsabs.harvard.edu/abs/2018ApJS..234...39S}
}

@ARTICLE{2007ApJS..173..267Salim,
       author = {{Salim}, Samir and {Rich}, R. Michael and {Charlot}, St{\'e}phane and {Brinchmann}, Jarle and {Johnson}, Benjamin D. and {Schiminovich}, David and {Seibert}, Mark and {Mallery}, Ryan and {Heckman}, Timothy M. and {Forster}, Karl and {Friedman}, Peter G. and {Martin}, D. Christopher and {Morrissey}, Patrick and {Neff}, Susan G. and {Small}, Todd and {Wyder}, Ted K. and {Bianchi}, Luciana and {Donas}, Jos{\'e} and {Lee}, Young-Wook and {Madore}, Barry F. and {Milliard}, Bruno and {Szalay}, Alex S. and {Welsh}, Barry Y. and {Yi}, Sukyoung K.},
        title = "{UV Star Formation Rates in the Local Universe}",
      journal = {\apjs},
         year = 2007,
        month = dec,
       volume = {173},
       number = {2},
        pages = {267-292},
          doi = {10.1086/519218},
archivePrefix = {arXiv},
       eprint = {0704.3611},
 primaryClass = {astro-ph},
       adsurl = {https://ui.adsabs.harvard.edu/abs/2007ApJS..173..267S}
}

@ARTICLE{1998ApJ...500..525Schlegel,
       author = {{Schlegel}, David J. and {Finkbeiner}, Douglas P. and {Davis}, Marc},
        title = "{Maps of Dust Infrared Emission for Use in Estimation of Reddening and Cosmic Microwave Background Radiation Foregrounds}",
      journal = {\apj},
         year = 1998,
        month = jun,
       volume = {500},
       number = {2},
        pages = {525-553},
          doi = {10.1086/305772},
archivePrefix = {arXiv},
       eprint = {astro-ph/9710327},
 primaryClass = {astro-ph},
       adsurl = {https://ui.adsabs.harvard.edu/abs/1998ApJ...500..525S}
}

@ARTICLE{1955ApJ...121..161Salpeter,
       author = {{Salpeter}, Edwin E.},
        title = "{The Luminosity Function and Stellar Evolution.}",
      journal = {\apj},
         year = 1955,
        month = jan,
       volume = {121},
        pages = {161},
          doi = {10.1086/145971},
       adsurl = {https://ui.adsabs.harvard.edu/abs/1955ApJ...121..161S}
}

@ARTICLE{2021arXiv211207639Seebeck,
       author = {{Seebeck}, Jerome and {Ravi}, Vikram and {Connor}, Liam and {Law}, Casey and {Simard}, Dana and {Uzgil}, Bade},
        title = "{The Effects of Selection Biases on the Analysis of Localised Fast Radio Bursts}",
      journal = {arXiv e-prints},
         year = 2021,
        month = dec,
          eid = {arXiv:2112.07639},
        pages = {arXiv:2112.07639},
          doi = {10.48550/arXiv.2112.07639},
archivePrefix = {arXiv},
       eprint = {2112.07639},
 primaryClass = {astro-ph.HE},
       adsurl = {https://ui.adsabs.harvard.edu/abs/2021arXiv211207639S}
}

@ARTICLE{2025ApJ...979L..21Shah,
       author = {{Shah}, Vishwangi and {Shin}, Kaitlyn and {Leung}, Calvin and {Fong}, Wen-fai and {Eftekhari}, Tarraneh and {Amiri}, Mandana and {Andersen}, Bridget C. and {Andrew}, Shion and {Bhardwaj}, Mohit and {Brar}, Charanjot and {Cassanelli}, Tomas and {Chatterjee}, Shami and {Curtin}, Alice and {Dobbs}, Matt and {Dong}, Yuxin and {Dong}, Fengqiu Adam and {Fonseca}, Emmanuel and {Gaensler}, B.~M. and {Halpern}, Mark and {Hessels}, Jason W.~T. and {Ibik}, Adaeze L. and {Jain}, Naman and {Joseph}, Ronniy C. and {Kaczmarek}, Jane and {Kahinga}, Lordrick A. and {Kaspi}, Victoria M. and {Kharel}, Bikash and {Landecker}, Tom and {Lanman}, Adam E. and {Lazda}, Mattias and {Main}, Robert and {Mas-Ribas}, Lluis and {Masui}, Kiyoshi W. and {Mckinven}, Ryan and {Mena-Parra}, Juan and {Meyers}, Bradley W. and {Michilli}, Daniele and {Nimmo}, Kenzie and {Pandhi}, Ayush and {Patil}, Swarali Shivraj and {Pearlman}, Aaron B. and {Pleunis}, Ziggy and {Prochaska}, J. Xavier and {Rafiei-Ravandi}, Masoud and {Sammons}, Mawson and {Sand}, Ketan R. and {Scholz}, Paul and {Smith}, Kendrick and {Stairs}, Ingrid},
        title = "{A Repeating Fast Radio Burst Source in the Outskirts of a Quiescent Galaxy}",
      journal = {\apjl},
         year = 2025,
        month = feb,
       volume = {979},
       number = {2},
          eid = {L21},
        pages = {L21},
          doi = {10.3847/2041-8213/ad9ddc},
archivePrefix = {arXiv},
       eprint = {2410.23374},
 primaryClass = {astro-ph.HE},
       adsurl = {https://ui.adsabs.harvard.edu/abs/2025ApJ...979L..21S}
}

@ARTICLE{2021A&A...648A...6Smith,
       author = {{Smith}, D.~J.~B. and {Haskell}, P. and {G{\"u}rkan}, G. and {Best}, P.~N. and {Hardcastle}, M.~J. and {Kondapally}, R. and {Williams}, W. and {Duncan}, K.~J. and {Cochrane}, R.~K. and {McCheyne}, I. and {R{\"o}ttgering}, H.~J.~A. and {Sabater}, J. and {Shimwell}, T.~W. and {Tasse}, C. and {Bonato}, M. and {Bondi}, M. and {Jarvis}, M.~J. and {Leslie}, S.~K. and {Prandoni}, I. and {Wang}, L.},
        title = "{The LOFAR Two-metre Sky Survey Deep Fields. The star-formation rate-radio luminosity relation at low frequencies}",
      journal = {\aap},
         year = 2021,
        month = apr,
       volume = {648},
          eid = {A6},
        pages = {A6},
          doi = {10.1051/0004-6361/202039343},
archivePrefix = {arXiv},
       eprint = {2011.08196},
 primaryClass = {astro-ph.GA},
       adsurl = {https://ui.adsabs.harvard.edu/abs/2021A&A...648A...6S}
}

@ARTICLE{2024Natur.635...61Sharma,
       author = {{Sharma}, Kritti and {Ravi}, Vikram and {Connor}, Liam and {Law}, Casey and {Ocker}, Stella Koch and {Sherman}, Myles and {Kosogorov}, Nikita and {Faber}, Jakob and {Hallinan}, Gregg and {Harnach}, Charlie and {Hellbourg}, Greg and {Hobbs}, Rick and {Hodge}, David and {Hodges}, Mark and {Lamb}, James and {Rasmussen}, Paul and {Somalwar}, Jean and {Weinreb}, Sander and {Woody}, David and {Leja}, Joel and {Anand}, Shreya and {Das}, Kaustav Kashyap and {Qin}, Yu-Jing and {Rose}, Sam and {Dong}, Dillon Z. and {Miller}, Jessie and {Yao}, Yuhan},
        title = "{Preferential occurrence of fast radio bursts in massive star-forming galaxies}",
      journal = {\nat},
         year = 2024,
        month = nov,
       volume = {635},
       number = {8037},
        pages = {61-66},
          doi = {10.1038/s41586-024-08074-9},
archivePrefix = {arXiv},
       eprint = {2409.16964},
 primaryClass = {astro-ph.HE},
       adsurl = {https://ui.adsabs.harvard.edu/abs/2024Natur.635...61S}
}

@ARTICLE{2006AJ....131.1163S_2mass,
       author = {{Skrutskie}, M.~F. and {Cutri}, R.~M. and {Stiening}, R. and {Weinberg}, M.~D. and {Schneider}, S. and {Carpenter}, J.~M. and {Beichman}, C. and {Capps}, R. and {Chester}, T. and {Elias}, J. and {Huchra}, J. and {Liebert}, J. and {Lonsdale}, C. and {Monet}, D.~G. and {Price}, S. and {Seitzer}, P. and {Jarrett}, T. and {Kirkpatrick}, J.~D. and {Gizis}, J.~E. and {Howard}, E. and {Evans}, T. and {Fowler}, J. and {Fullmer}, L. and {Hurt}, R. and {Light}, R. and {Kopan}, E.~L. and {Marsh}, K.~A. and {McCallon}, H.~L. and {Tam}, R. and {Van Dyk}, S. and {Wheelock}, S.},
        title = "{The Two Micron All Sky Survey (2MASS)}",
      journal = {\aj},
         year = 2006,
        month = feb,
       volume = {131},
       number = {2},
        pages = {1163-1183},
          doi = {10.1086/498708},
       adsurl = {https://ui.adsabs.harvard.edu/abs/2006AJ....131.1163S}
}

@ARTICLE{2014MNRAS.440..889Schawinski,
       author = {{Schawinski}, Kevin and {Urry}, C. Megan and {Simmons}, Brooke D. and {Fortson}, Lucy and {Kaviraj}, Sugata and {Keel}, William C. and {Lintott}, Chris J. and {Masters}, Karen L. and {Nichol}, Robert C. and {Sarzi}, Marc and {Skibba}, Ramin and {Treister}, Ezequiel and {Willett}, Kyle W. and {Wong}, O. Ivy and {Yi}, Sukyoung K.},
        title = "{The green valley is a red herring: Galaxy Zoo reveals two evolutionary pathways towards quenching of star formation in early- and late-type galaxies}",
      journal = {\mnras},
         year = 2014,
        month = may,
       volume = {440},
       number = {1},
        pages = {889-907},
          doi = {10.1093/mnras/stu327},
archivePrefix = {arXiv},
       eprint = {1402.4814},
 primaryClass = {astro-ph.GA},
       adsurl = {https://ui.adsabs.harvard.edu/abs/2014MNRAS.440..889S}
}

@ARTICLE{2021PhRvD.103h3017Sudoh,
       author = {{Sudoh}, Takahiro and {Linden}, Tim and {Beacom}, John F.},
        title = "{Millisecond pulsars modify the radio-star-formation-rate correlation in quiescent galaxies}",
      journal = {\prd},
         year = 2021,
        month = apr,
       volume = {103},
       number = {8},
          eid = {083017},
        pages = {083017},
          doi = {10.1103/PhysRevD.103.083017},
archivePrefix = {arXiv},
       eprint = {2005.08982},
 primaryClass = {astro-ph.GA},
       adsurl = {https://ui.adsabs.harvard.edu/abs/2021PhRvD.103h3017S}
}

@software{larry_bradley_2023_7946442,
  author       = {Larry Bradley},
  title        = {astropy/photutils: 1.8.0},
  month        = may,
  year         = 2023,
  publisher    = {Zenodo},
  version      = {1.8.0},
  doi          = {10.5281/zenodo.7946442},
  url          = {https://doi.org/10.5281/zenodo.7946442},
}



\appendix

\section{SED Fitting Data and Results}
\label{sec:appendixa}

We compile multi-wavelength photometry from several surveys. GALEX ultraviolet fluxes are from the AIS catalogue \citep{2011Ap&SS.335..161B_galex}, using the calibrated pipeline fluxes. Optical measurements are obtained from SDSS DR16 \citep{2020ApJS..249....3A_sdss16}, adopting the cModel fluxes.
Near-infrared photometry is taken from 2MASS \citep{2006AJ....131.1163S_2mass}, using the fluxes derived from the extrapolated profile fits. 
WISE mid-infrared fluxes are taken from the AllWISE catalogue \citep{2013wise.rept....1Cutri_wise}, using values derived from fitting to the imaging maps.
Far-infrared measurements from Herschel \citep{2024yCat.8112....0Herschel}, correspond to the average of the TML-derived 250 and 350 $\rm \mu m$ fluxes over all contributing sources.  All data in the table have been corrected for extinction. The photometric measurements originate from surveys with different angular resolutions and photometric methods, and therefore the apertures are not matched across bands. However, the observed spectral energy distribution shows no discontinuities or unphysical jumps, indicating that the fluxes are mutually consistent and can be reasonably compared.

This appendix presents the SED fitting result of the target galaxy using the \texttt{CIGALE} software. The figure below shows the best-fit SED model compared with the multi-wavelength photometric observations.

\begin{table}
\centering
\caption{Data used in the SED fitting for SDSS J114335.11+600334.9}
\label{tab:seddata1}
\begin{adjustbox}{width=0.5\textwidth}
\begin{tabular}{llcccccc}
\toprule
Instrument &Filter   & Effective Wavelength (\AA)&  Flux$^{\dagger}$ (mJy)     \\
\midrule
SDSS& $u$-band& 3546&0.0019$\pm$0.0015\\ 
& $g$-band& 4670 &0.0040$\pm$0.0007 \\ 
& $r$-band& 6156 &0.0063$\pm$0.0010 \\ 
& $i$-band& 7472 &0.0062$\pm$0.0016 \\ 
& $z$-band& 8917  &0.0079$\pm$0.0062 \\ 
WISE & W1 & 33461 &0.0070$\pm$0.0018 \\ 
& W2 & 45952 &0.0062$\pm$0.0036\\

\bottomrule
\end{tabular}
\end{adjustbox}
\vspace{2pt}
\raggedright
\footnotesize{
$^{\dagger}$ All fluxes have been corrected for Galactic extinction.}
\end{table}

\begin{table}
\centering
\caption{Data used in the SED fitting for SDSS J083444.42+660026.5}
\label{tab:seddata2}
\begin{adjustbox}{width=0.5\textwidth}
\begin{tabular}{llcccccc}
\toprule
Instrument &Filter& Effective Wavelength (\AA)& Flux$^{\dagger}$ (mJy)\\
\midrule
GALEX & FUV & 1549& 0.2306$\pm$0.0140\\
&NUV&2304  & 0.3197$\pm$0.0110\\
SDSS & $u$-band& 3546&0.2787$\pm$0.0064\\
& $g$-band& 4670 &1.0928$\pm$0.0010\\
& $r$-band& 6156 &1.9283$\pm$0.0010\\
& $i$-band& 7472 &2.6225$\pm$0.0010\\
& $z$-band& 8917 &3.2575$\pm$0.0283\\
2MASS & $J$ & 12319 & 2.8348$\pm$0.0750\\
& $H$ &16420 & 3.7904$\pm$0.1118\\
& $Ks$ &21567& 2.9201$\pm$0.1005\\
WISE & W1&33461  & 1.1981$\pm$0.0274\\ 
& W2 &45952 & 0.7822$\pm$0.0215\\
& W3 &115526  & 3.9434$\pm$0.1526\\
& W4 &220783& 6.9022$\pm$0.9721\\
Herschel/ & PMW &2500000 & 68.2003$\pm$11.5000\\
SPIRE&PSW &3630000 &126.7008$\pm$9.9000\\

\bottomrule
\end{tabular}
\end{adjustbox}
\raggedright
\footnotesize{
$^{\dagger}$ All fluxes have been corrected for Galactic extinction.}
\end{table}

\begin{table}
\centering
\caption{Data used in the SED fitting for SDSS J110020.55+595454.4}
\label{tab:seddata3}
\begin{adjustbox}{width=0.5\textwidth}
\begin{tabular}{llcccccc}
\toprule
Instrument &Filter& Effective Wavelength (\AA)& Flux$^{\dagger}$ (mJy)\\
\midrule
SDSS & $u$-band& 3546 &0.0010$\pm$0.0016\\
& $g$-band & 4670 & 0.0225$\pm$0.0007\\
& $r$-band& 6156 & 0.0511$\pm$0.0010\\
& $i$-band& 7472 &0.0757$\pm$0.0019\\
& $z$-band& 8917 & 0.1039$\pm$0.0064\\
WISE & W1 &33461 & 0.1184$\pm$0.0052\\ 
& W2 & 45952 &0.0844$\pm$0.0089\\
& W3 & 115526 & 0.4998$\pm$0.1091\\
& W4 & 220783& $<2.3102$\\

\bottomrule
\end{tabular}
\end{adjustbox}
\raggedright
\footnotesize{
$^{\dagger}$ All fluxes have been corrected for Galactic extinction.}
\end{table}

\begin{figure}
    \centering
    \includegraphics[width=0.5\textwidth]{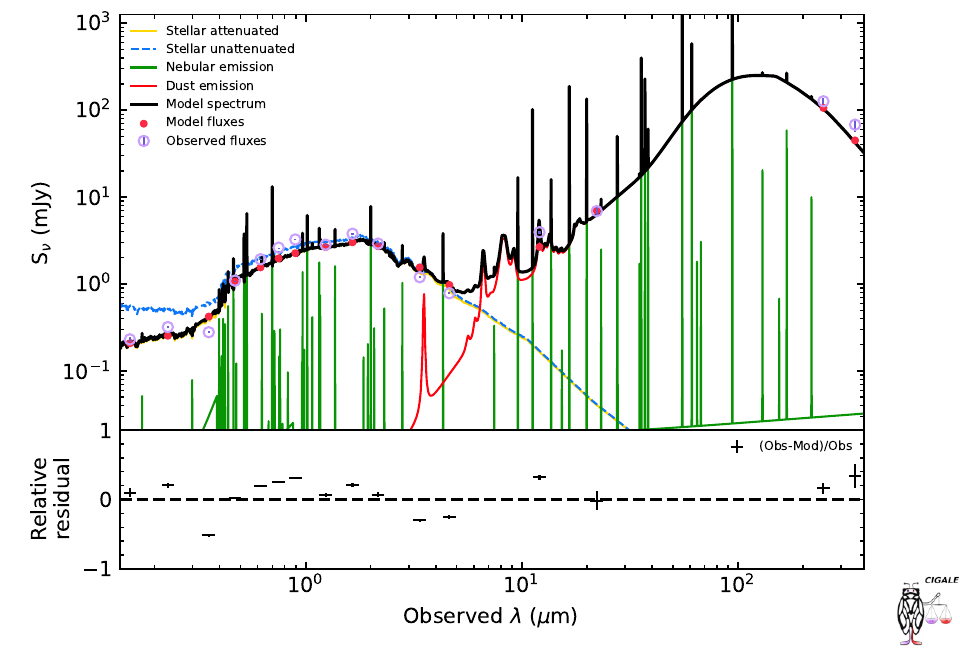} 
    \caption{Spectral Energy Distribution (SED) fitting result for the source MCG+11-11-014 using \texttt{CIGALE}. The purple circles represent observed fluxes, the black curve shows the best-fit model, and the red points denotes the model flux.}
    \label{fig:sed_fit_2}
\end{figure}

\begin{figure}
    \centering
    \includegraphics[width=0.5\textwidth]{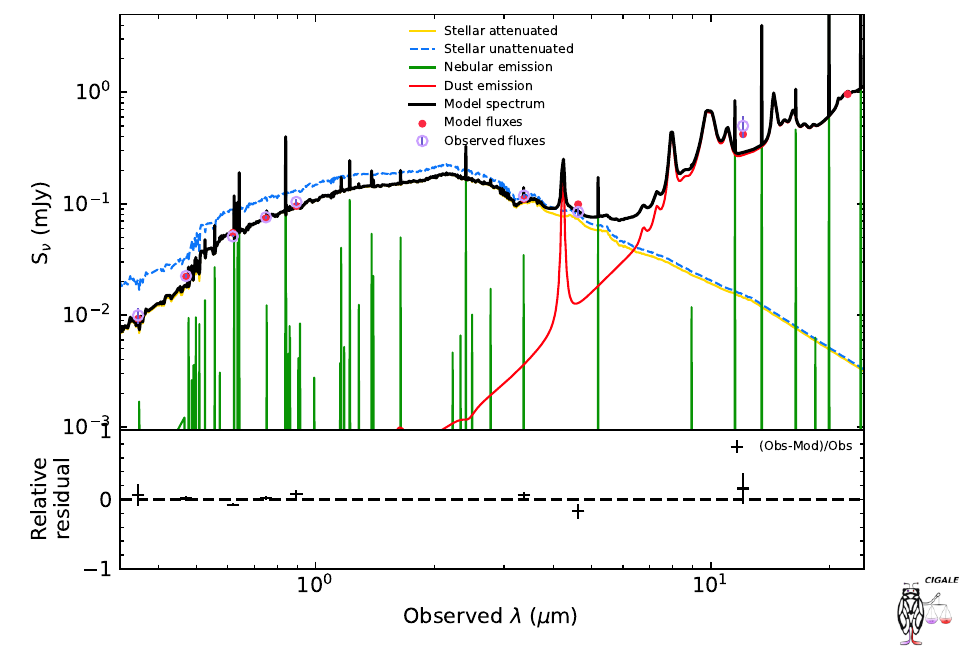} 
    \caption{Spectral Energy Distribution (SED) fitting result for the source SDSS J110020.55+595454.4 using \texttt{CIGALE}.}
    \label{fig:sed_fit_3}
\end{figure}

\begin{figure}
    \centering
    \includegraphics[width=0.5\textwidth]{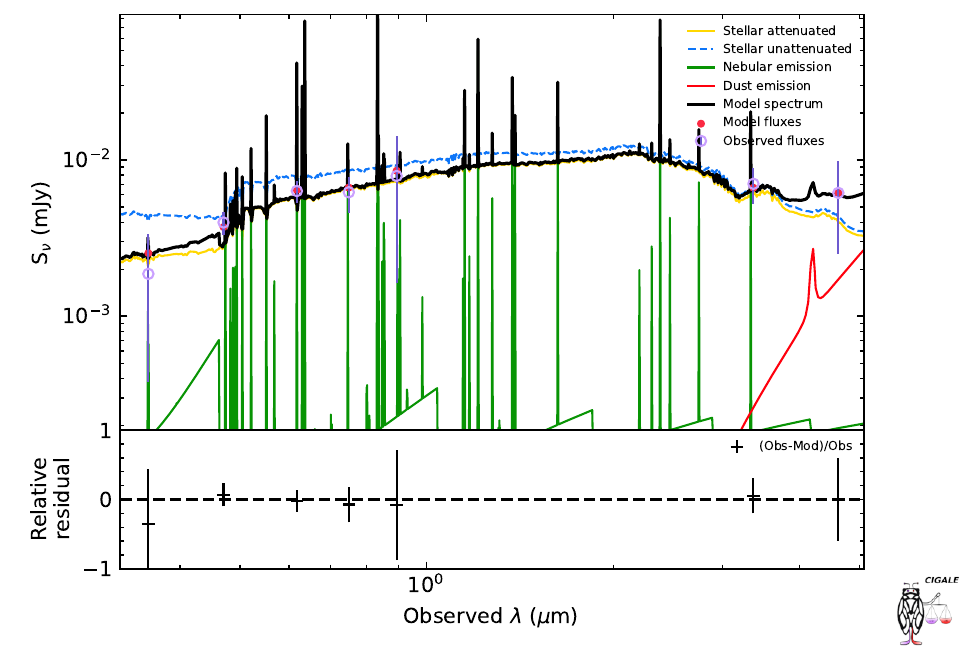} %
    \caption{Spectral Energy Distribution (SED) fitting result for the source SDSS J114335.11+600334.9 using \texttt{CIGALE}.}
    \label{fig:sed_fit_1}
\end{figure}

\section{Spectroscopic Analysis Results}

We obtained an optical spectrum of the candidate host galaxy of FRB~20190303B, MCG+11-11-014, using the SPRAT instrument on the Liverpool Telescope. The spectrum shows a clear detection of the H$\alpha$ emission line, which we fitted with \texttt{pPXF} \citep{2012ascl.soft10002Cappellari} to estimate its flux.

\begin{figure}
    \centering
    \includegraphics[width=0.5\textwidth]{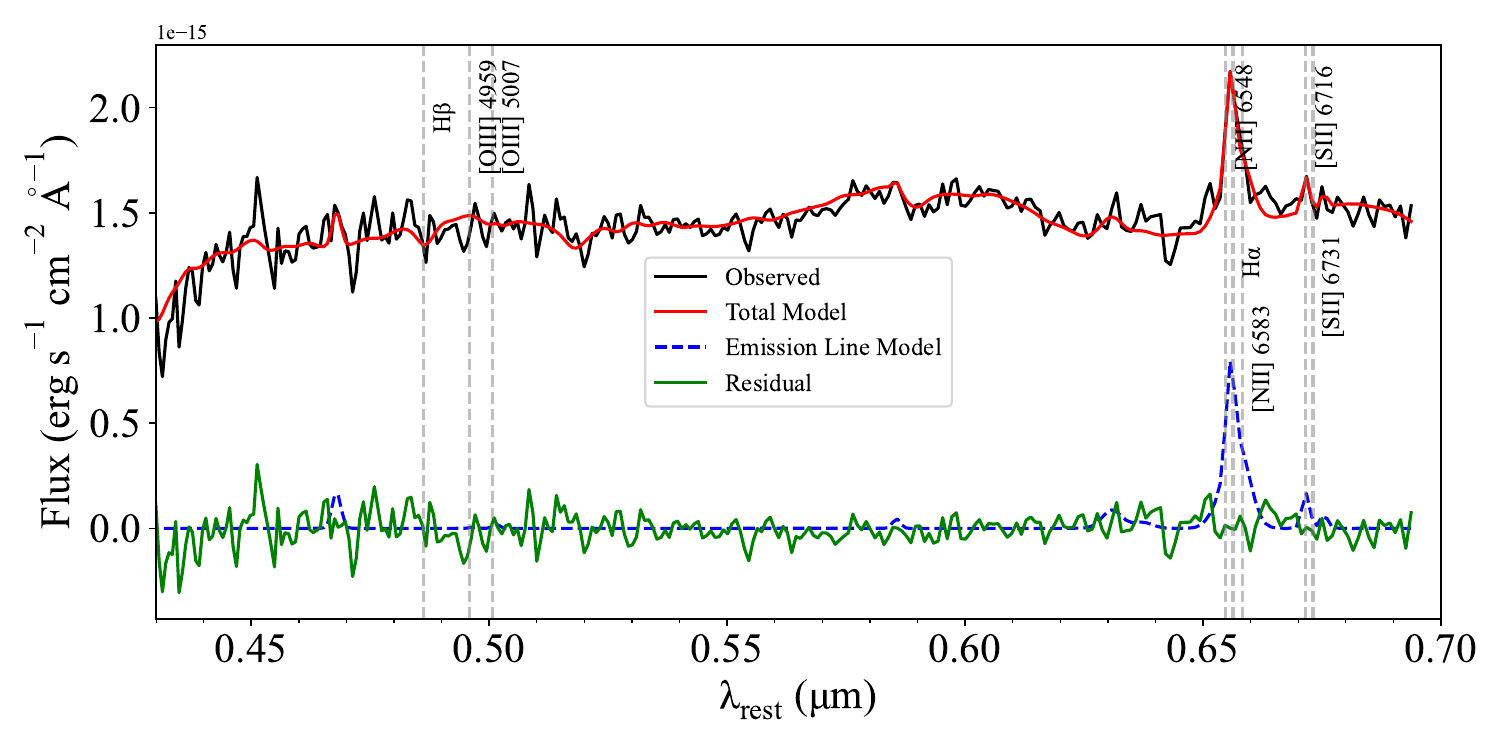}
    \caption{Optical spectrum obtained from the SPRAT instrument on the Liverpool Telescope of the host galaxy candidate MCG+11-11-014 for FRB 20190303B showing a clear detection of the H$\alpha$ emission line. The line is fitted with a single Gaussian profile to estimate its flux, which is later used to derive the star formation rate.
    }
    \label{fig:ppxf}
\end{figure}


\bsp	
\label{lastpage}
\end{document}